\documentclass[twocolumn]{aastex63}
\usepackage{amsmath}

\newcommand{\RomanNumeral}[1]{\scshape{\romannumeral #1}}

\graphicspath{{./}{figures/}}

\received{........}
\revised{........}
\accepted{........}

\submitjournal{ApJ Supplement Series}

\shorttitle{Light curves of core collapse supernovae}
\shortauthors{Limongi and Chieffi}

\begin{document}

\title{Hydrodynamical modeling of the light curves of core collapse supernovae with HYPERION. I. The mass range $\rm 13-25~M_\odot$, the metallicities $\rm -3\leq [Fe/H]\leq 0$ and the case of SN1999em}

\author[0000-0003-0636-7834]{Marco Limongi}
\affiliation{INAF - Osservatorio Astronomico di Roma Via Frascati 33, I-00040, Monteporzio Catone, Italy}
\affiliation{Kavli Institute for the Physics and Mathematics of the Universe Todai Institutes for Advanced Study \\ the University of Tokyo, Kashiwa, Japan 277-8583 (Kavli IPMU, WPI)} \affiliation{INFN. Sezione di Perugia, via A. Pascoli s/n, 06125 Perugia, Italy}

\author[0000-0002-3589-3203]{Alessandro Chieffi}
\affiliation{INAF - Istituto di Astrofisica e Planetologia Spaziali, Via Fosso del Cavaliere 100, I-00133, Roma, Italy}
\affiliation{Monash Centre for Astrophysics (MoCA),
School of Mathematical Sciences, Monash University, Victoria 3800, Australia}
\affiliation{INFN. Sezione di Perugia, via A. Pascoli s/n, 06125 Perugia, Italy}

\correspondingauthor{Marco Limongi}
\email{marco.limongi@inaf.it}

\begin{abstract}
We present the last version of {\scshape{Hyperion}} (HYdrodynamic Ppm Explosion with Radiation diffusION), a hydrodynamic code designed to calculate the explosive nucleosynthesis, remnant mass and light curve associated to the explosion of a massive star. By means of this code we compute the explosion of a subset of red supergiant models, taken from the database published by \cite{lc18}, for various explosion energies in the range $\rm \sim 0.20-2.00~10^{51}~erg$. 
The main outcomes of these simulations, i.e., remnant mass, $\rm ^{56}Ni$ synthesized, luminosity and length of the plateau of the bolometric light curve, are analyzed as a function of the initial parameters of the star (mass and metallicity) and of the explosion energy. As a first application of {\scshape{Hyperion}} we estimated the mass and the metallicity of the progenitor star of SN 1999em, a well studied SN IIP, by means of the light curve fitting. In particular, if the adopted distance to the host galaxy NGC 1637 is $\rm 7.83~Mpc$, the properties of the light curve point toward a progenitor with an initial mass of $\rm 13~M_\odot$ and a metallicity [Fe/H]=-1. If, on the contrary, the adopted distance modulus is $\rm 11.7~Mpc$, all the models with initial mass $13\leq M/M_\odot\leq 15$ and metallicities $\rm -1\leq [Fe/H] \leq 0$ are compatible with the progenitor of SN 1999em.
\end{abstract}


\section{Introduction} \label{sec:intro}
Type II supernovae are the endpoint of the evolution of massive stars that retain an H rich envelope. Depending on the morphology of their associated light curve (LC), they are generally classified into two broad classes: SNII-Plateau (or SNe IIP), that show a "plateau" phase lasting typically $\sim 100$ days where the optical luminosity remains almost constant, and SNII-Linear (or SN IIL) that, on the contrary, show a linear decline of the luminosity after the maximum light. Since the mass of the H-rich envelope is the main responsible of the length of the plateau phase \citep{grassberg+71,falkarnett77}, it has been recently proposed that the transition from the SNe IIP to the SNe IIL is a continuous process that depends on the mass size of the H-rich envelope, rather than the result of the evolution of two distinct categories of TypeII SNe \citep{anderson+14}.

The light curves of the SNe IIP are sistematically studied for a number of reasons among which: a) they have been proposed as distance indicators \citep{kirshner+74,eastman+96,jones+09} with possible use for cosmology, similar to the Type Ia SNe, once their basic properties and empirical correlations are known \citep{chieffi+03,nugent+06,poznanski+10,maguire+10}; (b) the comparison between the theoretical light curves and the observed ones allows to derive information on the properties of the progenitor stars \citep{tomasella18,tomasella13,utrobin07,bersten11,martinez+19}, in particular the initial mass and radius. Within the last context, it has been found in the literature, the existence of a tension between the masses and radii derived from the light curve fitting and those obtained from the analysis of the archival images acquired prior to the supernova explosion \citep{davies+18,martinez+19}. In general, the masses estimated from the fitting of the light curve are larger than those determined from the analysis of the archival images \citep{utrobin+08,utrobin+09,maguire+10,morozova+18}. However, in a recent paper, \cite{martinez+19} found that, for a number of SNe IIP, the masses determined from their hydrodynamical modeling are not sistematically larger than those previously found in literature. As a result, the existence or not of this tension is still debated. Studies on this subject are ongoing and new developments on both the detection of presupernova progenitors as well as light curve modelling are continuously achieved.

From the theoretical side there are a number of codes, more or less sophisticated, that are currently used to compute the theoretical light curve of a SN IIP. Most of them use as starting models a polytrope, or adopt some kind of parametric procedure \citep{baklanov+05, utrobin07, bersten11, pumo+11, martinez+19}. In this way the various properties of the progenitor star (like, e.g., the total mass, the envelope mass, the radius and so on) are assumed as free parameters that may be varied in an independent way. Others codes, on the contrary, follow a more auto consistent approach since they adopt as starting model the one that has passed through the whole presupernova evolution. This obviously means that the various properties of the progenitor star are not free parameters but the result of the presupernova evolution that, in turn, depends on the initial mass, metallicity and rotation velocity \citep{chieffi+03, SNEC, gengis+16, utrobin17, paxton+18, dessart+19, morozova+20}. Note that in the majority of the above mentioned studies, the explosive nucleosynthesis is not taken account, and the amount of $\rm ^{56}Ni$, that powers the light curve starting from the plateau phase until the radioactive tail, is assumed as a free parameter and deposited by hand in the progenitor model.

This paper is part of the series of works devoted to the study of the presupernova evolution, explosion and nucleosynthesis of massive stars \cite{cls98, lsc00, lc03, cl04, lc06, lc12, cl13, cl17,lc18}. In these works a great effort has been devoted to the predictions of the chemical composition of the ejecta after the supernova explosion. Since the explosive nucleosynthesis plays a crucial role for the determination of the abundance of most of the isotopes in the ejecta, we developed, in the course of the years, a hydro code capable to simulate the ejection of the mantle of a massive star due to the explosion and to compute simultaneously the explosive nucleosynthesis. Because of the rapid rise and fall of the temperature during the explosion and because of the high dependence of the cross sections on the temperature, the explosive nuclesynthesis occurs within the first few (1-2) seconds after the core bounce. For this reason, the adoption of the adiabatic approximation is well suited to follow the explosive nucleosynthesis.

In this paper we present the latest version of this hydro code, that is now named {\scshape{Hyperion}} (HYdrodynamic Ppm Explosion with Radiation diffusION). The most important upgrade of this code is the inclusion of the treatment of the radiation transport in the flux limited diffusion approximation. This makes this new version of the code well suited for the calculation of the bolometric light curves of core collapse supernovae, as well as the explosive nucleosynthesis and remnant mass determination. We use {\scshape{Hyperion}} to compute the explosions of a subset of models taken from \cite{lc18} that explode as red supergiants with a H-rich envelope. In particular, we consider the mass range $\rm 13-25~M_\odot$ and the initial metallicities [Fe/H]=0, -1, -2 and -3. In this way we derive the main properties of the light curve (luminosity and length of the plateau, radioactive tail, transition phase and so on) and the nature of the remnant mass as a function of the properties of the progenitor star (initial mass and metallicity) and of the explosion energy. Finally, as a possible application of {\scshape{Hyperion}} we fit the observed bolometric light curve of SN 1999em, a well studied SN IIP, in order to derive the basic properties of its progenitor star.

\section{The Code} \label{sec:code}
In this section we describe in detail the construction and the implementation of {\scshape{Hyperion}}.

The full system of the hydrodynamic equations (written in conservative form), supplemented by the radiative diffusion and by the equations describing the temporal variation of the chemical composition due to the nuclear reactions are written as:

\begin{eqnarray}
\label{fullsystem1}
\displaystyle{{\frac{\partial \rho}{\partial t}}}&=&\displaystyle{-{4 \pi \rho^2}{\frac{\partial r^2 v}{\partial m}}}  \\
 & & \nonumber \\
\label{fullsystem2} 
\displaystyle{{\partial v \over \partial t}} &=&\displaystyle{- {A}{\partial P \over \partial m}-{Gm \over r^2}} \\
 & & \nonumber \\
\label{fullsystem3} 
\displaystyle{\frac{\partial E}{\partial t}}&=&\displaystyle{-{\partial  \over \partial m}\left(A v P +L\right)+\epsilon} \\
 & & \nonumber \\
\label{fullsystem4} 
{\partial Y_i \over \partial t}&=&\sum_j c_i(j)\Lambda_j Y_j \nonumber \\
&+& \sum_{j,k} c_i(j,k)\rho N_A \langle \sigma v \rangle_{j,k}Y_j Y_k \nonumber \\
&+& \sum_{j,k,l} c_i(j,k,l) \rho^2 N_{A}^2 \langle \sigma v \rangle_{j,k,l}Y_j Y_k Y_l \nonumber \\
& &~~~i=1,....N
\end{eqnarray}

where $\rho$ is the density, $r$ is the radius, $v$ is the velocity, $m$ is the mass, $P$ is the pressure, $A=4 \pi r^2$,
$E$ is the total energy per unit mass (including the kinetic, internal and gravitational ones), $L$ is the radiative luminosity and $\epsilon$ is any source and/or sink of energy (e.g., nuclear energy production, neutrino losses, and so on). In the last set of $N$ equations, $N$ is the number of nuclear species followed in detail in the calculations, $Y_i$ is the abundance by number of the $i$-th nuclear species. The different terms in these equations refer to (1) $\beta$-decays, electron captures and photodisintegrations, (2) two-body reactions and (3) three-body reactions. The coefficients $c_i$ are given by $c_i(j)=\pm N_i$, $c_i(j,k)=\pm N_i/(N_j! N_k!)$, $c_i(j,k,l)=\pm N_i/(N_j! N_k! N_l!)$, where $N_i$ refers to the number of particles $i$ involved in the reaction, and $N_i$! prevents double counting for reactions involving identical particles. The sign depends on whether the particle $i$ is produced $(+)$ or destroyed $(-)$. $\Lambda$ refers to the weak interaction or the photodisintegration rate, while $\langle {\sigma v} \rangle$ refers to the two- or three-body nuclear cross section. The nuclear network adopted in these calculations includes 335 isotopes (from neutrons to $\rm ^{209}Bi$ (see Table \ref{tab:network}) linked by more than 3000 nuclear reactions.

\begin{deluxetable}{lrrlrr}
\tablewidth{0pt}
\tablecaption{Nuclear network adopted in the present calculations\label{tab:network}}
\tablehead{
\colhead{Element} & \colhead{$\rm A_{\rm min}$} & \colhead{$\rm A_{\rm max}$} &
\colhead{Element} & \colhead{$\rm A_{\rm min}$} & \colhead{$\rm A_{\rm max}$}
}
\startdata
n........  &   1   &   1   & Co.......  &  54   &  61   \\  
H........  &   1   &   3   & Ni.......  &  56   &  65   \\  
He.......  &   3   &   4   & Cu.......  &  57   &  66   \\  
Li.......  &   6   &   7   & Zn.......  &  60   &  71   \\  
Be.......  &   7   &  10   & Ga.......  &  62   &  72   \\  
B........  &  10   &  11   & Ge.......  &  64   &  77   \\
C........  &  12   &  14   & As.......  &  71   &  77   \\
N........  &  13   &  16   & Se.......  &  74   &  83   \\
O........  &  15   &  19   & Br.......  &  75   &  83   \\
F........  &  17   &  20   & Kr.......  &  78   &  87   \\
Ne.......  &  20   &  23   & Rb.......  &  79   &  88   \\
Na.......  &  21   &  24   & Sr.......  &  84   &  91   \\
Mg.......  &  23   &  27   & Y........  &  85   &  91   \\
Al.......  &  25   &  28   & Zr.......  &  90   &  97   \\
Si.......  &  27   &  32   & Nb.......  &  91   &  97   \\
P........  &  29   &  34   & Mo.......  &  92   &  98   \\
S........  &  31   &  37   & Xe.......  & 132   & 135   \\
Cl.......  &  33   &  38   & Cs.......  & 133   & 138   \\
Ar.......  &  36   &  41   & Ba.......  & 134   & 139   \\
K........  &  37   &  42   & La.......  & 138   & 140   \\
Ca.......  &  40   &  49   & Ce.......  & 140   & 141   \\
Sc.......  &  41   &  49   & Pr.......  & 141   & 142   \\
Ti.......  &  44   &  51   & Nd.......  & 142   & 144   \\
V........  &  45   &  52   & Hg.......  & 202   & 205   \\
Cr.......  &  48   &  55   & Tl.......  & 203   & 206   \\
Mn.......  &  50   &  57   & Pb.......  & 204   & 209   \\
Fe.......  &  52   &  61   & Bi.......  & 208   & 209   \\
\enddata                                              
 \end{deluxetable}       
 
The nuclear cross sections and the weak interactions rates are the ones adopted in \citet{lc18} (see their Tables 3 and 4).

In the diffusion approximation, the radiative luminosity is given by:

\begin{equation}
\label{eq:lumdiffrad}
L=-(4\pi r^2)^2 \frac{\lambda ac}{3 \kappa}\frac{\partial T^4}{\partial m} 
\end{equation}

where $a$ is the radiation constant, $c$ is the speed of the light, $\kappa$ is the Rosseland mean opacity and $\lambda$ is the flux limiter. For this last quantity we use the expression provided by \cite{LP81}: 

\begin{equation}\label{fluxlim1}
\lambda=\frac{6+3R}{6+3R+R^2}
\end{equation}

where

\begin{equation}\label{fluxlim2}
R=\frac{4\pi r^2}{kT^4}\bigg\vert\frac{\partial T^4}{\partial m}\bigg\vert
\end{equation}

The Rosseland mean opacities are calculated assuming a scaled solar distribution of all the elements, which for the solar metallicity corresponds to $Z=1.345\cdot 10^{-2}$ according to \citet{Asplund09}. At metallicities lower than solar ([Fe/H]=0) we consider an enhancement with respect to Fe of the elements C, O, Mg, Si, S, Ar, Ca and Ti, which is derived from the observations \citep{cayrel+04,spite+05}. As a result of these enhancements the total metallicity corresponding to [Fe/H]=-1, -2 and -3 is, $Z=3.236\cdot 10^{-3}$, $Z=3.236\cdot 10^{-4}$ and $Z=3.236\cdot 10^{-5}$, respectively. 
For the opacity tables we use three different sources: in the low temperature regime ($2.75<{\rm Log}T<4.5$) we use the tables of \cite{Fetal05} while in the intermediate temperature regime ($4.5<{\rm Log}T<8.7$) we adopt the OPAL tables \cite{OPAL}. In the high temperature regime ($8.7<{\rm Log}T<10.0$) we use the Los Alamos Opacity Library \citep{losalamos}. Although they are negligible for these calculations, let us mention, for the sake of completeness, that the opacity coefficients due to the thermal conductivity are derived from \citet{Itoh83}. The opacity floor has been computed according to \cite{SNEC}.

The equation of state (EOS) adopted is the same as described in \cite{SNEC}. It is based on the analytic EOS provided by \citet{Pac83}, that takes into account radiation, ions and electrons in an arbitrary (approximated) degree of degeneracy. We account for the H and He recombination by solving the Saha equations as proposed by \citet{Zagh00} and assume all other elements fully ionized.

The nuclear energy generation due to the nuclear reactions has been neglected, in the assumption that this is negligible compared to the other energy components. The energy deposition due to the $\gamma$-rays emitted by the radioactive decays $\rm ^{56}Ni \rightarrow  ^{56}Co \rightarrow ^{56}Fe$, on the contrary, is taken into account following the scheme proposed by \citet{Swartz95} and \cite{SNEC}. 

The hydrodynamic equations \ref{fullsystem1}, \ref{fullsystem2}, \ref{fullsystem3} are solved by means of the fully Lagrangian scheme of the Piecewise Parabolic Method described by \cite{cw84}. This is done in the following three steps: (1) first, we interpolate the profiles of the variables $\rho$, $v$ and $P$ as a function of the mass coordinate by means of the interpolation algorithm described in \cite{cw84}; (2) then, we solve appropriate Riemann problems at the cell interfaces in order to calculate the time-averaged values of the pressure and the velocity at the zone edges; (3) finally, we update the conserved quantities by applying the forces due to the time-averaged pressures and velocities at zone edges. In the following we will describe the step 3 in detail.

Let us assume that $\bar{v}_{j+1/2}$ and $\bar{P}_{j+1/2}$ are the solutions of the Riemann problem at the interface between the zones $j$ and $j+1$, then we first update the radius of the interface $j+1/2$ in the timestep $\Delta t=t^{n+1}-t^{n}$ as:

\begin{equation}\label{rnew}
    r^{n+1}_{j+1/2}=r^{n}_{j+1/2}+\bar{v}_{j+1/2} \Delta t
\end{equation}

Once we know this quantity we update the time averaged surface at the zone interface $j+1/2$ according to:

\begin{equation}\label{amid}
    \bar{A}_{j+1/2} =  {4\over 3}\pi{\left(r^{n+1}_{j+1/2}\right)^3-\left(r^n_{j+1/2}\right)^3\over r_{j+1/2}^{n+1}-r_{j+1/2}^{n}}
\end{equation}

The density and velocity of zone $j$ are then updated according to

\begin{equation}\label{rhonew}
    \rho^{n+1}_j =  {3 \Delta m_j\over  4 \pi \left[ \left(r^{n+1}_{j+1/2}\right)^3-\left(r^{n+1}_{j-1/2}\right)^3\right]}
\end{equation}

\begin{equation}\label{vnew}
\begin{split}
    v^{n+1}_j = & v^{n}_j \\
                & + {1\over 2} \left(\bar{A}_{j+1/2}+\bar{A}_{j-1/2}\right){\Delta t \over \Delta m_j}\left(\bar{P}_{j+1/2}-\bar{P}_{j-1/2}\right) \\
                & + {\Delta t \over 2}\left(g_j^{n+1}+g_j^n\right) 
\end{split}    
\end{equation}

where $\Delta m_j=m_{j+1/2}-m_{j-1/2}$ is the mass size of the zone $j$, $g_j= G m_j/r^2_j$ is the gravity, $m_j $ and $r_j$ are the mass and radius of the zone $j$, this last quantitiy given in general by:

\begin{equation}
    r_j = \left[ {1\over 3}{\left(r_{j+1/2}\right)^3-\left(r_{j-1/2}\right)^3\over r_{j+1/2}-r_{j-1/2}}\right]^{1/2}
\end{equation}

The equation of the conservation of the total energy is linearized as:

\begin{equation}\label{energy1}
\begin{split}
  E_j^{n+1}= & E_j^{n} \\
             & - {\Delta t \over \Delta m_j}\left(\bar{A}_{j+1/2}\bar{v}_{j+1/2}\bar{P}_{j+1/2}-\bar{A}_{j-1/2}\bar{v}_{j-1/2}\bar{P}_{j-1/2} \right) \\
             & - {\Delta t\over \Delta m_j}\left(L_{j+1/2}^{n+1}-L_{j-1/2}^{n+1}\right)+\epsilon_{j}^{n+1} {\Delta t} 
\end{split}  
\end{equation}

this equation cannot be solved directly because $L^{n+1}$ and $\epsilon^{n+1}$ depend on the updated values of the temperature $T^{n+1}$ (e.g., eq. \ref{eq:lumdiffrad}), that is still unknown at this stage. However, since $E=E_{\rm kin}+E_{\rm int}+E_{\rm grav}$, equation \ref{energy1} can be rewritten as:

\begin{equation}\label{energy2}
\begin{split}
 E_{{\rm int},j}^{n+1} = & E_{{\rm int},j}^{n}+\left(E_{{\rm kin},j}^{n}+E_{{\rm grav},j}^{n}-E_{{\rm kin},j}^{n+1}+E_{{\rm grav},j}^{n+1}\right) \\
                        & + \epsilon^{n+1}_{j} {\Delta t} \\
                        & - {\Delta t \over \Delta m_j}\left(\bar{A}_{j+1/2}\bar{v}_{j+1/2}\bar{P}_{j+1/2}-\bar{A}_{j-1/2}\bar{v}_{j-1/2}\bar{P}_{j-1/2} \right) \\ 
                        & - {\Delta t\over \Delta m_j}\left(L_{j+1/2}^{n+1}-L_{j-1/2}^{n+1}\right) 
\end{split}
\end{equation}

the first two terms of equation \ref{energy2} are known and do not depend on the updated temperature. In fact, $E_{{\rm kin},j}^n,~E_{{\rm int},j}^n~,E_{{\rm grav},j}^n$ are the values corresponding to the previous model, while $E_{{\rm kin},j}^{n+1}={1/2}\left(v_j^{n+1}\right)^2$ and $E_{{\rm grav},j}^{n+1}=-G m_j/r^{n+1}_j$ depend on variables that are already updated. Also the third term depends on variables that are already updated and therefore it is known. In general $\epsilon=\epsilon_{\rm nuc}+\epsilon_{\nu}$, where $\epsilon_{\rm nuc}$ is the energy generated by the nuclear reactions while $\epsilon_{\nu}$ is the energy loss due to neutrino produced by both thermal processes and weak interactions. In this version of the code we neglect the neutrino losses and the energy produced by nuclear reactions with the exception of the energy produced by the radioactive decay of $\rm ^{56}Ni\rightarrow ^{56}Co \rightarrow ^{56}Fe$, i.e. $\epsilon_{\rm nuc}=\epsilon_{\rm ^{56}Ni}$. This last quantity is computed as mentioned above and does not depend on the updated value of the temperature.

Thus, defining the quantities

\begin{equation}
\begin{split}
    C_j= & E_{{\rm int},j}^{n}+\left(E_{{\rm kin},j}^{n}+E_{{\rm grav},j}^{n}-E_{{\rm kin},j}^{n+1}+E_{{\rm grav},j}^{n+1}\right) \\
         & - {\Delta t \over \Delta m_j}\left(\bar{A}_{j+1/2}\bar{v}_{j+1/2}\bar{P}_{j+1/2}-\bar{A}_{j-1/2}\bar{v}_{j-1/2}\bar{P}_{j-1/2} \right)
\end{split}    
\end{equation}

and 

\begin{equation}
    G_j=E_{{\rm int},j}^n+C_j+\epsilon_{\rm ^{56}Ni,j}^{n+1} {\Delta t}
\end{equation}

equation \ref{energy2} can be rewritten as:

\begin{equation}\label{energy3}
 E_{{\rm int},j}^{n+1}=G_j-{\Delta t\over \Delta m_j}\left(L_{j+1/2}^{n+1}-L_{j-1/2}^{n+1}\right)
\end{equation}

with $G_j$ constant and defined at the zone center.

According to equation \ref{eq:lumdiffrad} the luminosity, defined at the zone interfaces, can be linearized as:

\begin{equation}\label{difrad1}
\begin{split}
L_{j+1/2}^{n+1}= & -\bar{A}^{2}_{j+1/2}\bigg(\frac{1}{\kappa_{j+1/2}}\bigg)^{n+1}\frac{ac\lambda_{j+1/2}^{n+1}}{3} \\
 & \times \frac{(T_{j+1}^{n+1})^4-(T_{j}^{n+1})^4}{m_{j+1}-m_{j}}
\end{split}
\end{equation}

The opacity $\kappa$ depends on the temperature and density and therefore it is naturally defined at the zone center. For this reason we define the value $\kappa_{j+1/2}$ of the opacity at the zone interface as described in \cite{SNEC}:

\begin{equation}\label{difrad2}
\bigg(\frac{1}{\kappa_{j+1/2}}\bigg)^{n+1}=\frac{(T_{j+1}^{n+1})^4/\kappa_{j+1}^{n+1}+(T_{j}^{n+1})^4/\kappa_{j}^{n+1}}{(T_{j+1}^{n+1})^4+(T_j^{n+1})^4}
\end{equation}

According to equations \ref{fluxlim1} and \ref{fluxlim2}, the flux limiter $\lambda$ is given by:

\begin{equation}\label{difrad3}
\lambda_{j+1/2}^{n+1}=\frac{6+3R^{n+1}_{j+1/2}}{6+3R^{n+1}_{j+1/2}+(R^{n+1}_{j+1/2})^2}
\end{equation}

where

\begin{equation}\label{difrad4}
\begin{split}
R^{n+1}_{j+1/2}= & \frac{2\bar{A}_{j+1/2}}{m_{j+1}-m_j} \\
                 & \times \frac{|(T_{j+1}^{n+1})^4-(T_{j}^{n+1})^4|}{(T_{j+1}^{n+1})^4+(T_{j}^{n+1})^4}\bigg(\frac{1}{\kappa_{j+1/2}}\bigg)^{n+1}
\end{split}
\end{equation}

By means of equations \ref{difrad1}, \ref{difrad2}, \ref{difrad3} and \ref{difrad4}, and since $E_{{\rm int},j}^{n+1}$ depends on $\rho_j^{n+1}$ and $T_j^{n+1}$, it is easy to verify that equation \ref{energy3} depends only on $T^{n+1}_{j-1}$, $T^{n+1}_{j}$ and $T^{n+1}_{j+1}$.
If the number of zones is $M$, assuming for the boundary conditions that $L_{1-1/2}=0$ and $L_{M+1/2}=L_{M-1/2}$, equation \ref{energy3} written for all the zones produces a system of $M$ equations for the $M$ unknowns $T_j^{n+1}~~(j=1,...,M)$, that is solved by means of a Newtown-Raphson method. In particular, assuming a trial values for the temperature $T^{n+1}_j~~(j=1,...,M)$, this algorithm implies the solution of the following system:

\begin{equation}\label{energy4}
\begin{split}
{\partial E_{{\rm int},j}^{n+1}\over \partial T_j}\Delta T_j  
  -{\Delta t\over \Delta m_j}\left({\partial L_{j+1/2}^{n+1} \over \partial T_{j+1}}\Delta T_{j+1}+{\partial L_{j+1/2}^{n+1} \over \partial T_{j}}\Delta T_{j}\right) \\
 -\left({\partial L_{j-1/2}^{n+1} \over \partial T_{j}}\Delta T_{j}-{\partial L_{j-1/2}^{n+1} \over \partial T_{j-1}}\Delta T_{j-1}\right) =-\delta_j &
\end{split}
\end{equation}

where

\begin{equation}\label{energy5}
 \delta_j = E_{{\rm int},j}^{n+1}-G_j+{\Delta t\over \Delta m_j}\left(L_{j+1/2}^{n+1}-L_{j-1/2}^{n+1}\right)
\end{equation}

The derivative of the internal energy with respect to the temperature is obtained from the equation of state while the derivatives of the luminosity can be computed according to equation \ref{difrad1}

\begin{equation}
\begin{split}
{\partial L_{j+1/2}^{n+1} \over \partial T_{j+1}} = & -\bar{A}^{2}_{j+1/2}\bigg(\frac{1}{\kappa_{j+1/2}}\bigg)^{n+1} \\
& \times \frac{4ac\lambda_{j+1/2}^{n+1}}{3}\frac{(T_{j+1}^{n+1})^3}{m_{j+1}-m_{j}}
\end{split}
\end{equation}

\begin{equation}
\begin{split}
{\partial L_{j+1/2}^{n+1} \over \partial T_{j}}= & +\bar{A}^{2}_{j+1/2}\bigg(\frac{1}{\kappa_{j+1/2}}\bigg)^{n+1} \\ & \times \frac{4ac\lambda_{j+1/2}^{n+1}}{3}\frac{(T_{j}^{n+1})^3}{m_{j+1}-m_{j}}
\end{split}
\end{equation}

\begin{equation}
\begin{split}
{\partial L_{j-1/2}^{n+1} \over \partial T_{j}}= & -\bar{A}^{2}_{j-1/2}\bigg(\frac{1}{\kappa_{j-1/2}}\bigg)^{n+1} \\ & \times \frac{4ac\lambda_{j-1/2}^{n+1}}{3}\frac{(T_{j}^{n+1})^3}{m_{j}-m_{j-1}}
\end{split}
\end{equation}

\begin{equation}
\begin{split}
{\partial L_{j-1/2}^{n+1} \over \partial T_{j}}= & \bar{A}^{2}_{j-1/2}\bigg(\frac{1}{\kappa_{j-1/2}}\bigg)^{n+1} \\ & \times \frac{4ac\lambda_{j-1/2}^{n+1}}{3}\frac{(T_{j-1}^{n+1})^3}{m_{j}-m_{j-1}}
\end{split}
\end{equation}

We neglect in this case the derivatives of the opacity as a function of the temperature.

Therefore, the matrix of the coefficient of the system \ref{energy5}, rewritten as follows,

\begin{equation}\label{energy6}
\begin{split}
& {\Delta t \over \Delta m_j} {\partial L_{j-1/2}^{n+1} \over \partial T_{j-1}}\Delta T_{j-1} \\ 
& +\left[{\partial E_{{\rm int},j}^{n+1}\over \partial T_j}-{\Delta t\over \Delta m_j}\left({\partial L_{j+1/2}^{n+1} \over \partial T_{j}}-{\partial L_{j-1/2}^{n+1}\over \partial T_{j}}\right)\right]\Delta T_j \\
& -{\Delta t\over \Delta m_j}{\partial L_{j+1/2}^{n+1} \over \partial T_{j+1}}\Delta T_{j+1}=-\delta_j
\end{split}
\end{equation}

is a tridiagonal band matrix like

\begin{equation}
\left[
\begin{array}{ccccc}
b_1 &	 c_1 & 0 & \cdots	&  0    \\
a_2 &  b_2 & c_2 &        & \vdots \\
0 & \ddots&	\ddots&	\ddots & 0\\
\vdots	& & a_{M-1} & b_{M-1} & c_{M-1} \\
0 & \cdots & 0 & a_{M} & b_{M}
\end{array}
\right]
\end{equation}

where

\begin{equation}
 a_j={\Delta t \over \Delta m_j} {\partial L_{j-1/2}^{n+1} \over \partial T_{j-1}}
\end{equation}

\begin{equation}
 b_j={\partial E_{{\rm int},j}^{n+1}\over \partial T_j}-{\Delta t\over \Delta m_j}\left({\partial L_{j+1/2}^{n+1} \over \partial T_{j}}-{\partial L_{j-1/2}^{n+1}\over \partial T_{j}}\right)
\end{equation}

\begin{equation}
 c_j=-{\Delta t\over \Delta m_j}{\partial L_{j+1/2}^{n+1} \over \partial T_{j+1}}    
\end{equation}

To invert this matrix we use the {\scshape{SPARSEKIT2}} package (Yousef Saad webpage https://www-users.cs.umn.edu/~saad/software/SPARSKIT/). Once the system is solved, the initial trial values of the temperature are updated, e.g., $T^{n+1}_j\rightarrow T^{n+1}_j+\Delta T_{j+1}$, and all the process is repeated until both the equations and the normalized corrections $\Delta T\over T$ become less than a chosen tolerance.

By means of the updated values of the temperature and density in each zone, the system of equations \ref{fullsystem4} is solved with a Newton-Raphson method in order to compute the updated values of the abundances of all the nuclear species included in the nuclear network (Table \ref{tab:network}).

The PPM algorithm described above assumes the presence of six ghost zones at the inner and outer boundaries of the computation domain. At the inner edge we impose reflecting boundary conditions, that means that all the various quantities in the ghost zones are defined as:

\begin{equation}
    a_{7-j}=\pm a_{7+j-1}~~~~~j=1,...,6
\end{equation}

where the sign is negative for the velocity and positive for all the other quantities. 
At the outer edge of the computation domain we assume that all the quantities in the ghost zones are kept constant and equal to the values of the last "real" zone, with the exception of the pressure which is set to a fixed value corresponding to $\rm 10^{-24}~dyne~cm^{-2}$.

\section{Explosion and Light Curve of a typical case}\label{sec:exp15a}
\begin{figure}[ht!]
\epsscale{1.2}
\plotone{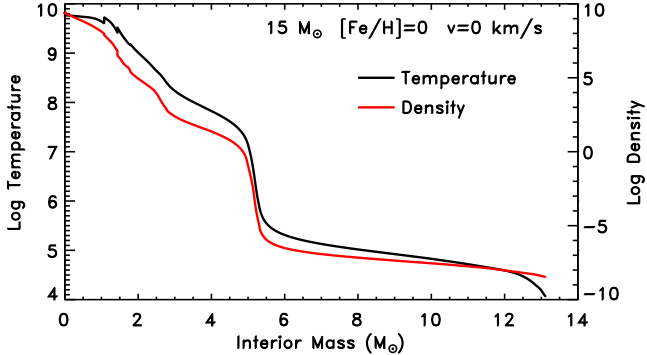}
\caption{Temperature and densities profiles as a function of the interior mass at the presupernova stage of model 15a \label{fig:15aprefis}}
\end{figure}

\begin{figure}[ht!]
\epsscale{1.2}
\plotone{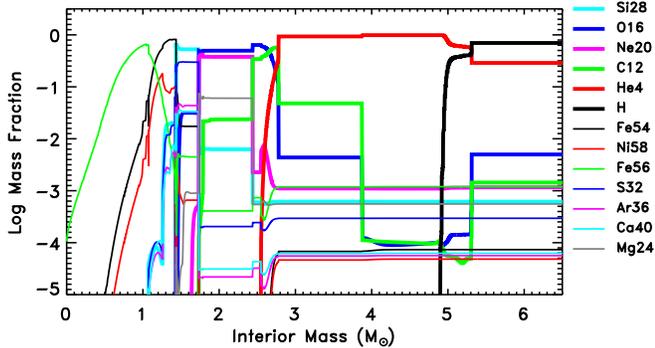}
\caption{Chemical composition as a function of the interior mass at the presupernova stage of model 15a. The most abundant nuclei in inner zones of the Fe core, like, e.g., the neutron rich isotopes $\rm ^{50}Ti$, $\rm ^{54}Cr$, $\rm ^{60}Fe$, $\rm ^{64}Ni$ and so on, are not shown.\label{fig:15aprechim}}
\end{figure}

In this section we describe in detail the main properties of the explosion and of the light curve of a model that we consider as typical, i.e. a solar metallicity non rotating $\rm 15~M_\odot$ (model 15a). The explosion, computed by means of {\scshape{Hyperion}} (section \ref{sec:code}), is induced by removing the inner $\rm 0.8~M_\odot$ of the presupernova model and by depositing instantaneously a given amount of thermal energy in the inner $\rm 0.1~M_\odot$ (i.e., in the region between 0.8 and $\rm 0.9~M_\odot$) . The energy deposited is chosen in order to have a final explosion energy (mainly in form of kinetic energy of the ejecta) $E_{\rm expl}\simeq1.0~{\rm foe}$ ($\rm 1~foe=10^{51}~erg$). Such an artificially way of inducing the explosion is due to the lack of a routinely way of computing a self consistent multi dimensions explosion of a massive star and it constitutes the typical technique, with few small variations, adopted to calculate explosive nucleosynthesis and remnant masses of core collapse supernovae \citep{WW95, TNH96,UN02,lc03,HW10}. A detailed explanation on how the nucleosynthesis as well as the remnant masses depend on the explosion parameters can be found in \citet{ABT91} and in \citet[and references therein]{UY17}. Let us only remark that, at variance with the other similar calculations, we choose an initial mass cut internal enough such that the properties of the shock wave, at the time it reaches the iron core edge, mildly depends on the initial conditions.
Figures \ref{fig:15aprefis} and \ref{fig:15aprechim} show, respectively, the temperature plus density profiles and the chemical composition of the star at the presupernova stage.

\subsection{Propagation of the shock wave, explosive nucleosynthesis, fallback and shock breakout}\label{sec:expl_nuc}
\begin{figure*}[ht!]
\epsscale{0.95}.  
\plotone{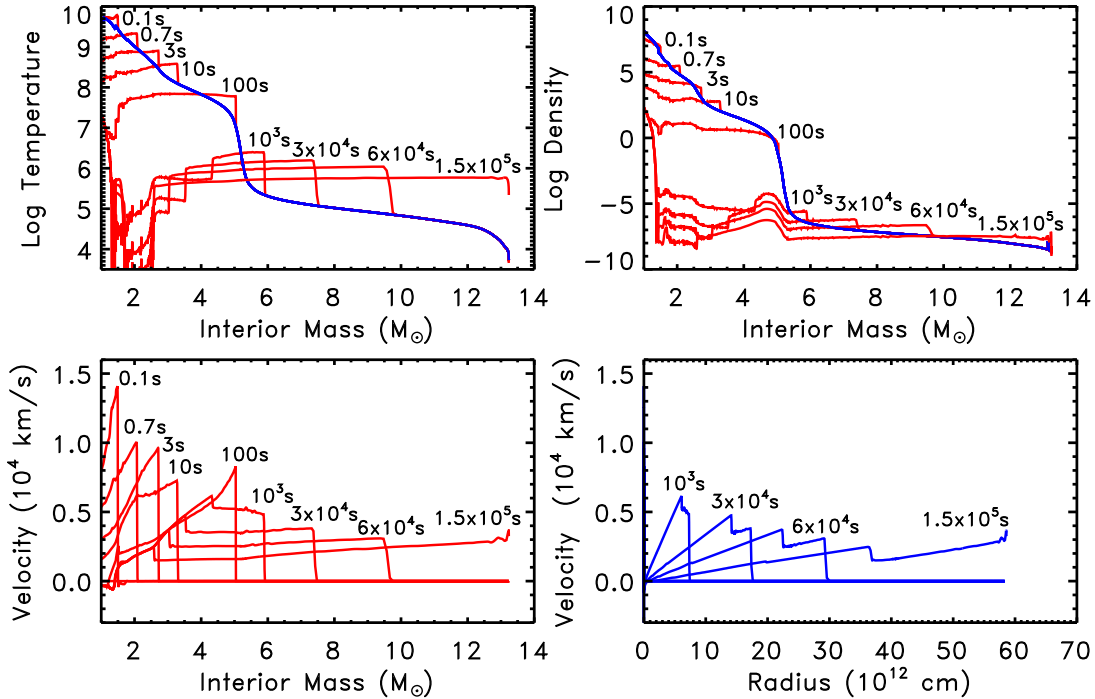}
\caption{Main physical and chemical properties of model 15a at various times during the explosion. \label{fig:15a1foeexpl}}
\end{figure*}

The injection of thermal energy into the model, heats, compresses and accelerates the overlying layers inducing a progressive conversion of the internal energy into kinetic energy, so that a shock wave forms and begins to propagate outward. The temperature behind the shock is almost constant, as expected when radiation dominates the energy budget, and reaches values high enough ($\rm \gtrsim 7 \cdot 10^{9}~K$) to trigger explosive nucleosynthesis (Figure \ref{fig:15a1foeexpl}, upper left and right panels).

\begin{figure}[ht!]
\epsscale{1.15}
\plotone{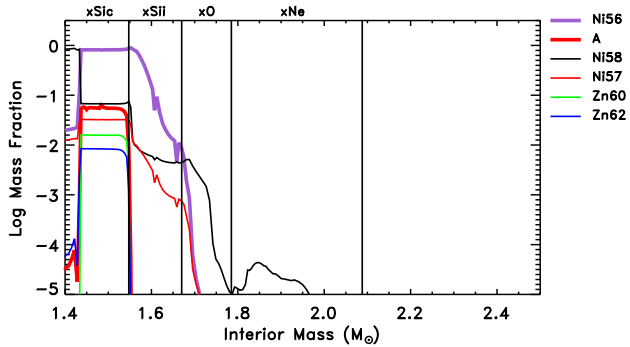}
\caption{Profile of the most abundant isotopes produced by the explosive Si burning with complete Si exhaustion in the zone between the edge of the iron core and $\rm \sim1.55~M_\odot$   \label{fig:chemsic}}
\end{figure}

The inner zone between the edge of the iron core and $\rm \sim1.55~M_\odot$ is the one exposed to the highest temperature ($\rm T \geq 5~GK$) and undergoes explosive Si burning with complete Si exhaustion and is dominated by $\rm ^{56}Ni$ ($\rm ^{56}Fe$), which is by far the most abundant nuclear species (the total $\rm ^{56}Ni$ ejected in this model is $\rm 0.126~M_\odot$). Other abundant isotopes in this zone are $\rm ^{58}Ni$, $\rm ^{57}Ni$ ($\rm ^{57}Fe$), $\rm ^{60}Zn$ ($\rm ^{60}Ni$), $\rm ^{62}Zn$ ($\rm ^{62}Ni$) and $\rm ^{4}He$ (the unstable nuclei will decay at late times into their parent stable isotopes reported in parenthesis) (Figure \ref{fig:chemsic}). 
\begin{figure}[ht!]
\epsscale{1.15}
\plotone{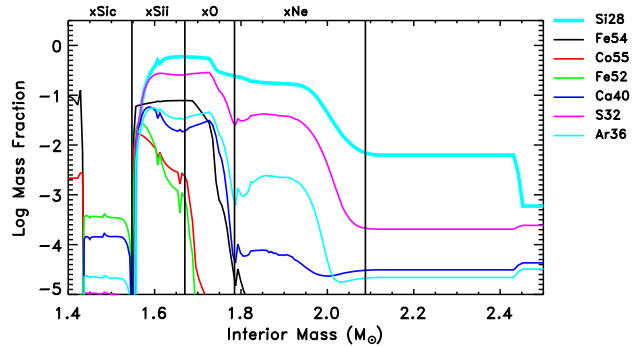}
\caption{Profile of the most abundant isotopes produced by the explosive Si burning with incomplete Si exhaustion in the zone between $\rm \sim1.55~M_\odot$ and $\rm \sim1.69~M_\odot$ \label{fig:chemsii}}
\end{figure}

The layers between $\rm \sim1.55~M_\odot$ and $\rm \sim1.69~M_\odot$ undergo explosive Si burning with incomplete Si exhaustion (peak temperature $\rm 5 \geq T \geq 4~GK$) and are mainly loaded with the iron peak elements $\rm ^{56}Ni$, $\rm ^{58}Ni$, $\rm ^{57}Ni$ ($\rm ^{57}Fe$), $\rm ^{54}Fe$, $\rm ^{55}Co$ ($\rm ^{55}Mn$), $\rm ^{52}Fe$ ($\rm ^{52}Cr$), and the $\alpha$ nuclei $\rm ^{32}S$, $\rm ^{40}Ca$, $\rm ^{36}Ar$ and $\rm ^{28}Si$ (the one that remains partially unburnt) (Figure \ref{fig:chemsii}).

\begin{figure}[ht!]
\epsscale{1.15}
\plotone{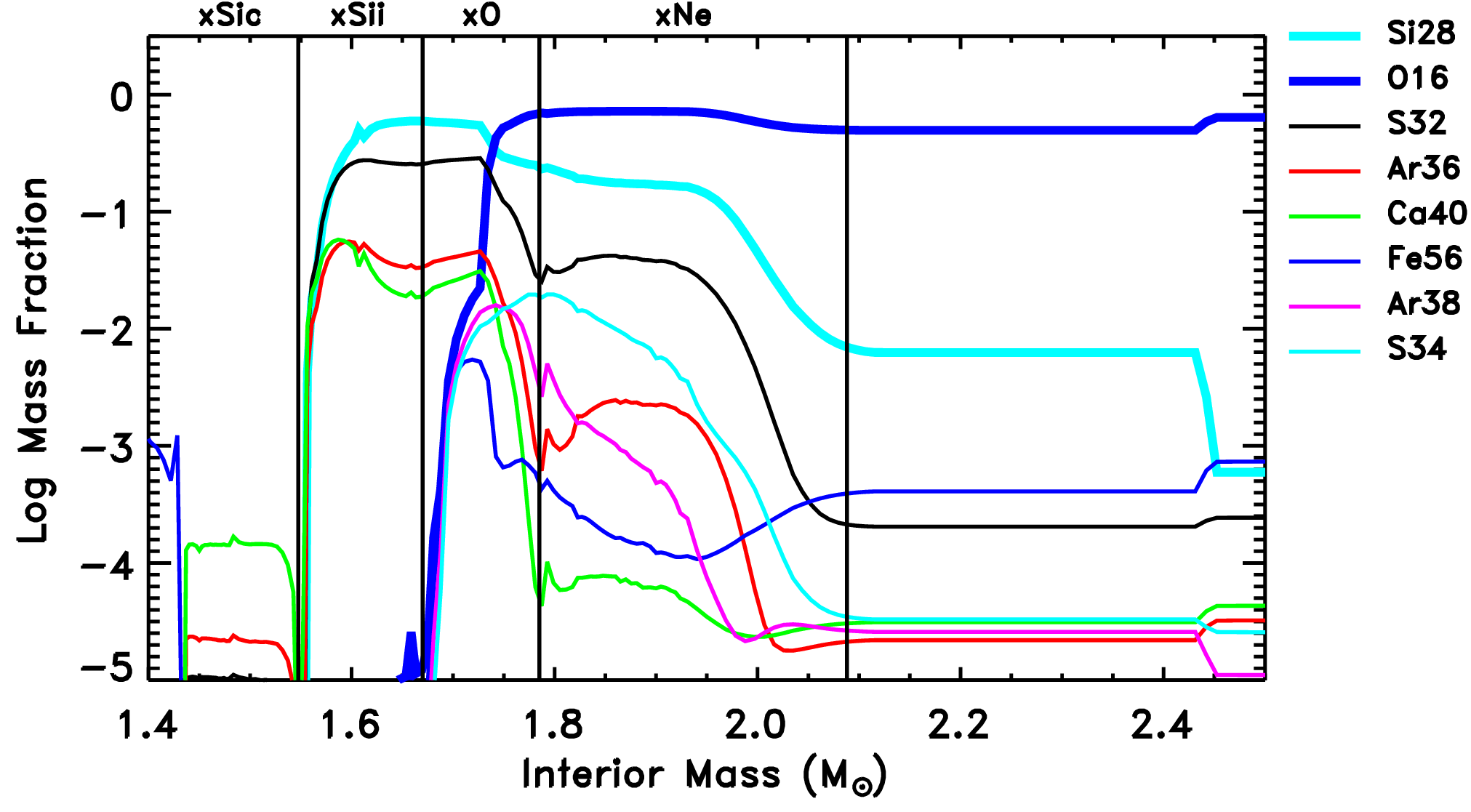}
\caption{Profile of the most abundant isotopes produced by the explosive O burning in the zone between $\rm \sim1.70~M_\odot$ and $\rm \sim1.77~M_\odot$ \label{fig:chemOx}}
\end{figure}

Explosive O burning occurs in the region between $\rm \sim 1.70~M_\odot$ and $\rm \sim1.77~M_\odot$ (peak temperature $\rm 4 \geq T \geq 3.2~GK$) and produces mainly the $\alpha$ nuclei $\rm ^{28}Si$, $\rm ^{32,34}S$, $\rm ^{36,38}Ar$ and $\rm ^{40}Ca$ (Figure \ref{fig:chemOx}).

\begin{figure}[ht!]
\epsscale{1.15}
\plotone{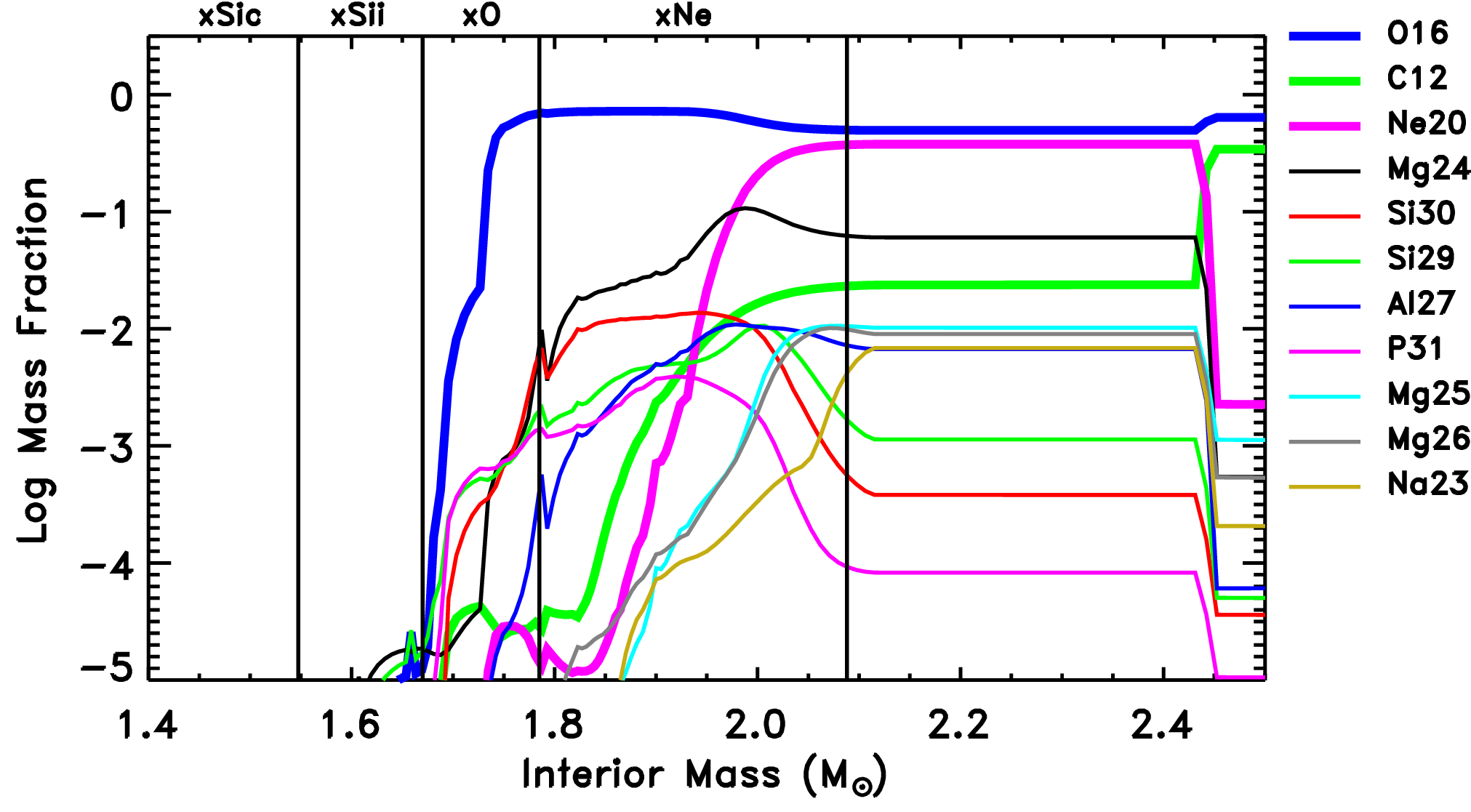}
\caption{Profile of the most abundant isotopes produced by the explosive Ne burning in the zone between $\rm \sim1.77~M_\odot$ and $\rm \sim2.10~M_\odot$ \label{fig:chemNex}}
\end{figure}

Explosive Ne burning occurs in the zones between $\rm \sim 1.77~M_\odot$ and $\rm \sim 2.10~M_\odot$ (peak temperature $\rm 3.2 \geq T \geq 2.0~GK$) and produces or partially modifies (destroys or produces) the pre explosive abundances of $\rm ^{16}O$, $\rm ^{20}Ne$, $\rm ^{23}Na$, $\rm ^{24,25,26}Mg$, $\rm ^{27}Al$, $\rm ^{28,29,30}Si$ and $\rm ^{31}P$ (Figure \ref{fig:chemNex}).

Explosive C burning occurs where the peak temperature of the shock wave reaches $\rm \sim  2.0\cdot 10^{9}~K$ (Figure \ref{fig:15a1foeexpl}) and this happens at the mass coordinate of $\rm \sim 2.1~ M_\odot$ (Figure \ref{fig:15aprechim}). Note that the products of this explosive burning are almost negligible, in this specific case, because of the very low $\rm ^{12}C$ mass fraction present in the C convective shell. This mass coordinate is reached by the shock wave $\rm \sim 0.7~s$ after the start of the explosion, and this time marks the end of the explosive burning since beyond this mass the peak temperature of the shock wave becomes too low to trigger additional burning. At this time the velocity of the shocked zones ranges between $\sim 0.6~10^{4}$ and $\rm \sim 10^{4}~km/s$ (Figure \ref{fig:15a1foeexpl}, lower left panel).

\begin{figure*}[ht!]
\epsscale{0.95}
\plotone{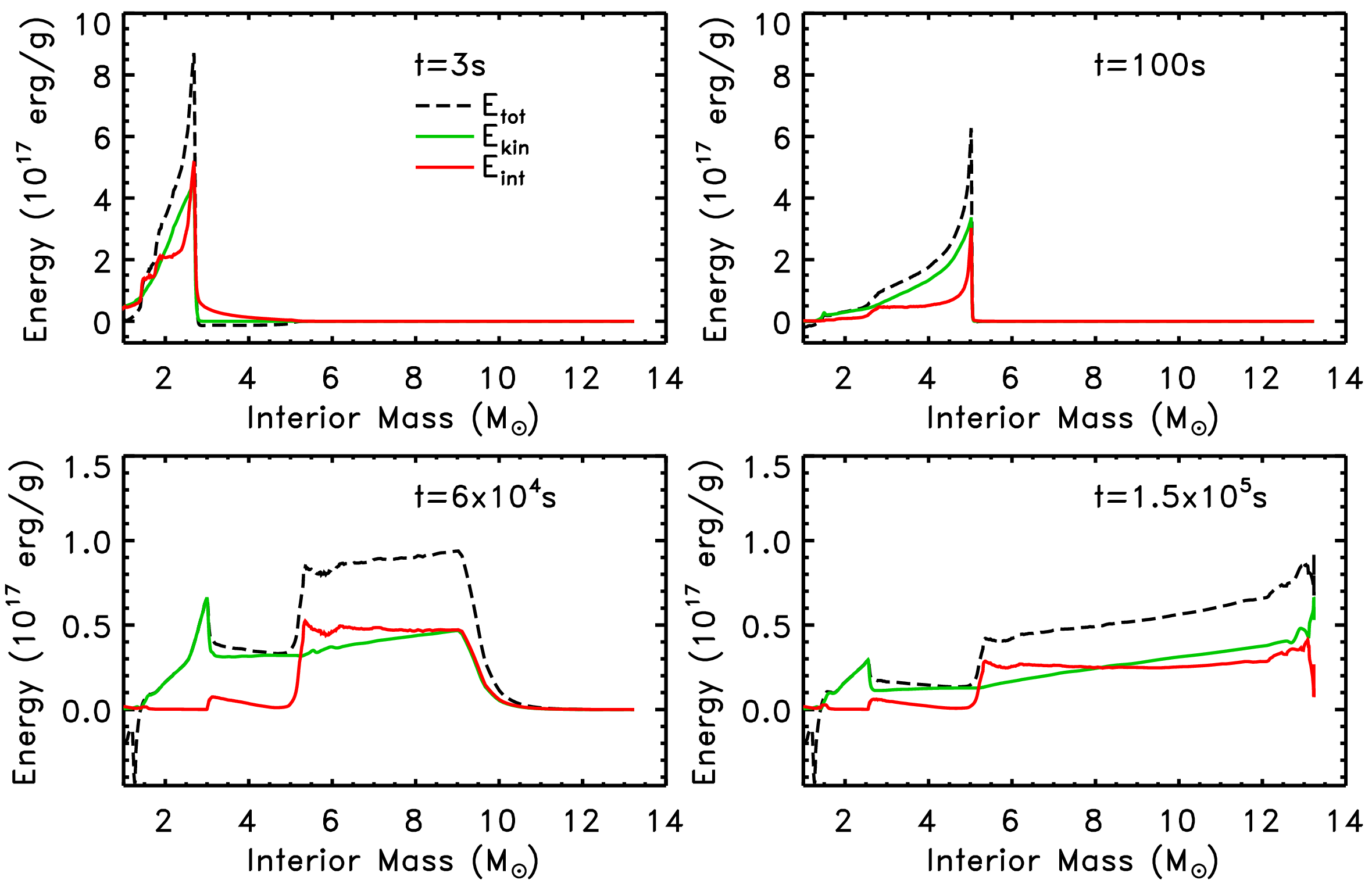}
\caption{Interior profiles of the total energy (black dashed line), kinetic energy (green line) and internal energy (red line) at various times during the explosion.\label{fig:15a1foeenergy}}
\end{figure*}

Roughly $\rm 3~s$ after the beginning of the explosion, the shock wave reaches the edge of the CO core. At this time the temperature and the density of the shock have decreased to $\rm \sim 6\cdot 10^{8}~K$ and $\rm \sim 10^{4}~g~cm^{-3}$, respectively, while the velocity of the shocked layers ranges between $\sim 0.3\cdot 10^{4}$ and $\rm \sim 10^{4}~km/s$ (Figure \ref{fig:15a1foeexpl}, lower left panel). Figure \ref{fig:15a1foeenergy} shows the run of the internal (red), kinetic (green) and total (black) energy within the expanding ejecta at four key points. The upper left panel in the Figure refers to $\rm t=3~s$. Already at this point the kinetic energy of many layers becomes comparable to, or even larger than, the internal one.

Roughly $\rm 10~s$ after the beginning of the explosion, some of the most internal layers revert their velocity because are not able to reach the escape velocity and therefore they fall back onto the compact remnant. Almost $\rm 0.45~M_\odot$ of the initial ejecta collapse back in the initial remnant, increasing the mass cut, i.e. the mass that divides the remnant from the ejecta, to $\rm \sim 1.25~M_\odot$. 

In $\rm \sim 100~s$ the shock wave reaches the He/H interface where a strong density gradient is present (Figure \ref{fig:15aprefis}). Most of the internal energy behind the shock has been converted into kinetic energy that now dominates the total energy (Figure \ref{fig:15a1foeenergy}, upper right panel), while the gravitational energy becomes negligible in this region. The presence of the strong density gradient at the He/H interface induces the formation of a reverse shock, see e.g. \citet{WW95}, so that from this time onward the explosion is characterized by a forward shock that continues to propagate outward and by a reverse shock that propagates inward in mass and that slows down the material previously accelerated by the forward shock (Figure \ref{fig:15a1foeexpl}). 

As the two shocks move away from each other, the temperature remains almost constant in the region between the two, while the density shows a bump close to the H/He interface that will persist up to the late stages and that will have some important consequences on the features of the light curve during the transition from the plateau phase to the radioactive tail (see below). Both the temperature and the density decrease maintaining their shape as the time goes by. During this phase additional internal zones fall back onto the compact remnant because of the interaction with the reverse shock. This process eventually ends $\rm \sim10^{4}~s$ after the onset of the explosion, leaving a final compact remnant of $\rm \sim 1.42~M_\odot$. Note that such a fall back brings back part of the matter where explosive Si burning with complete Si exhaustion occurred, and where most of the $\rm ^{56}Ni$ and many iron peak nuclei are synthesized (Figure \ref{fig:chemsic}), preventing their ejection into the interstellar medium.

\begin{figure}[ht!]
\epsscale{1.15}
\plotone{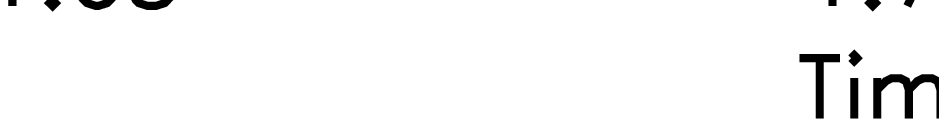}
\caption{Evolution of the bolometric luminosity and effective temperature of model 15a with $E_{\rm rm net}=1.0$ foe during the breakout.\label{fig:15abol}}
\end{figure}

The forward shock eventually reaches the surface of the star $\rm \sim 1.5\cdot 10^{5}~s$ ($\rm \sim 1.7~days$) after the onset of the explosion and at this stage the reverse shock has moved down to  $\rm \sim 2.5~M_\odot$. When the shock wave reaches the surface both the temperature and the bolometric luminosity increase up to $\rm \sim 2 \cdot 10^{5}~K$ and $\rm \sim 2 \cdot 10^{45}~erg/s$, respectively (Figure \ref{fig:15abol}) and all the expanding mantle is totally ionized since the temperature exceeds $\rm \sim 10^{5}~K$ everywhere.

Before closing this section let us remark that once the main shock wave overturn the H/He interface, the total energy in the shocked part of the H rich mantle is dominated by the kinetic energy, while it is basically equiparted between internal and kinetic within the He core (Figure \ref{fig:15a1foeenergy}, lower right panel).

\subsection{Adiabatic cooling}\label{sec:adcoooling}

The first phase of expansion of the ejecta (i.e between 1.7 and 18 days) is characterized by a few phenomena worth being reminded. 

\begin{figure}[ht!]
\epsscale{1.15}
\plotone{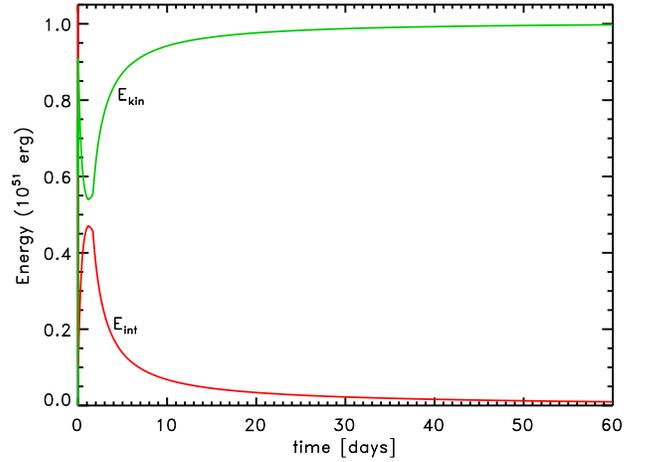}
\caption{Evolution of the total internal (red line) and kinetic (green line) energies. The maximum of the internal energy (minimum of the kinetic energy) corresponds to the time of the shock breakout. Note that, as it is mentioned in the text, at the time of the shock breakout, the total energy ($\rm \sim 10^{51}~erg=1~foe$) is roughly divided in equal proportions between internal and kinetic energy.\label{fig:enertime}}
\end{figure}

First of all the velocity of the various layers after the break out (Figure \ref{fig:15a1foeexpl}, lower right panel) does not remain frozen because the internal energy still feeds the kinetic one. Figure \ref{fig:enertime} shows the temporal evolution of both the kinetic and internal energies of the ejecta. The kinetic energy increases from 0.6 (the value at the break out) to 0.9 foe in the first 3.5 days after the break out ($\rm \sim 5$ days from the explosion), increasing up to almost the final value of 1 foe in other 13 days ($\rm \sim 18$ days from the explosion). 

\begin{figure}[ht!]
\epsscale{1.15}
\plotone{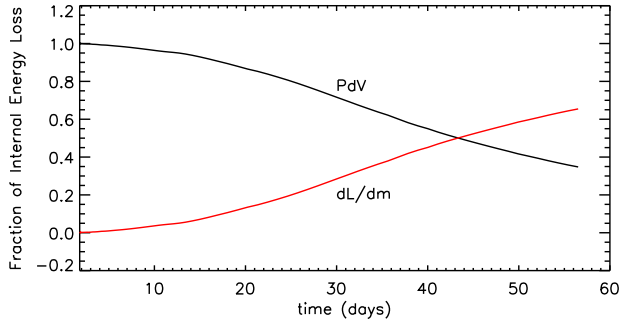}
\caption{Fraction of internal energy loss due to adiabatic cooling ($P\dot{V}/\dot{E}$), black line, and radiative losses ($\partial L/\partial m/\dot{E}$), red line.\label{fig:movie_ener}}
\end{figure}

The second thing worth being reminded is that the decrease of the internal energy is initially due almost exclusively to an adiabatic expansion while the radiative losses prevail at later time. This is well visible in Figure \ref{fig:movie_ener}. From the first law of thermodynamic we have $\dot{E}=-P\dot{V}-\partial L/\partial m$, where $E$ is the internal energy per unit mass and where we have for the moment neglected any other source term, the other terms having the usual meaning. The Figure shows clearly that more than 90\% of the internal energy losses are due to the $P\dot{V}$ term, at least up to day $\sim 18$, and therefore that the expansion is essentially adiabatic, i.e. $\dot{E}\simeq -P\dot{V}$, in this phase.

\begin{figure}[ht!]
\epsscale{1.15}
\plotone{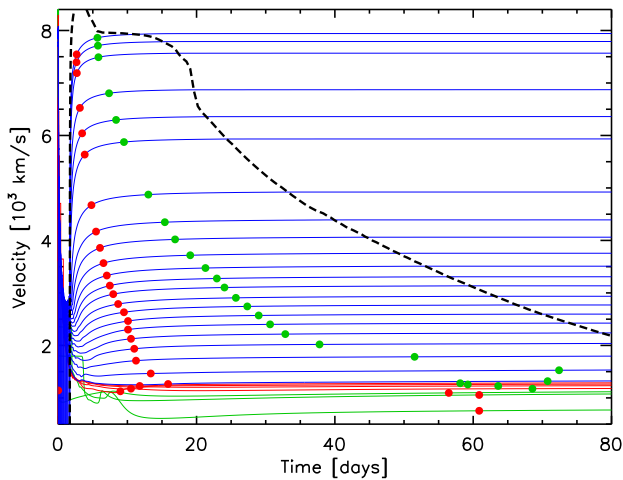}
\caption{Evolution of the velocity of selected layers with time. The red and green filled dots mark the times when the layer reaches 95\% and 99\% of its terminal velocity, respectively. The blue, red and green lines mark the H-rich, He-rich and CO-rich layers, respectively. The black dashed line is the velocity of the photosphere. \label{fig:velstru}}
\end{figure}

\begin{figure}[ht!]
\epsscale{1.15}
\plotone{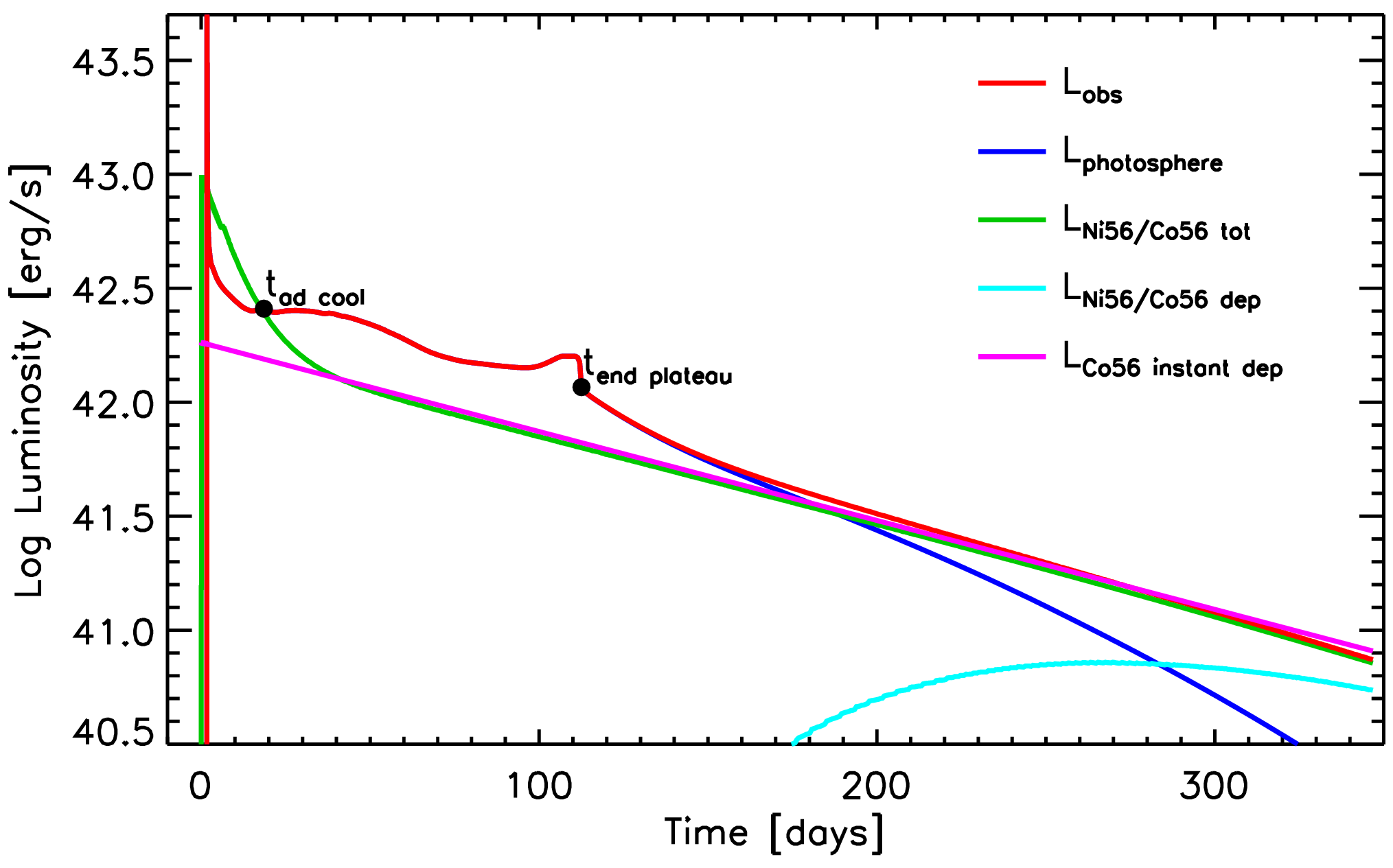}
\caption{Bolometric observed luminosity (red line), luminosity at the photosphere (blue line), total luminosity due to the $\rm ^{56}Ni/^{56}Co$ decay (green line),  luminosity due to the absorption of $\gamma$-rays from $\rm ^{56}Ni/^{56}Co$ decay above the photosphere (cyan line), instantaenous energy deposition by the radioactive $\rm ^{56}Co$ decay (magenta line) corresponding to $\rm 0.126~M_\odot$ of $\rm ^{56}Ni$ initially synthesized, as a function of the time elapsed from the explosion \citep[eq.19]{nad94}.  
\label{fig:template_lc}}
\end{figure}

The temperature within the whole expanding mantle is well above $\rm 10^{5}~K$, so that matter is fully ionized everywhere, occurrence that prevents it from becoming transparent to the radiation. As a consequence, the surface of the expanding mantle (defined as the mass coordinate where $\rm \tau=2/3$) remains basically anchored at the same mass coordinate. Figure \ref{fig:velstru} shows the velocity of selected layers together to the location of the photosphere. It is well visible that in the first 18 days or so the mass coordinate of the photosphere does not change significantly. The temporal evolution of the surface luminosity (by definition the luminosity of the photosphere) follows the behavior of the photosphere itself. $ L$ is approximately proportional to $R^2~T^4$ (where $R$ and $T$ refer to the photosphere); since in a adiabatic expansion of a radiation dominated gas $\dot{E}\simeq -P\dot{V}$ implies $T\propto R^{-1}$, $T^4$ scales as $R^{-4}$ and hence $L\propto 1/R^2$.This explains the initial decline of the luminosity after the break out. Figure \ref{fig:template_lc} shows the evolution of the bolometric luminosity as a red line. The adiabatic cooling phase goes from the break out to the beginning of a phase in which the surface luminosity is roughly constant (marked by the black dot). This change of behavior will be discussed in the next section. 

As already mentioned above, at day $\sim 18$ all the ejecta have almost reached their terminal velocity (Figure \ref{fig:velstru}), and hence the following evolution is characterized by a free expansion where forces due to pressure gradient and gravitation are now negligible. In this regime the expansion becomes homologous, i.e., characterized by a constant velocity of each layer that scales linearly with the radius. Note that the more internal zones are the last to achieve this stage because the reverse shock reaches the base of the expanding envelope only $\sim 10$ days after the explosion.

\subsection{Recombination front and Plateau Phase}\label{sec:recombinationwave}
The phase of adiabatic expansion of the ejecta ends at day $\sim 18$, i.e.,
when the temperature of the photosphere drops to $\rm \sim 5500~K$ and the H recombines. Since the opacity is mainly due to the electron scattering, it decreases dramatically in these zones, increasing their transparency to radiation. As a consequence, the internal energy is radiated away very efficiently and the temperature drops abruptly at the recombination front. Such an occurrence marks the end of the phase in which the photosphere remains anchored to the most external layer of the ejecta. In fact, as the expansion proceeds, the temperature of a progressively increasing number of (more internal) zones drops below the critical value for the H recombination and, as a consequence a cooling wave, due to the transparency induced by the recombination front, progressively penetrates inward in mass. The strong reduction of the opacity implies a strong reduction of the optical depth, therefore the location of the photosphere, defined as the first layer where $\tau=2/3$, closely follows the recombination wave. For sake of simplicity, in the following, we will consider the recombination front and the photosphere, coincident in mass. 

\begin{figure}[ht!]
\epsscale{1.15}
\plotone{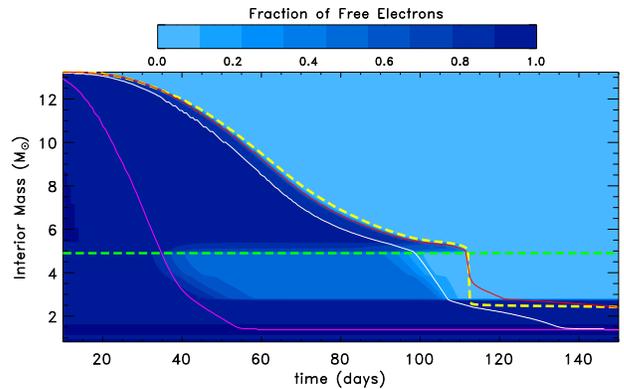}
\caption{Fraction of free electrons as a function of the time (x-axis) and the interior mass (y-axis) according to the color coding reported in the color bar on top the plot. The yellow dashed line marks the location of the photosphere, the red, white and magenta lines mark the layers where the recombination temperatures for H {\RomanNumeral{2}}, He {\RomanNumeral{2}} and He {\RomanNumeral{3}} are achieved, respectively. The horizontal dashed green line marks the H/He interface. \label{fig:frac_ele_contour}}
\end{figure}

\begin{figure}[ht!]
\epsscale{1.15}
\plotone{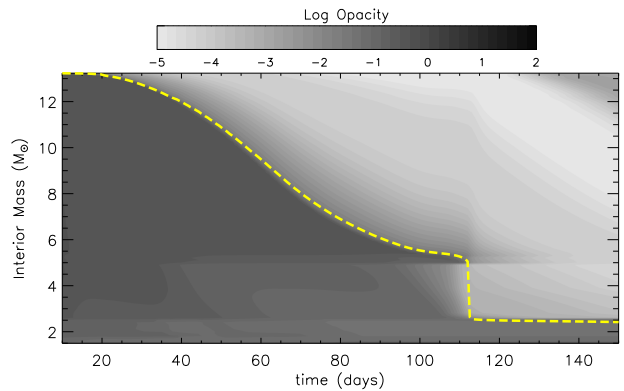}
\caption{Opacity as a function of the time (x-axis) and the interior mass (y-axis) according to the color coding reported in the color bar on top the plot. The yellow dashed line marks the location of the photosphere. \label{fig:opa_contour}}
\end{figure}

\begin{figure}[ht!]
\epsscale{1.15}
\plotone{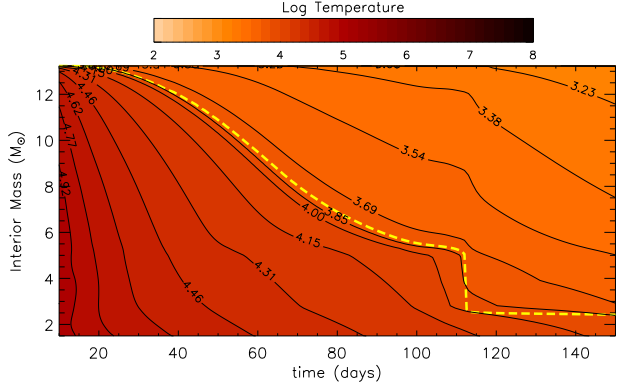}
\caption{Temperature as a function of the time (x-axis) and the interior mass (y-axis) according to the color coding reported in the color bar on top the plot. The contour levels are also plotted. The yellow dashed line marks the location of the photosphere. \label{fig:temp_contour}}
\end{figure}

Figures \ref{fig:frac_ele_contour}, \ref{fig:opa_contour} and \ref{fig:temp_contour}
map the temporal evolution of the fraction of free electrons, the opacity and the temperature inside the star, respectively. These first two plots clearly show that the opacity drops whenever the fraction of free electrons reduces.
Morevore the three solid lines in Figure \ref{fig:frac_ele_contour}, marking the location where He {\RomanNumeral{3}} (magenta), He {\RomanNumeral{2}} (white) and H {\RomanNumeral{2}} (red) recombine, show that the recombination of He {\RomanNumeral{3}} obviously occurs first. Such an occurrence, however, does not affect appreciably the fraction of free electrons (and hence the opacity) in the H-rich envelope because in this zone that fraction is mainly determined by the hydrogen itself. For this reason, in the first $\sim 18$ days the fraction of free electrons does not change appreciably in any layer of the star. Roughly at day 18, H begins to recombine and the photosphere starts moving inward, leaving outside matter with a very low fraction of free electrons and hence with a low opacity. It is worth noting that in this phase the photosphere (yellow dashed line) closely follow the isothermal corresponding to the H recombination temperature (Figure \ref{fig:temp_contour}). 

Around day 40, the temperature in the He core drops below the threshold value for the He {\RomanNumeral{3}} recombination first (and for the He {\RomanNumeral{2}} later) and this determines a strong reduction of the number of free electrons (and of the opacity) within the He core (see Figures \ref{fig:frac_ele_contour} and \ref{fig:opa_contour}). Once the photosphere reaches the H/He interface (at day $\rm \sim110$), very quickly shifts down to the CO core because of the very low opacity between the CO core and the H/He interface. The fraction of free electrons remains equal to one within the CO core because we assume matter to remain fully ionized in the He exhausted zone (see section \ref{sec:code}).

\begin{figure}[ht!]
\epsscale{1.15}
\plotone{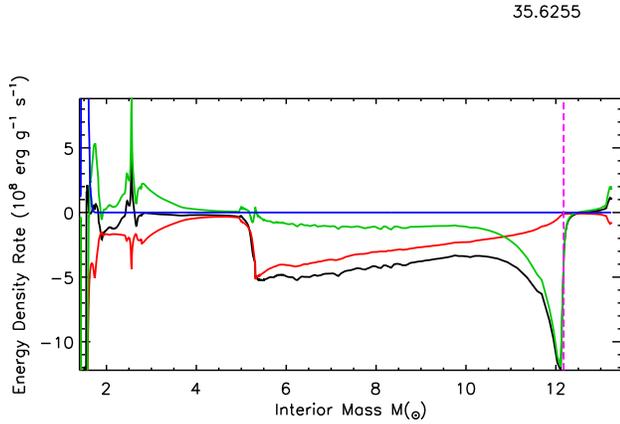}
\caption{Time derivative of the internal energy per unit mass ($\dot{E}$) as a function of the interior mass (black line) for a model at day $\sim 35$. The contribution due to the adiabatic expansion ($P\dot{V}$), to the radiative diffusion ($dL/dm$) and to the $\rm ^{56}Ni$ radioactive decay are shown with the red, green and blue lines respectively. The dashed vertical magenta line marks the location of the photosphere. \label{fig:movie_ener_2}}
\end{figure}

Figure \ref{fig:movie_ener_2} shows the typical relative contributions of the adiabatic cooling (red line), of the radiative losses (green line) and of the $\rm ^{56}Ni$ radioactive decay (blue line) to the variation of the internal energy in the phase in which the recombination front moves within the H rich mantle (the figure is a snapshot taken at day $\sim 35$). The Figure shows very clearly that behind the recombination front (marked by the magenta vertical dashed line) the cooling due to the adiabatic expansion (red line) dominates the energy losses up to $\rm \sim 10~M_\odot$ while the radiative ones dominate close to the photosphere and beyond.

\begin{figure}[ht!]
\epsscale{1.15}
\plotone{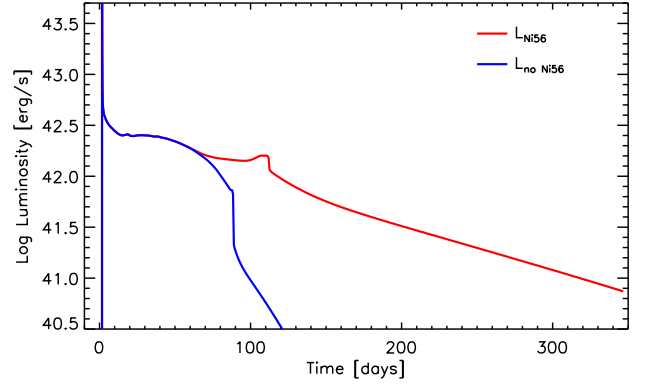}
\caption{Comparison between the reference light curve (red line) and the one obtained by setting artificially to zero the abundance of $\rm ^{56}Ni$ (blue line). \label{fig:compare_lc_ni_noni}}
\end{figure}

The surface luminosity levels off after the first phase of adiabatic expansion and maintains a roughly flat profile until the recombination front reaches the H/He discontinuity (Figure \ref{fig:template_lc}). The reason is that both its radius and temperature do not vary significantly in this phase. Since the expansion of the mantle behind the recombination front is almost adiabatic (see Figure \ref{fig:movie_ener_2}), the temperature of each layer scales as $T\simeq R^{-1}$ and since the recombination temperature is roughly fixed ($\rm at \sim 5500~K)$, also the recombination radius remains practically frozen at a constant value. It must be noted that the release of energy coming from the cascade decay of $\rm ^{56}Ni$ contributes to determine the duration of the plateau phase. The contribution of $\rm ^{56}Ni-^{56}Co$ decay in sculpting the shape of the light curve in the plateau phase, and in particular its duration, is clearly shown in Figure \ref{fig:compare_lc_ni_noni}, where the light curve of the reference model (red line) is compared to one computed switching artificially off the cascade decay of $\rm ^{56}Ni$ (blue line).

The luminosity profile in the transition from the plateau phase to the radioactive tail depends on a complex interplay among the temporal evolution of temperature, density and chemical composition. We will discuss how this interplay affects both the slope of the luminosity profile and the formation of a luminosity bump in this transition phase in section \ref{sec:bump}.

\subsection{Radioactive tail}\label{sec:rad_tail}
Once the photosphere reaches the H/He interface $\rm (t\sim 112~days)$, its backward velocity speeds up because of the sudden reduction of the opacity (see above) and it reaches the border of the CO core in roughly 1 day. The penetration of the recombination front in the He core causes a sharp drop in the luminosity because the amount of energy stored in the He core is much less than the one present in the H rich mantle (see the lower two panels in Figure \ref{fig:15a1foeenergy}). After this sharp drop, the release of energy coming from the stored energy reduces progressively and the luminosity declines approaching gradually the one produced by the $\rm ^{56}Co$ decay (green line in Figure \ref{fig:template_lc}). A refined temporal evolution of the luminosity provided by the cascade decay of $\rm ^{56}Ni$ as a function of the amount of $\rm ^{56}Ni$ synthesized in the explosion, may be found in \cite{nad94}, eq. (19). In this phase the light curve is clearly a direct measure of the amount of $\rm ^{56}Ni$ synthesized during the explosion.

\begin{figure}[ht!]
\epsscale{1.15}
\plotone{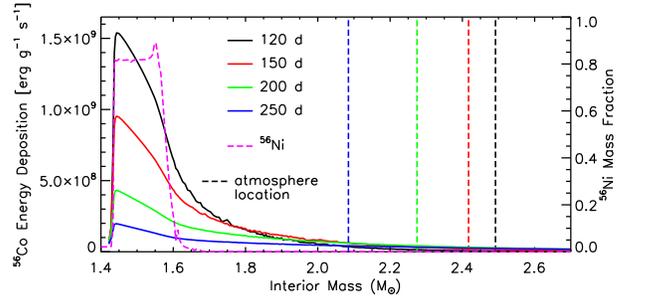}
\caption{Energy deposition ($\rm erg~g^{-1}~s^{-1}$) due to the $\rm ^{56}Co$ radioactive decay as a function of the interior mass at various times (solid lines, primary y-axis) after the explosion: 120 days (black line), 150 days (red line), 200 days (green line), 250 days (blue line). Location of the atmosphere corresponding to the selected times (dashed lines). $\rm ^{56}Ni$ mass fraction resulting from the explosive nucleosynthesis (section \ref{sec:expl_nuc}) as a function of the interior mass (magenta dashed line).\label{fig:nideposit_snap}}
\end{figure}

It is eventually worth noting that the $\gamma$-ray photons released by the radioactive material are not 100\% trapped locally but, as times goes by, a fraction of them are absorbed by more external layers, even outside the formal photosphere. Figure \ref{fig:nideposit_snap} shows the energy deposition function, i.e. the amount of $\gamma$-ray photons absorbed by each layer, at various times. Starting from day 200, a fraction of the energy released by the $\rm ^{56}Co$ radioactive decay is deposited outside the photosphere, (cyan line in Figure \ref{fig:template_lc}).

Finally, Figure \ref{fig:template_lc} shows that, in this phase, the total luminosity corresponds to the total instantaneous rate of energy deposition by the radioactive decay of $\rm ^{56}Co$ (magenta line). This is due to the fact that the envelope remains optically thick to the $\gamma$-rays until late times. If, on the contrary, the envelope would have become partially thin to them (e.g., because of a lower $\gamma$-ray opacity), a fraction of these $\gamma$-rays would have escaped freely and the slope of the light curve would have become steeper.

\subsection{The transition from the Plateau to the radioactive tail and the formation of a luminosity bump}\label{sec:bump}

We left this part of the temporal evolution of the light curve as the last subsection of this chapter because it deserves not just the description of what happens but also the presentation of some tests that allow us to identify the physical keys that control the luminosity profile in this phase. 

\begin{figure}[ht!]
\epsscale{1.15}
\plotone{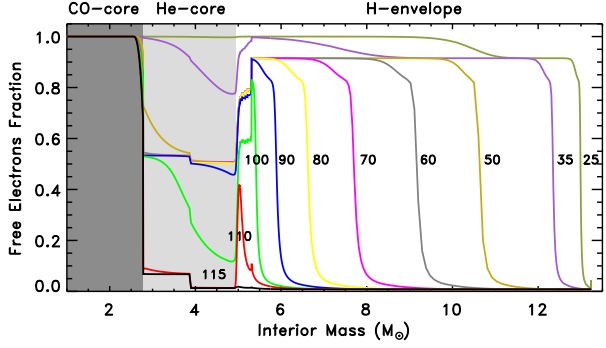}
\caption{Free electrons fraction as a function of the interior mass at various times (shown in days close to the various lines). The dark and light grey areas mark the CO core and He core respectively. \label{fig:fracele_multi_snap}}
\end{figure}

\begin{figure}
\epsscale{1.15}
\plotone{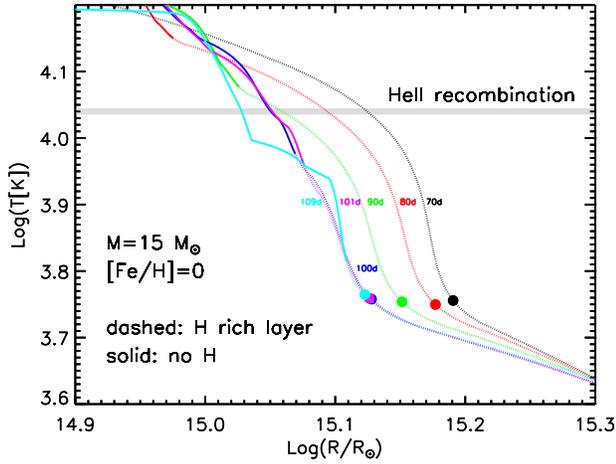}
\caption{Temperature profile as a function of the radius at 70 (black line), 80 (red line), 90 (green line), 100 (blue line), 101 (magenta line) and 109 days (cyan), respectively. The dashed part of each line refers to the H rich matter while the solid part refers to the region within the He core. The horizontal grey line marks the critical temperature below which He {\RomanNumeral{2}} recombines. The filled dots represent the position of the photosphere.
\label{fig:teradstd}}
\end{figure}

\begin{figure}
\epsscale{1.15}
\plotone{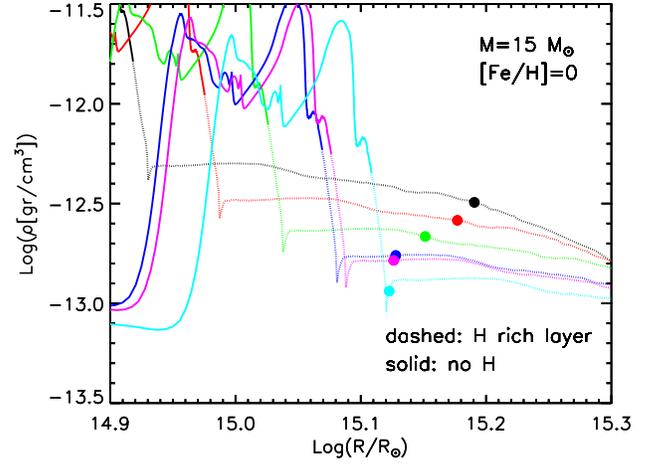}
\caption{Density profile as a function of the radius at 70 (black line), 80 (red line), 90 (green line), 100 (blue line), 101 (magenta line) and 109 days (cyan), respectively. The dashed part of each line refers to the H rich matter while the solid part refers to the region within the He core. The filled dots represent the position of the photosphere.
\label{fig:roradstd}}
\end{figure}

\begin{figure}
\epsscale{1.15}
\plotone{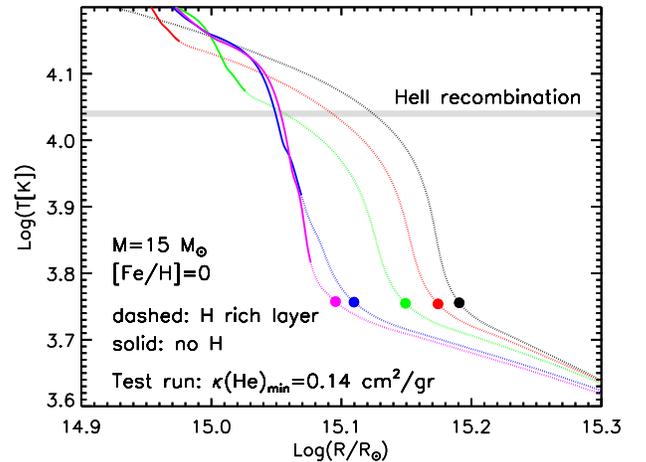}
\caption{Same as Figure \ref{fig:teradstd} but for the model in which we have artificially inhibited the opacity to decrease below $\rm 0.14~cm^2~g^{-1}$ within the He core.\label{fig:teradfloor}}
\end{figure}

In addition to the critical temperature that controls the H recombination $\rm (\sim 5600~K)$ and hence the position of the photosphere, there is another key temperature: the recombination temperature of He {\RomanNumeral{2}} $\rm (\sim 11000~K)$. This is a crucial temperature because it sharply changes the fraction of free electrons and hence the opacity within the He core. Figure \ref{fig:fracele_multi_snap} shows the fraction of free electrons at various times: within the He core (light grey area) He {\RomanNumeral{3}} recombines very early (within the first 50 days or so) when the photosphere is still very far from the H/He discontinuity and reduces the fraction of free electrons from $\sim 1$ to $\sim0.5$. He {\RomanNumeral{2}} begins to recombine roughly at day 90 and in 20 days or so most of the He core is recombined. It is important to note that such a recombination occurs when the photosphere is quite close to the H/He discontinuity. Figures \ref{fig:opa_contour} and \ref{fig:temp_contour} show very clearly what happens when He {\RomanNumeral{2}} recombines. A low opacity region begins to form around day 90 in the He core while on top of it the H rich matter is still ionized and hence it still has a quite high opacity. Figure \ref{fig:teradstd} shows how the temperature profile changes in time: the dashed part of each line refers to the H rich matter while the solid part refers to the region within the He core. The horizontal grey line shows the critical temperature below which He {\RomanNumeral{2}} recombines. The filled dots represent the position of the photosphere. Within the first 100 days or so, the radial temperature profile preserves its shape (lines black, red, green and blue in Figure \ref{fig:teradstd}). Up to this time all (or most of) the He core is at temperature higher than $11000~K$. But between day 100 and 110, He {\RomanNumeral{2}} recombines, the opacity drops down and a significant fraction of the energy stored in the He core flows outward up to the high opacity region where this extra energy is absorbed. Such a sudden injection of energy keeps the temperature of these H rich  layers quite high in spite of the continuous expansion. Lines blue, magenta and cyan in Figure \ref{fig:teradstd} clearly show that up to day 109 or so the temperature profile remains roughly constant in the region where $\tau=2/3$, i.e. between ${\rm Log}(R/R_\odot)=15.10-15.14$. Only when all this extra energy is radiated away the temperature profile will start moving inward again, and hence the photosphere as well. Three days later (day 112) the recombination front has moved down to the CO core and since at this point the energy stored in the He core is too low to maintain the luminosity level of the plateau phase, the light curve bends down landing on the radioactive tail that from now on dominates the light curve.

The sharp release of energy from the He core explains why the light curve does not bend down when He {\RomanNumeral{2}} recombines but it does not explain, by itself, why the luminosity actually increases for a while creating a bump. We must remind at this point that the temporal evolution of the density does not depend on the temperature or the position of the photosphere since the expansion is homologous in this phase, but it only depends on the expansion velocity of the various layers. So we are facing in the region of interest, i.e. around ${\rm Log}(R/R_\odot)=15.10-15.14$, a situation in which the density lowers progressively (Figure \ref{fig:roradstd}) while the temperature does not. Since $\tau$ scales directly with both the opacity and the density $\rm (\tau=\int\kappa\rho dr)$, a reduction of the density requires an increase of the opacity to keep the photosphere at $\tau=2/3$. But the opacity scales mainly with the temperature (only very mildly with the density in these conditions) so that $\tau=2/3$ requires a temperature higher if the density reduces. In addition to this, also the radius of the photosphere slightly increases between day 100 and 109 $(\rm \Delta Log(R)\sim0.005)$. Quantitatively, the luminosity increase at the bump is of the order of $\Delta {\rm Log} (L)=0.05$ (by the way a very modest increase!) and the temperature of the photosphere increases by $\Delta {\rm Log}(T_{\rm eff})\sim 0.01$. Since $\Delta {\rm Log}(L)=2\Delta {\rm Log}(R) + 4\Delta {\rm Log}(T_{\rm eff})$, the temperature increase explains $\sim 80\%$ of the luminosity increase, the remaining $\sim 20\%$ being due to the small increase of the radius of the photosphere.

\begin{figure}[ht!]
\epsscale{1.15}
\plotone{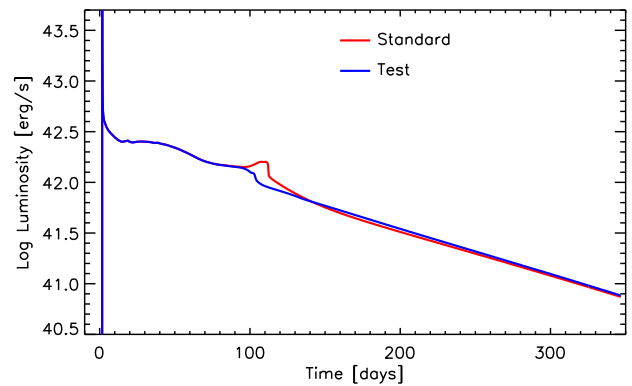}
\caption{Comparison between the bolometric observed light curve of the standard model (red line) and of  a test model in which the opacity floor is set artificially to $\rm 0.14~cm^2~g^{-1}$ within the He core (blue line) \label{fig:compare_lc_opafloor}}
\end{figure}

In order to verify the role played by the opacity drop due to the recombination of the He {\RomanNumeral{2}} on the light curve, we have computed a test model in which we have artificially inhibited the opacity to drop below $\rm 0.14~cm^2~g^{-1}$, i.e. the value of the opacity before the He {\RomanNumeral{2}} recombination, within the He core.
The evolution of the temperature profile of this test run is shown in Figure \ref{fig:teradfloor}. This Figure is the analogous of Figure \ref{fig:teradstd}. Of course only after day $\sim 90$ the standard and test run start to be  different. The most striking difference between Figures \ref{fig:teradfloor} and \ref{fig:teradstd} is that now the temperature increase in the region between ${\rm Log}(R/R_\odot)=10.10-15.14$ is not present any more and the magenta line (in both cases it refers to day 101) is now free to move leftward, which means that this region can now cool down. So in this case both the temperature and the density drop down and the position of the recombination front may recede in radius forcing the luminosity of the photosphere to decrease. Once the photosphere reaches the H/He discontinuity also in this case the luminosity quickly drops until the radiactive tail shows up.
Figure \ref{fig:compare_lc_opafloor} shows a comparison between the light curves of the reference and test run. By the way note that the high opacity in the He core produces also a slightly shorter plateau.

Though this test clearly confirms our analysis of the reference run, it is obviously an unphysical way to remove the bump. Since, as far as we know, this feature is not observed in the SNe IIP light curves, it is important to try to identify which {\it real} phenomenon (or phenomena) controls the presence of the bump but also the shape of the light curve while it bends towards the radioactive tail. \citet{utrobin17} studied in detail such a problem and showed that it depends in general on different factors like, e.g., the presence of both a density "bump" in the He core, the sharp change of chemical composition close to the H/He interface and also on the spatial distribution of the $\rm ^{56}Ni$ produced during the explosion. They concluded that a proper combination of an artificial smoothing of the density gradient and of the chemical composition at the H/He interface and also of the $\rm ^{56}Ni$ profile, prevents the formation of the luminosity bump in the transition phase from the plateau to the radioactive tail. Such a kind of smoothing and mixing should mimic, indeed, multidimensional effects in spherical symmetry.

\begin{figure}
\epsscale{1.15}
\plotone{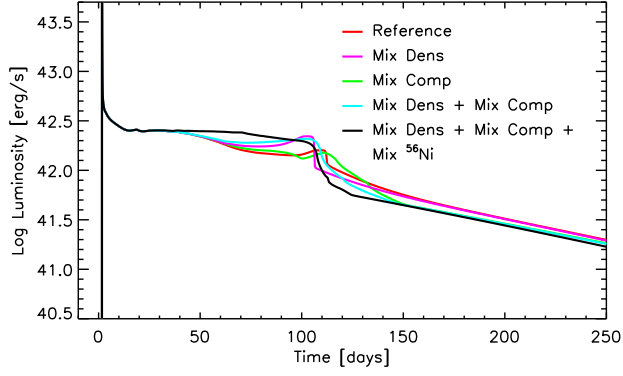}
\caption{Light curve of the reference model (red line) compared to the ones obtained for various assumptions: the density around the H/He interface is artificially smoothed (magenta line); the chemical composition is artificially mixed, keeping the original density gradient (green line); both the density and the composition are smoothed (cyan line)   \label{fig:compare_lc_varitest}}
\end{figure}

We made some tests analogous to those presented by \cite{utrobin17} and we basically confirm their finding. Figure \ref{fig:compare_lc_varitest} summarizes our tests.

\begin{figure}[ht!]
\epsscale{1.15}
\plotone{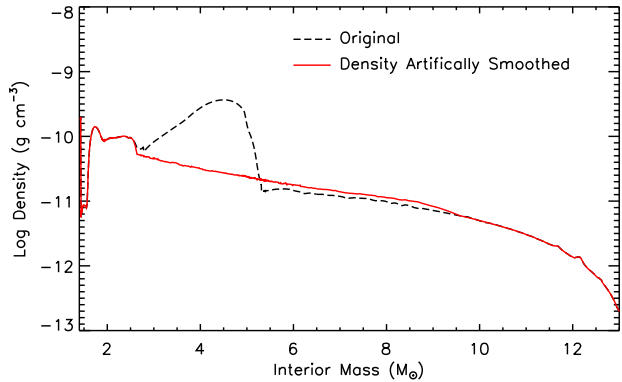}
\caption{Density profile as a function of the interior mass for the reference model $\rm \sim 2\cdot 10^{6}~s$ after the onset of the explosion. The black dahsed line refers to the original model while the red solid line refers to the model in which the density is artificially smoothed (see text). \label{fig:density_smooth}}
\end{figure}

The first test is the one in which we artificially smooth the density gradient around the H/He interface $\rm \sim 2\cdot 10^{6}~s$ after the explosion. As it is shown in Figure \ref{fig:density_smooth}, the density is smoothed between $\rm \sim 2.6~M_\odot$ and $\rm \sim 9.2~M_\odot$. Figure \ref{fig:compare_lc_varitest} (magenta line) shows that, as it has been also found by \citet{utrobin17}, such a smoothing implies a shorter plateau and a more pronounced bump in the transition phase from the plateau to the radioactive tail, compared to the standard model. 

\begin{figure}[ht!]
\epsscale{1.15}
\plotone{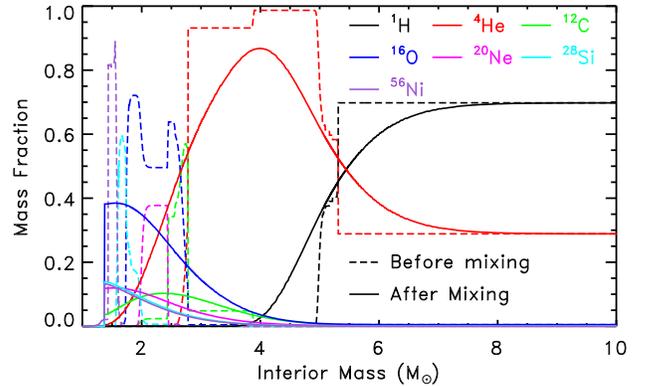}
\caption{Interior chemical composition of the reference model before and after mixing of chemical composition (see text).\label{fig:mix_chim}}
\end{figure}

In the second test, we keep the original density profile while we smooth artificially the chemical composition, when the elapsed time after the onset of the explosion is $\rm ~5\cdot 10^{5}~s$, by means of a "boxcar" averaging \citep{kw09} with a boxcar mass width of $\rm \Delta m=0.4~M_\odot$ (see Figure \ref{fig:mix_chim}). 
More specifically, the abundance of each nuclear species $k$ in each zone $j$ is defined as:

\begin{equation}
    X_{k,j}={1\over \Delta m}\sum_{i=j}^{j_{\Delta m}} X_{k,i}~~~j=1,N
\end{equation}

where $j_{\Delta m}$ is the zone such that $m(j_{\Delta m})-m(j)=\Delta m$ and $N$ is the total number of zones. This calculation is then repeated $n=4$ times.

In this case the spike is still present and the main effect of such a mixing is that of making the transition between the plateau phase to the radioactive tail smoother (green line in Figure \ref{fig:compare_lc_varitest}). Note that, the radioactive tail is slightly less luminous compared to the reference one because of a general decrease of the electron fraction in the ejecta that implies a decrease of the $\gamma$-ray opacity ($\kappa_\gamma=0.06~Y_{\rm e}~{\rm cm^2~g^{-1}}$). 

\begin{figure}[h]
\epsscale{1.15}
\plotone{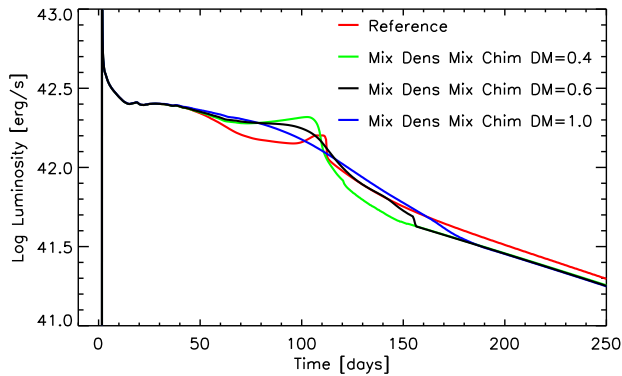}
\caption{ Light curve of the reference model (red line) compared to the ones obtained for various assumptions on the boxcar mass width: $\Delta m=0.4$ (green line), $\Delta m=0.6$ (black line), $\Delta m=1.0$ (blue line).\label{fig:compare_lc_varitest_mix}}
\end{figure}

In the third test we apply both a smoothing of the density and a mixing of the composition, as described above. In this case the two effects discussed in the previous two tests add up to each other (cyan line in Figure \ref{fig:compare_lc_varitest}). In this case, the impact of a difference choice of the boxcar mass width is shown in Figure \ref{fig:compare_lc_varitest_mix}. In general the thicker the boxcar mass, the flatter the plateau and the smoother the transition from the plateau phase to the radioactive tail. Note that, in none of these cases both a flat plateau and a rapid decline of the luminosity from the plateau phase to the radioactive tail have been obtained.

The last test is similar to the third test but with an additional homogeneous mixing of the $\rm ^{56}Ni$ produced during the explosion, from the inner edge of the exploding mantle ($\rm 1.4~M_\odot$) to about half of the H-rich envelope ($\rm 9.0~M_\odot$). This additional mixing of the $\rm ^{56}Ni$ produces an early contribution of the $\gamma$ rays to the luminosity, and this implies a flatter plateau, the disappearance of the spike and a rapid decline of the luminosity in the transition phase from the plateau to the radioactive tail (black line in Figure \ref{fig:compare_lc_varitest}). A similar result has been also obtained by \cite{bersten11} (see their Figure 12). Let us eventually mention that an efficient mixing of $\rm ^{56}Ni$ into the H-rich layer is not unreasonable and it has been confirmed by studies on SN1987A 
\citep{w88,a88,blin20}.

\begin{figure}[h]
\epsscale{1.15}
\plotone{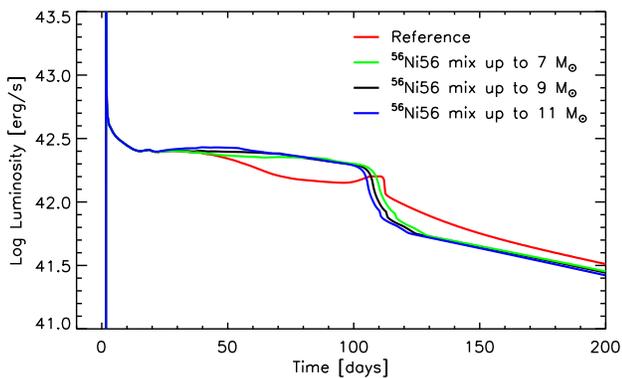}
\caption{ Light curve of the reference model (red line) compared to the ones obtained for various assumptions on the out mass coordinate up to which the $\rm ^{56}Ni$ is homogeneously mixed.\label{fig:compare_lc_ni_nimix}}
\end{figure}

As a final comment let us note that a variation of the outer mass coordinate up to which $\rm ^{56}Ni$ is homogeneously mixed affects only mildly the overall shape of the light curve, i.e., it slightly changes the length of the plateau (Figure \ref{fig:compare_lc_ni_nimix}). Therefore the choice of this quantity is not crucial in deriving the physical parameters of the progenitor star from the light curve fitting (see the next section).

\section{Explosions and Light Curves of Red Super Giant models} \label{sec:results}
In the previous section we described in detail the explosion of a star that may be considered typical, i.e. a non rotating, solar metallicity $\rm 15~M_\odot$ model with $E_{\rm expl}=1.0$ foe and its associated bolometric light curve. 

In this section we study and discuss how the bolometric light curve depends on the explosion energy. The reason for such a parametric study is that, while in a "real" core collapse supernova the energy of the explosion is a natural outcome of the explosion itself (and it is uniquely determined by the initial mass, metallicity and eventually the initial rotational velocity of the progenitor star), our modeling of the explosion requires the injection of some arbitrary amount of energy to generate the shock wave (like the vast majority of similar computations available in the literature, see section \ref{sec:exp15a}). This is the reason why we are forced to compute grid of simulations for different (arbitrary) amount of explosion energies that in most cases lead to results that do not correspond neither to more sophisticated multi-dimensions explosions nor to the typical observed values.

Since we are focusing on the bolometric light curves of the Type IIP supernovae, we computed the explosions only of the subset of models present in our database published in \cite{lc18} that reach the core collapse as red supergiant stars.

The main properties of these models, relevant for the light curve calculations, are reported in Table \ref{propsn}: model identifier (column 1); initial mass (column 2); initial metallicity (column 3); effective temperature (column 4); luminosity (column 5); mass at the time of the explosion (column 6); mass of the He core (column 7);  mass of the CO core (column 8);  mass of the Fe core (column 9); amount of H and He in the envelope (columns 10 and 11, respectively); binding energy of the material outside the Fe core (column 12); mass of the H envelope (defined as the difference between the total mass and the mass coordinate where the H mass fraction drops below $10^{-4}$) in units of $\rm 10~M_\odot$ (column 13); total radius in units of $\rm 500~R_\odot$ (column 14).

\floattable
\startlongtable
\begin{deluxetable*}{lccccccccccccc}
\tablecolumns{11}
\tablewidth{0pc}
\tablecaption{Properties of the presupernova models \label{propsn}}
\tablehead{
Model &  $\rm M_i$   & [Fe/H] &    Log($\rm T_{eff}$) & Log($\rm L/L_\odot$) &  $\rm M_f$           &  $\rm M_{He}$     &    $\rm M_{CO}$  &    $\rm M_{Fe}$  &      H           &        He        &  $\rm E_{bind}$ & $M_{10}$ & $R_{500}$\\
    Id.       & ($\rm M_\odot$) & &          (K)          &      & ($\rm M_\odot$)   & ($\rm M_\odot$)  &  ($\rm M_\odot$) &  ($\rm M_\odot$) &  ($\rm M_\odot$) &  ($\rm M_\odot$) & ($\rm 10^{51}~erg$) &  ($\rm 10~M_\odot$) &  ($\rm 500~R_\odot$)}
\startdata
13a & 13  & 0 &  3.55 &  4.82 &  11.9 &  4.08 &  2.03 &  1.36 &  5.37 &  4.31 &  0.65 & 0.784 & 1.324 \\ 
15a & 15  & 0 &  3.54 &  4.98 &  13.3 &  4.95 &  2.78 &  1.43 &  5.67 &  4.63 &  0.95 & 0.833 & 1.678 \\
    &     &   &       &       &       &       &       &       &       &       &       &       &       \\
13b & 13  &-1 &  3.60 &  4.85 &  12.5 &  4.26 &  2.13 &  1.19 &  5.83 &  4.44 &  1.08 & 0.826 & 1.125 \\
15b & 15  &-1 &  3.59 &  5.05 &  14.2 &  5.22 &  3.01 &  1.40 &  6.34 &  4.54 &  1.33 & 0.900 & 1.477 \\
20b & 20  &-1 &  3.59 &  5.26 &  18.4 &  7.52 &  4.21 &  1.43 &  7.47 &  6.50 &  1.79 & 1.090 & 1.862 \\
25b & 25  &-1 &  3.58 &  5.48 &  20.6 & 10.20 &  6.82 &  1.59 &  6.96 &  6.62 &  4.02 & 1.049 & 2.521 \\
    &     &   &       &       &       &       &       &       &       &       &       &       &       \\
13c & 13  &-2 &  3.65 &  4.88 &  13.0 &  4.34 &  2.14 &  1.40 &  6.23 &  4.44 &  0.85 & 0.866 & 0.935 \\
15c & 15  &-2 &  3.64 &  5.01 &  14.8 &  5.21 &  2.72 &  1.08 &  6.86 &  5.09 &  1.70 & 0.964 & 1.092 \\
20c & 20  &-2 &  3.64 &  5.27 &  19.7 &  7.49 &  4.23 &  1.43 &  8.60 &  6.71 &  1.83 & 1.230 & 1.483 \\
25c & 25  &-2 &  3.67 &  5.20 &  24.7 &  9.87 &  5.93 &  1.53 & 10.20 &  8.26 &  2.54 & 1.490 & 1.220 \\
    &     &   &       &       &       &       &       &       &       &       &       &       &       \\
13d & 13  &-3 &  3.66 &  4.88 &  13.0 &  4.22 &  2.15 &  1.15 &  6.22 &  4.44 &  1.30 & 0.878 & 0.857 \\
15d & 15  &-3 &  3.66 &  5.06 &  15.0 &  5.22 &  3.09 &  1.46 &  6.95 &  4.62 &  1.45 & 0.981 & 1.082 \\
20d & 20  &-3 &  3.66 &  5.26 &  19.8 &  7.42 &  4.35 &  1.44 &  8.64 &  6.63 &  1.86 & 1.252 & 1.358 \\
25d & 25  &-3 &  3.66 &  5.46 &  24.6 &  9.84 &  6.29 &  1.53 & 10.13 &  8.00 &  2.89 & 1.495 & 1.709 \\
\enddata
\end{deluxetable*}

For each presupernova progenitor (reported in Table \ref{propsn}) we computed a grid of different explosions for various explosion energies. All the explosions were computed by assuming a smoothing density profile as well as a mixing of the chemical composition and $\rm ^{56}Ni$ as described in the previous paragraph. In particular, $\rm ^{56}Ni$ is always mixed from the base of the ejecta, after the shock breakout and after the fallback is ended, up to half of the H-rich envelope. As Figure \ref{fig:compare_lc_ni_nimix} shows, this choice does not affect significantly the shape of the light curve in the late stages of the plateau and in the transition between the plateau and the radioactive tail.

The main results of these calculations are reported in Table \ref{tabexpl}: model identifier (column 1); explosion energy (column 2) (mainly the kinetic energy of the ejecta); time elapsed at the shock breakout (column 3); time to the end of the fall back of material onto the remnant (column 4); amount of $\rm ^{56}Ni$ ejected (column 5); mass of the remnant, including the fallback material (column 6); mass of the ejecta (column 7); bolometric luminosities 30 and 50 days after the shock breakout (columns 8 and 9, respectively); time duration of the plateau phase, in days, assuming that the beginning of the plateau coincides with the shock breakout and defining the end of the plateau when the radius of the photosphere reduces to $50\%$ of its maximum value (column 10).

\floattable
\startlongtable
\begin{deluxetable*}{lccccccccc}
\tablewidth{0pt}
\tablecaption{Main results of the explosion calculations\label{tabexpl}}
\tablehead{\colhead{Model} &  \colhead{$E_{\rm expl}$} &\colhead{$t_{\rm break~out}$} & \colhead{$t_{\rm fallback}$} & \colhead{$\rm ^{56}Ni_{\rm ejected}$} & \colhead{$M_{\rm rem}$} & \colhead{$M_{\rm ejecta}$} & \colhead{${\rm Log} (L_{30})$} & \colhead{${\rm Log} (L_{50})$} & \colhead{$t_{\rm plateau}$}\\
\colhead{Id.} & \colhead{($\rm erg$)} & \colhead{(s)} & \colhead{(s)} & \colhead{($\rm M_\odot$)} & \colhead{($\rm M_\odot$)} & \colhead{($\rm M_\odot$)} & \colhead{($\rm erg~s^{-1}$)} & \colhead{($\rm erg~s^{-1}$)} & \colhead{(days)}   
}
\startdata
  13a   &  1.99E+50  &   2.35E+05  &  2.92E+07 &  9.78E-40  &  2.15  &   9.71 &   41.703 & 41.658 & 1.27E+02 \\
  13a   &  2.50E+50  &   2.11E+05  &  1.73E+07 &  9.84E-40  &  2.03  &   9.83 &   41.796 & 41.752 & 1.17E+02 \\
  13a   &  5.34E+50  &   1.49E+05  &  1.29E+06 &  7.48E-05  &  1.60  &  10.26 &   42.087 & 42.018 & 1.07E+02 \\
  13a   &  1.05E+51  &   1.12E+05  &  1.21E+02 &  1.46E-01  &  0.86  &  11.00 &   42.342 & 42.339 & 1.12E+02 \\
  13a   &  1.56E+51  &   9.16E+04  &  7.79E+01 &  1.62E-01  &  0.84  &  11.02 &   42.502 & 42.478 & 9.93E+01 \\
  13a   &  2.08E+51  &   8.02E+04  &  0.00E+00 &  1.72E-01  &  0.81  &  11.05 &   42.606 & 42.576 & 9.08E+01 \\
        &            &             &           &            &        &        &          &        &          \\
  15a   &  2.17E+50  &   2.92E+05  &  1.79E+07 &  1.03E-39  &  3.00  &  10.23 &   41.808 & 41.789 & 1.30E+02 \\
  15a   &  2.43E+50  &   2.78E+05  &  1.53E+07 &  1.04E-39  &  2.89  &  10.35 &   41.853 & 41.833 & 1.27E+02 \\
  15a   &  2.74E+50  &   2.61E+05  &  2.55E+07 &  1.05E-39  &  2.89  &  10.34 &   41.905 & 41.880 & 1.23E+02 \\
  15a   &  5.88E+50  &   1.87E+05  &  1.62E+05 &  6.33E-17  &  2.14  &  11.09 &   42.192 & 42.141 & 1.12E+02 \\
  15a   &  1.05E+51  &   1.45E+05  &  1.07E+04 &  1.26E-01  &  1.41  &  11.82 &   42.403 & 42.395 & 1.15E+02 \\
  15a   &  1.55E+51  &   1.21E+05  &  2.22E+02 &  1.51E-01  &  0.89  &  12.35 &   42.555 & 42.534 & 9.99E+01 \\
  15a   &  2.07E+51  &   1.06E+05  &  1.51E+02 &  1.74E-01  &  0.85  &  12.38 &   42.670 & 42.624 & 9.13E+01 \\
        &            &             &           &            &        &        &          &        &          \\
  13b   &  1.88E+50  &   2.11E+05  &  2.19E+07 &  1.02E-39  &  2.32  &  10.17 &   41.621 & 41.588 & 1.17E+02 \\
  13b   &  2.12E+50  &   2.00E+05  &  2.25E+07 &  1.03E-39  &  2.25  &  10.24 &   41.668 & 41.634 & 1.14E+02 \\
  13b   &  2.41E+50  &   1.87E+05  &  1.75E+07 &  1.03E-39  &  2.22  &  10.26 &   41.722 & 41.683 & 1.12E+02 \\
  13b   &  5.29E+50  &   1.31E+05  &  1.59E+06 &  3.91E-13  &  1.76  &  10.73 &   42.020 & 41.960 & 1.04E+02 \\
  13b   &  1.07E+51  &   9.79E+04  &  3.17E+02 &  3.34E-01  &  0.85  &  11.63 &   42.278 & 42.389 & 1.31E+02 \\
  13b   &  1.59E+51  &   8.04E+04  &  5.93E+01 &  3.53E-01  &  0.83  &  11.66 &   42.446 & 42.534 & 1.14E+02 \\
  13b   &  2.12E+51  &   6.98E+04  &  0.00E+00 &  3.64E-01  &  0.81  &  11.68 &   42.567 & 42.637 & 1.03E+02 \\
        &            &             &           &            &        &        &          &        &          \\
  15b   &  2.19E+50  &   2.63E+05  &  2.27E+07 &  1.08E-39  &  3.46  &  10.71 &   41.762 & 41.749 & 1.33E+02 \\
  15b   &  2.44E+50  &   2.50E+05  &  1.95E+07 &  1.09E-39  &  3.46  &  10.71 &   41.805 & 41.789 & 1.23E+02 \\
  15b   &  5.91E+50  &   1.68E+05  &  1.93E+05 &  1.36E-23  &  2.47  &  11.71 &   42.146 & 42.102 & 1.17E+02 \\
  15b   &  1.06E+51  &   1.31E+05  &  2.80E+04 &  2.59E-02  &  1.60  &  12.58 &   42.358 & 42.316 & 1.00E+02 \\
  15b   &  1.56E+51  &   1.10E+05  &  2.03E+04 &  2.07E-01  &  1.33  &  12.85 &   42.511 & 42.516 & 1.15E+02 \\
  15b   &  2.08E+51  &   9.71E+04  &  7.93E+05 &  2.31E-01  &  1.31  &  13.28 &   42.625 & 42.672 & 8.75E+01 \\
        &            &             &           &            &        &        &          &        &          \\
  20b   &  2.35E+50  &   3.62E+05  &  4.26E+06 &  1.34E-39  &  5.02  &  13.34 &   41.804 & 41.801 & 1.40E+02 \\
  20b   &  2.63E+50  &   3.46E+05  &  4.17E+05 &  1.35E-39  &  4.87  &  13.48 &   41.845 & 41.845 & 1.38E+02 \\
  20b   &  2.90E+50  &   3.27E+05  &  6.51E+06 &  1.36E-39  &  4.83  &  13.52 &   41.894 & 41.897 & 1.36E+02 \\
  20b   &  5.93E+50  &   2.36E+05  &  9.30E+06 &  1.44E-39  &  4.01  &  14.34 &   42.182 & 42.163 & 1.15E+02 \\
  20b   &  1.10E+51  &   1.83E+05  &  2.13E+05 &  2.67E-17  &  2.53  &  15.83 &   42.390 & 42.348 & 1.06E+02 \\
  20b   &  1.60E+51  &   1.54E+05  &  2.21E+05 &  5.69E-09  &  1.95  &  16.41 &   42.525 & 42.474 & 9.17E+01 \\
  20b   &  2.12E+51  &   1.37E+05  &  3.83E+04 &  1.68E-01  &  1.57  &  16.79 &   42.628 & 42.606 & 9.92E+01 \\
        &            &             &           &            &        &        &          &        &          \\
  25b   &  4.39E+50  &   3.59E+05  &  1.99E+05 &  1.31E-39  &  7.54  &  13.04 &   42.143 & 42.189 & 1.26E+02 \\
  25b   &  1.15E+51  &   2.36E+05  &  1.68E+06 &  4.32E-19  &  5.67  &  14.91 &   42.513 & 42.515 & 1.04E+02 \\
  25b   &  1.62E+51  &   2.05E+05  &  1.36E+05 &  2.99E-18  &  4.04  &  16.54 &   42.635 & 42.622 & 9.49E+01 \\
  25b   &  2.12E+51  &   1.85E+05  &  1.09E+07 &  5.71E-12  &  3.24  &  17.34 &   42.726 & 42.708 & 8.42E+01 \\
        &            &             &           &            &        &        &          &        &          \\
  13c   &  2.37E+50  &   1.59E+05  &  2.82E+07 &  1.06E-39  &  2.39  &  10.57 &   41.639 & 41.606 & 1.08E+02 \\
  13c   &  5.62E+50  &   1.07E+05  &  9.29E+05 &  1.58E-13  &  1.94  &  11.03 &   41.974 & 41.916 & 9.65E+01 \\
  13c   &  1.06E+51  &   8.23E+04  &  1.84E+02 &  2.28E-01  &  0.90  &  12.07 &   42.202 & 42.292 & 1.24E+02 \\
  13c   &  1.59E+51  &   6.84E+04  &  1.66E+02 &  2.52E-01  &  0.85  &  12.11 &   42.381 & 42.449 & 1.08E+02 \\
  13c   &  2.11E+51  &   5.93E+04  &  6.46E+01 &  2.70E-01  &  0.84  &  12.13 &   42.491 & 42.552 & 9.87E+01 \\
        &            &             &           &            &        &        &          &        &          \\
  15c   &  2.02E+50  &   2.12E+05  &  1.31E+07 &  1.17E-39  &  3.17  &  11.62 &   41.601 & 41.578 & 1.18E+02 \\
  15c   &  2.28E+50  &   1.99E+05  &  1.89E+07 &  1.18E-39  &  3.07  &  11.72 &   41.650 & 41.625 & 1.15E+02 \\
  15c   &  5.69E+50  &   1.32E+05  &  5.91E+05 &  9.05E-25  &  2.50  &  12.29 &   42.011 & 41.959 & 1.02E+02 \\
  15c   &  1.09E+51  &   1.01E+05  &  1.88E+05 &  4.09E-01  &  0.98  &  13.81 &   42.228 & 42.331 & 1.49E+02 \\
  15c   &  1.61E+51  &   8.47E+04  &  1.60E+02 &  4.42E-01  &  0.87  &  13.92 &   42.383 & 42.499 & 1.27E+02 \\
  15c   &  2.14E+51  &   7.37E+04  &  2.39E+01 &  4.69E-01  &  0.84  &  13.95 &   42.504 & 42.612 & 1.16E+02 \\
        &            &             &           &            &        &        &          &        &          \\
  20c   &  2.33E+50  &   3.01E+05  &  2.92E+07 &  1.47E-39  &  5.27  &  14.46 &   41.700 & 41.703 & 1.38E+02 \\
  20c   &  2.61E+50  &   2.87E+05  &  7.84E+06 &  1.48E-39  &  4.99  &  14.73 &   41.744 & 41.739 & 1.35E+02 \\
  20c   &  5.94E+50  &   1.96E+05  &  1.53E+07 &  1.57E-39  &  4.08  &  15.64 &   42.085 & 42.058 & 1.17E+02 \\
  20c   &  1.09E+51  &   1.53E+05  &  5.12E+05 &  3.67E-16  &  2.58  &  17.14 &   42.293 & 42.248 & 1.02E+02 \\
  20c   &  1.59E+51  &   1.28E+05  &  3.00E+05 &  1.14E-11  &  1.99  &  17.74 &   42.428 & 42.377 & 9.14E+01 \\
  20c   &  2.12E+51  &   1.13E+05  &  2.67E+04 &  1.76E-01  &  1.58  &  18.14 &   42.531 & 42.511 & 1.04E+02 \\
        &            &             &           &            &        &        &          &        &          \\
  25c   &  2.38E+50  &   2.48E+05  &  7.03E+06 &  1.74E-39  &  7.42  &  17.23 &   41.680 & 41.593 & 1.30E+02 \\
  25c   &  2.66E+50  &   2.35E+05  &  2.15E+05 &  1.75E-39  &  7.22  &  17.44 &   41.717 & 41.629 & 1.29E+02 \\
  25c   &  2.98E+50  &   2.21E+05  &  5.30E+05 &  1.77E-39  &  7.03  &  17.62 &   41.762 & 41.668 & 1.27E+02 \\
  25c   &  1.08E+51  &   1.29E+05  &  9.02E+04 &  1.73E-16  &  4.17  &  20.48 &   42.198 & 42.048 & 9.20E+01 \\
  25c   &  1.58E+51  &   1.09E+05  &  3.97E+05 &  3.96E-16  &  2.86  &  21.80 &   42.320 & 42.148 & 8.21E+01 \\
  25c   &  2.08E+51  &   9.60E+04  &  1.96E+05 &  1.68E-11  &  2.38  &  22.28 &   42.408 & 42.214 & 5.06E+01 \\
        &            &             &           &            &        &        &          &        &          \\
  13d   &  2.17E+50  &   1.52E+05  &  2.81E+07 &  1.06E-39  &  2.44  &  10.54 &   41.571 & 41.538 & 1.06E+02 \\
  13d   &  5.38E+50  &   9.99E+04  &  2.76E+06 &  5.64E-14  &  1.90  &  11.07 &   41.924 & 41.861 & 1.00E+02 \\
  13d   &  1.07E+51  &   7.64E+04  &  1.63E+02 &  3.77E-01  &  0.86  &  12.12 &   42.187 & 42.342 & 1.37E+02 \\
  13d   &  1.60E+51  &   6.28E+04  &  4.07E+01 &  4.04E-01  &  0.83  &  12.15 &   42.359 & 42.501 & 1.21E+02 \\
        &            &             &           &            &        &        &          &        &          \\
  15d   &  2.95E+50  &   1.73E+05  &  1.84E+07 &  1.16E-39  &  3.51  &  11.44 &   41.763 & 41.735 & 1.14E+02 \\
  15d   &  6.05E+50  &   1.25E+05  &  1.07E+05 &  6.52E-17  &  2.70  &  12.25 &   42.033 & 41.987 & 1.16E+02 \\
  15d   &  1.07E+51  &   9.81E+04  &  3.10E+04 &  2.61E-02  &  1.70  &  13.25 &   42.237 & 42.201 & 1.03E+02 \\
  15d   &  2.10E+51  &   7.31E+04  &  9.60E+02 &  2.73E-01  &  0.87  &  14.08 &   42.500 & 42.542 & 1.07E+02 \\
        &            &             &           &            &        &        &          &        &          \\
  20d   &  6.18E+50  &   1.80E+05  &  2.25E+05 &  1.58E-39  &  4.03  &  15.76 &   42.054 & 42.018 & 1.14E+02 \\
  20d   &  1.10E+51  &   1.40E+05  &  2.61E+05 &  2.36E-16  &  2.57  &  17.22 &   42.261 & 42.210 & 1.04E+02 \\
  20d   &  1.59E+51  &   1.19E+05  &  1.58E+05 &  9.96E-10  &  2.03  &  17.76 &   42.398 & 42.340 & 8.87E+01 \\
  20d   &  2.12E+51  &   1.04E+05  &  2.89E+04 &  2.14E-01  &  1.56  &  18.23 &   42.501 & 42.484 & 5.04E+01 \\
        &            &             &           &            &        &        &          &        &          \\
  25d   &  1.09E+51  &   1.90E+05  &  1.87E+07 &  1.90E-16  &  4.65  &  19.98 &   42.298 & 42.273 & 1.17E+02 \\
  25d   &  1.59E+51  &   1.62E+05  &  7.22E+04 &  5.16E-16  &  3.13  &  21.50 &   42.429 & 42.404 & 1.03E+02 \\
  25d   &  2.11E+51  &   1.44E+05  &  2.19E+05 &  6.72E-12  &  2.55  &  22.08 &   42.531 & 42.500 & 9.50E+01 \\
\enddata                                              
 \end{deluxetable*}

\begin{figure}[h]
\epsscale{1.15}
\plotone{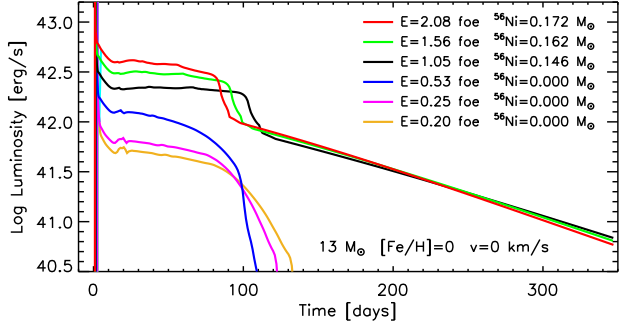}
\caption{Bolometric light curves of a non rotating, solar metallicity $\rm 13~M_\odot$, for different explosion energies. The $\rm ^{56}Ni$ shown in the legend is the one produced during the explosion. In each case we assume a smoothing of the density and a mixing of the chemical composition and $\rm ^{56}Ni$ as described in the text.\label{fig:lc_013a000}}
\end{figure}

\begin{figure}[h]
\epsscale{1.15}
\plotone{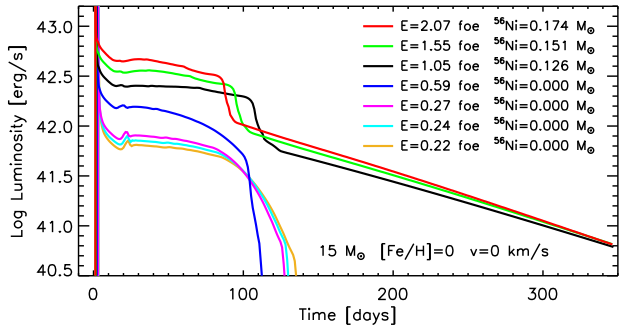}
\caption{As Figure \ref{fig:lc_013a000} but for a non rotating, solar metallicity $\rm 15~M_\odot$.\label{fig:lc_015a000}}
\end{figure}

\begin{figure}[h]
\epsscale{1.15}
\plotone{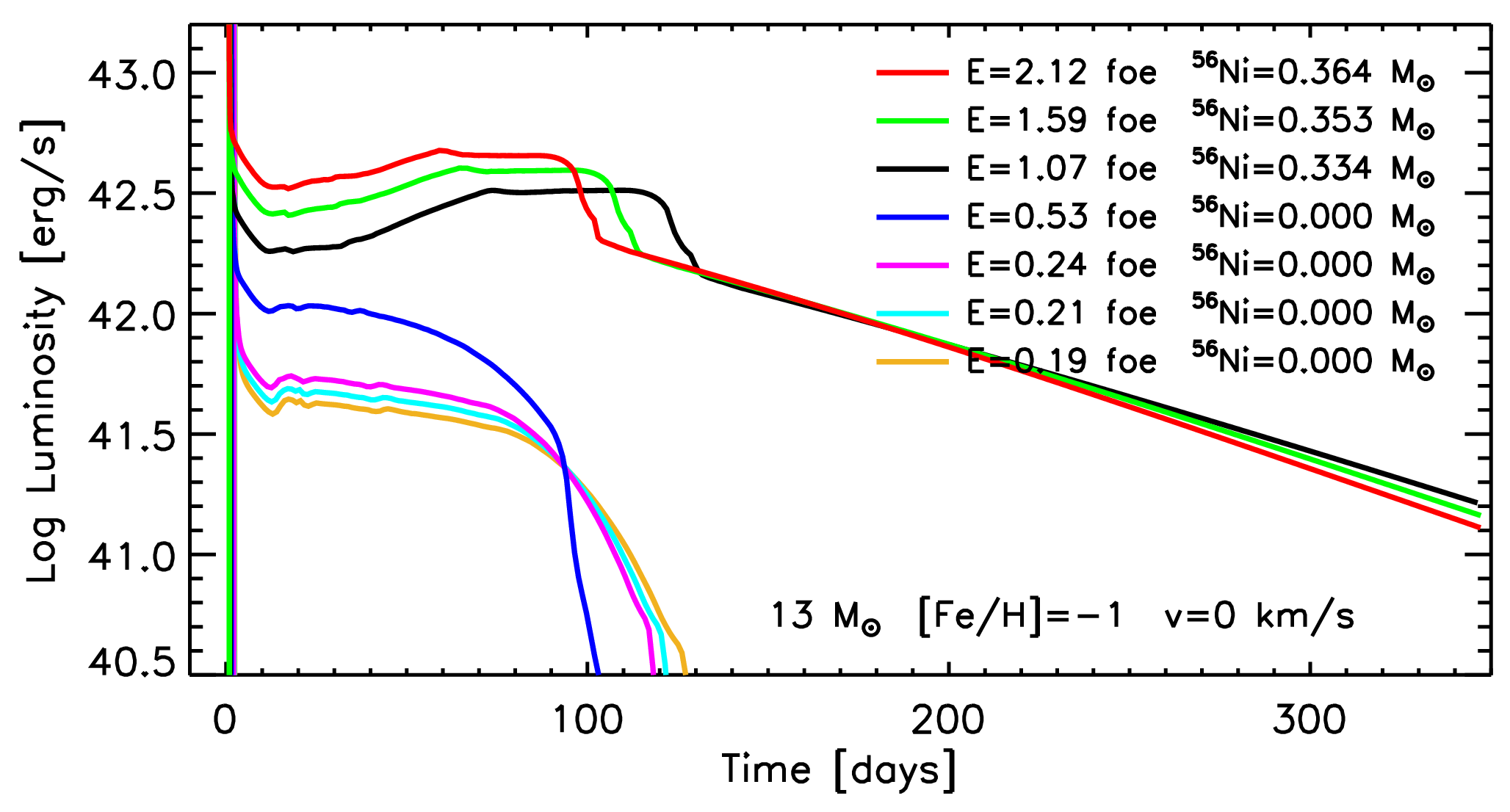}
\caption{As Figure \ref{fig:lc_013a000} but for a non rotating, $\rm 13~M_\odot$ with initial composition corresponding to [Fe/H]=-1 (see text).\label{fig:lc_013b000}}
\end{figure}

\begin{figure}[h]
\epsscale{1.15}
\plotone{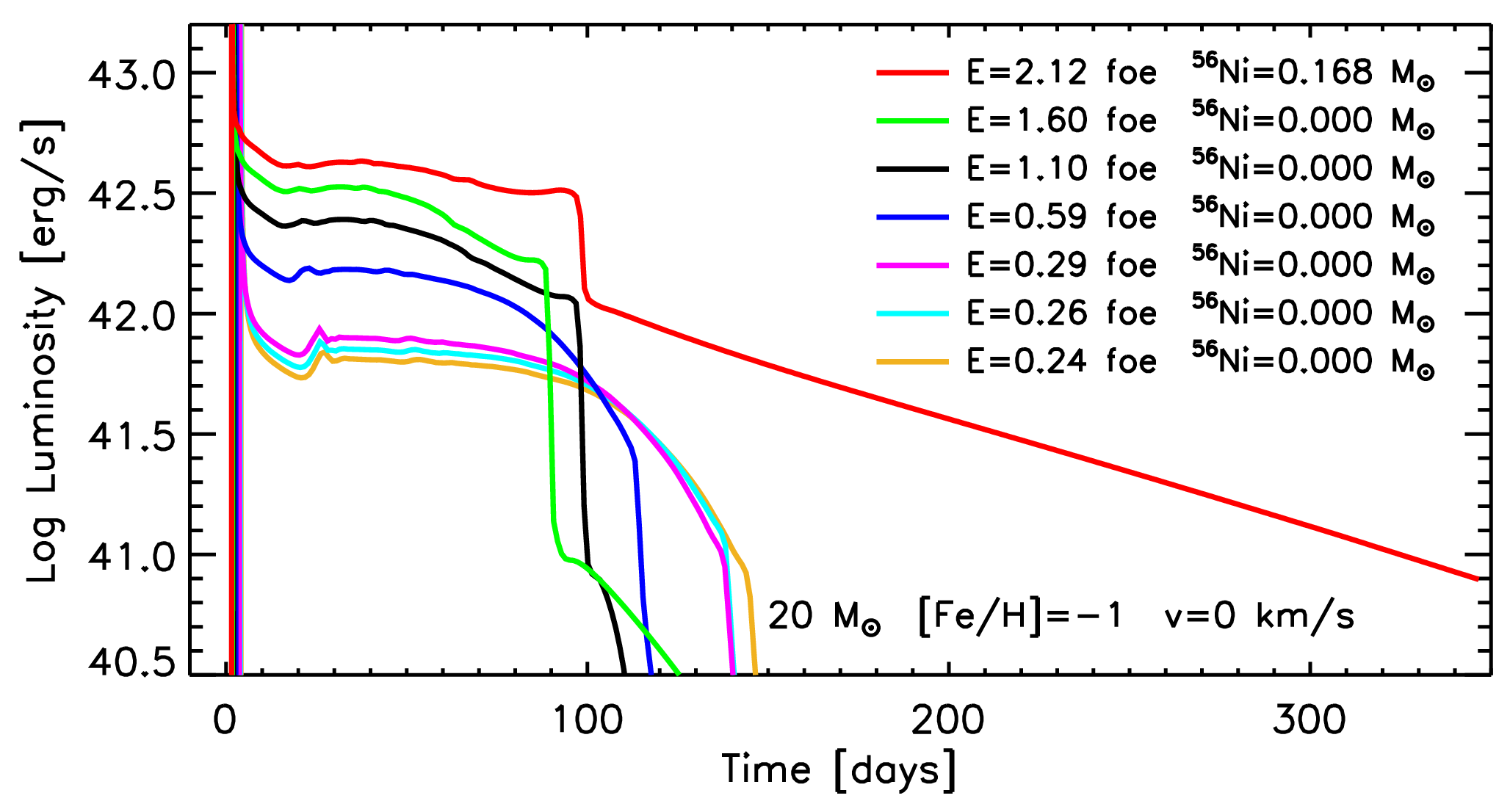}
\caption{As Figure \ref{fig:lc_013a000} but for a non rotating, $\rm 20~M_\odot$ with initial composition corresponding to [Fe/H]=-1 (see text).\label{fig:lc_020b000}}
\end{figure}

\begin{figure}[h]
\epsscale{1.15}
\plotone{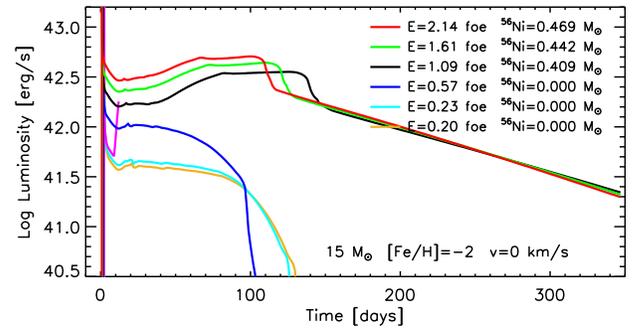}
\caption{As Figure \ref{fig:lc_013a000} but for a non rotating, $\rm 15~M_\odot$ with initial composition corresponding to [Fe/H]=-2 (see text).\label{fig:lc_015c000}}
\end{figure}

Figures \ref{fig:lc_013a000}, \ref{fig:lc_015a000}, \ref{fig:lc_013b000}, \ref{fig:lc_020b000} and \ref{fig:lc_015c000} show the light curves obtained for some selected progenitor models as a function of the explosion energy.

These figures visually show how the shape of the light curve depends on the progenitor mass, the initial metallicity and the explosion energy. In general, for the same progenitor star, an increase of the explosion energy implies an increase of the luminosity of the plateau, a decrease of its duration (in time), a decrease of the remnant mass and an increase of $\rm ^{56}Ni$ ejected. It goes without saying that the radioactive tail in the light curve disappears if the amount of $\rm ^{56}Ni$ ejected is negligible (see the legenda in Figures \ref{fig:lc_013a000}, \ref{fig:lc_015a000}, \ref{fig:lc_013b000}, \ref{fig:lc_020b000} and \ref{fig:lc_015c000}). 

\begin{figure}[h]
\epsscale{1.15}
\plotone{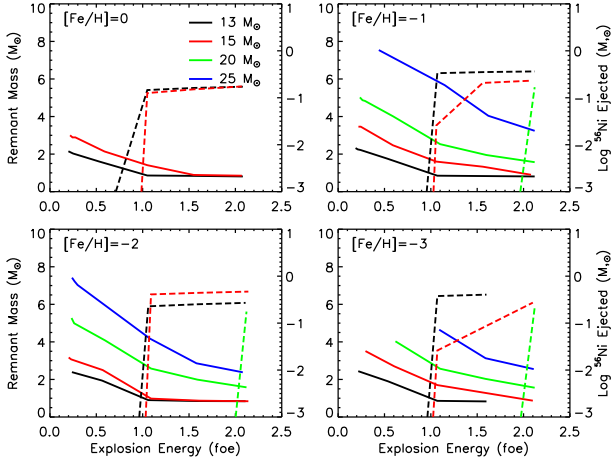}
\caption{Remnant mass (primary y-axis) and $\rm ^{56}Ni$ ejected (secondary y-axis) as a function of the explosion energy for the various progenitor masses (see colors in the legend) and for the various initial metallicities: [Fe/H]=0 (upper left panel), [Fe/H]=-1 (upper right panel), [Fe/H]=-2 (lower left panel) and [Fe/H]=-3 (lower right panel).\label{fig:multimrem_Z}}
\end{figure}

Figure \ref{fig:multimrem_Z} shows the remnant mass (on the primary y-axis) and the $\rm ^{56}Ni$ ejected (on the secondary y-axis) as a function of the explosion energy for the various progenitor masses, for each initial metallicity. In general, for any initial metallicity, the remnant mass scales inversely with the explosion energy and directly with the progenitor mass. The obvious reason of this behavior is that the larger the initial mass the larger the binding energy of the mantle of the star above the iron core (Table \ref{propsn}). As the metallicity decreases the dramatic reduction of the mass loss implies larger CO cores, for the same initial mass, and therefore a higher binding energy. Therefore at lower metallicities more massive remnants are obtained for the same progenitor mass and explosion energies. As discussed in section \ref{sec:results}, the amount of $\rm ^{56}Ni$ ejected depends on the remnant mass. In general the larger the remnant mass the smaller the $\rm ^{56}Ni$ ejected. For progenitor masses smaller than $\rm 20~M_\odot$, a sizeable amount of $\rm ^{56}Ni$ is ejected only for explosion energies larger than $\sim 0.5$ foe. In particular, the amount of $\rm ^{56}Ni$ increases rapidly for explosion energies in the range $\sim 0.5-1.0$ foe and then it remains almost constant for larger explosion energies. 
For progenitor stars with initial mass $\rm \sim 20~M_\odot$ a substantial amount of $\rm ^{56}Ni$ is ejected only for explosion energies larger $\sim 1.5$ foe. For more massive progenitors no $\rm ^{56}Ni$ is ejected in this range of explosion energies. As a final comment, let us note that the fallback occurs on rather long times, ranging from few dozens of seconds up to $\sim10^6-10^7$ seconds (see Table \ref{tabexpl}), and that in general the lower the explosion energy the longer the duration of the fallback.

\begin{figure}[h]
\epsscale{1.15}
\plotone{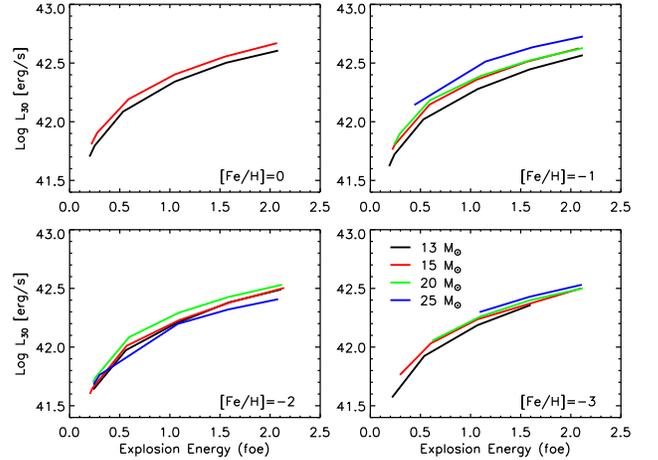}
\caption{Luminosity at $\sim 30$ days after the shock breakout ($L_{30}$) as a function of the explosion energy for the various progenitor masses (see colors in the legend) and for the various initial metallicities: [Fe/H]=0 (upper left panel), [Fe/H]=-1 (upper right panel), [Fe/H]=-2 (lower left panel) and [Fe/H]=-3 (lower right panel).\label{fig:multil30_Z}}
\end{figure}

As we have shown in section \ref{sec:results}, the luminosity of the plateau, at late stages, depends, among the other things, also on the mixing of $\rm ^{56}Ni$. Therefore the average luminosity of the plateau must be evaluated at early times if we want that it is not affected by the amount of $\rm ^{56}Ni$ ejected. For this reason, we choose to define the average luminosity of the plateau as the luminosity at $\sim 30$ days after the shock breakout ($L_{30}$), rather than the one evaluated after 50 days ($L_{50}$) \citep{gengis+16,kw09}. Figure \ref{fig:multil30_Z}, shows this quantity ($L_{30}$), as a function of the explosion energy, for the various progenitor stars and the various initial metallicities. Overall, ${\rm Log}L_{30}$ varies between $\sim 41.6$ and $\sim 42.7$, i.e. slightly more than  one order of magnitude. In general, for any initial metallicity,  $L_{30}$ increases significantly with the explosion energy. The reason is that the luminosity scales with $\sim R^2 T^4$, where both $R$ and $T$ are evaluated at the photosphere; the temperature is roughly constant since it corresponds to the one for the H recombination (see Figure \ref{fig:temp_contour}) while $R$ scales directly with the kinetic energy of the ejecta, that dominates the explosion energy. This last occurrence is due to the fact that, in order to obtain a higher final kinetic energy of the ejecta for any given progenitor mass, a larger amount of energy must be injected to start the explosion. Since, as it is discussed in section \ref{sec:results} (see also Figure \ref{fig:15a1foeenergy}), the internal energy in the H rich mantle at the time of the shock breakout is about half of the total energy (the remaining being the kinetic energy), the higher the amount of energy injected to start the explosion, the higher the internal energy in the envelope at the beginning of the adiabatic cooling (see section \ref{sec:adcoooling}). Since, as it is mentioned in section \ref{sec:adcoooling}, during the adiabatic cooling the radius scales as $R\propto 1/T$, in more energetic explosions the envelope will have to expand more (starting from a higher internal energy content) to reach the radius corresponding to the 
H recombination temperature. 

For a similar reason, in general, $L_{30}$ increases slightly also with the progenitor mass for the same explosion energy: first of all the amount of energy to be injected in a star to obtain the same final kinetic energy of the ejecta scales directly with the progenitor mass (actually the He core mass) and second, the radius of a star at the onset of the collapse scales directly with the initial mass (obviously we are considering only red supergiants stars here). 

\begin{figure}[h]
\epsscale{1.15}
\plotone{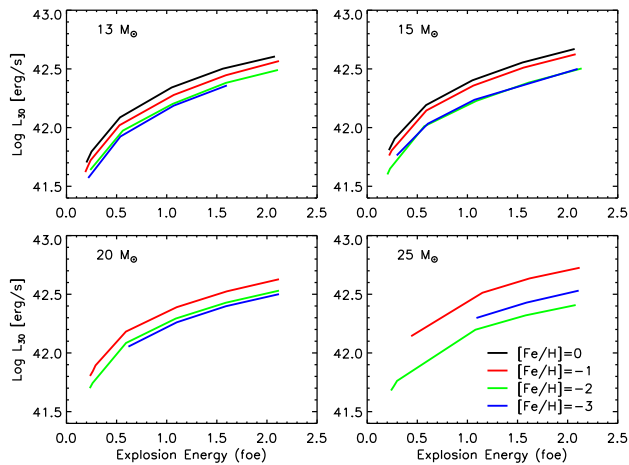}
\caption{Luminosity at $\sim 30$ days after the shock breakout ($L_{30}$) as a function of the explosion energy for the various metallicities (see colors in the legend) and for the various progenitor masses: $\rm 13~M_\odot$ (upper left panel), $\rm 15~M_\odot$ (upper right panel), $\rm 20~M_\odot$ (lower left panel) and $\rm 13~M_\odot$ (lower right panel).\label{fig:multil30_m}}
\end{figure}

The dependence of $L_{30}$ on the initial metallicity can be appreciated in Figure \ref{fig:multil30_m}, where it is shown $L_{30}$ as a function of the explosion energy for the various metallicities for each progenitor mass.
As it is expected, for a fixed explosion energy, $L_{30}$ decreases with decreasing the initial metalliciy because lower metallicity stars are in general more compact than the higher metallicity ones. This effect, however is modest for lower mass models and increases slightly for the more massive ones.

\begin{figure}[h]
\epsscale{1.15}
\plotone{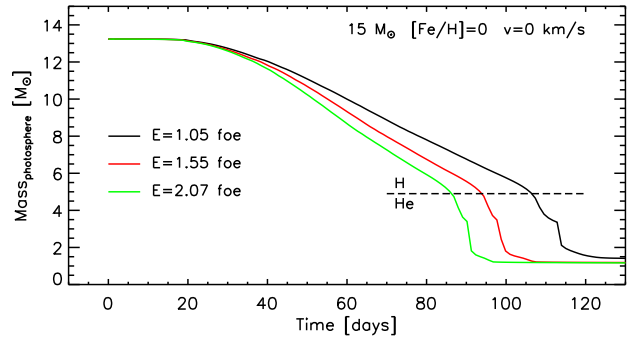}
\caption{Location in mass of the photosphere as a function of time, for the models 15a, for various explosion energies (see color legend). The horizontal black dashed line mark the H/He interface.
\label{fig:multimassphot_a}}
\end{figure}

\begin{figure}[h]
\epsscale{1.15}
\plotone{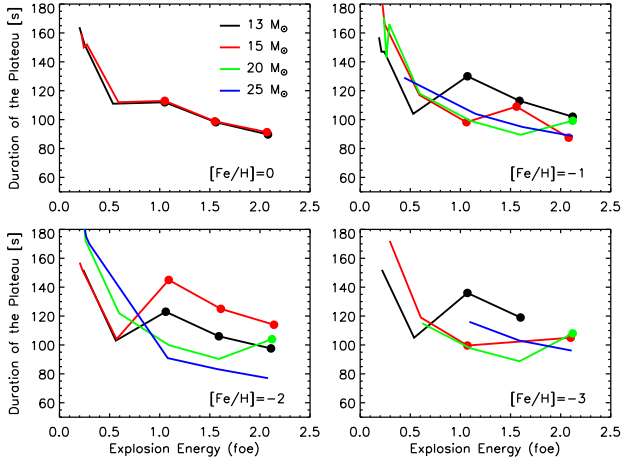}
\caption{Plateau duration as a function of the explosion energy for the various progenitor masses and initial metallicities. The filled dots marks the cases where enough $\rm ^{56}Ni$ is ejected (see Table \ref{tabexpl}). The plateau duration is defined when the radius of the photosphere decreases down to 50\% of its maximum value.
\label{fig:multiplateau_z}}
\end{figure}

As already discussed in sections \ref{sec:recombinationwave} and \ref{sec:rad_tail}, the plateau phase ends when the photosphere approaches the He core. In general, the time at which the photosphere reaches the H/He interface decreases with increasing the expansion velocity,  and therefore with the explosion energy (Figure \ref{fig:multimassphot_a}). As a consequence the duration of the plateau decreases with increasing the explosion energy.
However, since this quantity depends also on the amount of $\rm ^{56}Ni$ ejected and on the mass of the H-rich envelope, the trend is not monotonic over the whole range of explosion energies, progenitor masses and initial metallicities. In particular, Figure \ref{fig:multiplateau_z} shows the existence of two distinct behaviors as a function of the explosion energy, depending on the amount of $\rm ^{56}Ni$ ejected. The plateau duration initially decreases as the explosion energy increases as long as the amount of $\rm ^{56}Ni$ ejected is lower than $\rm \sim10^{-3}~M_\odot$. When this quantity increases enough the plateau duration start increasing until it reaches a local maximum and then decreases again for higher explosion energies. Note that in all the models where the $\rm ^{56}Ni$ ejected is negligible the plateau duration decreases monotonically with increasing the explosion energy.

\section{Comparison with observations}
As a first application of the simulations discussed in the previous sections we applied this database of explosions to derive the physical properties of the progenitor of the well known Type II Supernova SN1999em. A detailed and more extended study of a larger sample of supernovae will be presented in a subsequent paper. We have chosen SN1999em because it is a widely studied supernova, its bolometric light curve is available in literature and there are also high quality optical images of its host galaxy before its explosion \citep{elm03,smartt+02,sd98}.

\bigskip

SN1999em has been discovered on 1999 October 29 by the Lick Observatory Supernova Search in NGC 1637 \citep{li99} at an unfiltered CCD magnitude of $\rm \sim13.5~mag$. It was soon confirmed as a SNII and then, since it was a bright event, it has been well studied both spectroscopically and photometrically for more than 500 days \citep{hamuy+01,leonard+02,elm03}. It has been classified as a normal SN IIP due to the long plateau phase lasting $\sim90$ days \citep{leonard+01}. Observations in radio and X-ray wavelengths at early times provided information on the structure of the circumstellar material and are consistent with a mass loss rate of $\rm \sim 2\times 10^{-6}~M_\odot/yr$ and a wind velocity of $\rm \sim 10~km/s$ \citep{pooley+01}, i.e., consistent with a red supergiant progenitor. The nature of the progenitor has been discussed by \cite{smartt+02} who used high-resolution optical images of NGC 1637 taken several years before the SN 1999em event by \cite{sd98} at the Canada-France-Hawaii-Telescope (CFHT). In particular, due to the lack of point sources at the position corresponding to SN 1999em they derived bolometric luminosity limits and constrained the luminosity of the progenitor star as a function of the assumed effective temperature (see their figure 4).

The determination of the distance is obviously fundamental to compare the theoretical light curve with the observed one. Unfortunately there is no agreement on this point. Using the expanding photospheric method (EPM) \citep{kk74} the following values have been obtained: $\rm 7.5\pm 0.1~Mpc$ \citep{hamuy+01}, $\rm 8.2\pm 0.6~Mpc$ \citep{leonard+02} and $\rm 7.83\pm 0.3~Mpc$ \citep{elm03}. On the other hand \cite{leonard+03} identified 41 Cepheid variable stars in NGC1637, the host galaxy of SN1999em, and derived a Cepheid distance to this galaxy of $\rm 11.7\pm 1.0~Mpc$, which is $\sim50\%$ higher than the one derived with the EPM. \cite{sd98} studied the bright stellar content in NGC 1637 and estimated a distance of $\rm 7.8\pm 1.0~Mpc$ using the brightest red supergiants method, value close to the one obtained with the EPM. On the other hand, \cite{baron+04} obtained a distance to SN1999em of $\rm 12.5\pm 1.8~Mpc$ by means of the spectral-fitting expanding atmospheric model (SEAM), value in agreement with the Cepheid distance obtained by \cite{leonard+03}. \cite{dh06}, improving the EPM found a value of $\rm 11.5\pm 1.0~Mpc$, which is consistent with the SEAM and Cepheid distances.

Since the distance to SN1999em is still under debate, we present a comparison between the observed and the theoretical bolometric light curves for the two extreme values of the distance reported in literature. In particular we will consider the bolometric light curve based on the photometry of \cite{elm03,leonard+01,hamuy+01,leonard+03} (Benetti, S. private communication) for the two different adopted distances, i.e., $\rm 7.83~Mpc$ (LD) and $\rm 11.7~Mpc$ (HD). In both cases the total extinction adopted is $A_V=0.31$.

In general, the comparison between the observations and the models proceeds through the following steps. First of all we select the models, among those reported in Table \ref{tabexpl}, with a metallicity similar to the one of the SN host galaxy. Second, we consider only those for which $L_{30}$ is close to the observed one. Third, we modify the ejected $\rm ^{56}Ni$, and rerun the simulation, in order to fit the radioactive tail. Finally, we fit the shape of the light curve in the transition phase between the plateau and the radioactive tail, by changing the efficiency and the extension of the mixing of both the chemical composition and the $\rm ^{56}Ni$ (also in this case, this final step requires additional simulations). 

It is worth noting that, in general, the database of light curves reported in Table \ref{tabexpl}, cannot be used "sic et simpliciter" but they must be complemented by additional simulations in order to really constrain the fit of the SN light curve under exam ($L_{30}$, $\rm ^{56}Ni$ and the shape of the transition phase between the plateau and the radioactive tail). Hence, the calculations reported in Table \ref{tabexpl} must be seen as a basic database useful to study the general dependence of of the light curves on the initial progenitor parameters (mass and metallicity), and the features of the explosion itself.

Let us also stress that, if we know only the metallicity of the host galaxy and the bolometric light curve of a given supernova, we cannot disentangle between the progenitor mass and the kinetic energy of the ejecta. In fact, for a given metallicity, we can obtain the same value of ${\rm Log} L_{30}$ by changing both the the progenitor mass and the kinetic energy of the ejecta (see Figure \ref{fig:multil30_Z}). Only the independent knowledge of one of the two would fix the other.

Having said this, let us turn to the fit to SN1999em. According to the relation between the absolute magnitude and the metallicity for external galaxies \citep{bh91}, \cite{sd98} derived for NGC 1637 a metallicity of $\rm [Fe/H]\sim-0.33$. Since this metallicity falls between the two grid points, i.e.  [Fe/H]=0 and [Fe/H]=-1, we consider these two set of models. 

In the LD case, the observed ${\rm Log} L_{30}$ is $\sim 41.7$, therefore, from Figure \ref{fig:multil30_Z} and Table \ref{tabexpl}, we select the models 13a, with $E_{\rm expl}=1.99\cdot 10^{50}~{\rm erg}$ (13a1 in the following), and 13b, with $E_{\rm expl}=2.41\cdot 10^{50}~{\rm erg}$ (13b3 in the following). For all the other computed models, ${\rm Log} L_{30}$ is larger than the observed value. Therefore both these progenitor masses and explosion energies should be considered as upper limits (see Figure \ref{fig:multil30_Z}). 

\begin{figure}[ht!]
\epsscale{1.16}
\plotone{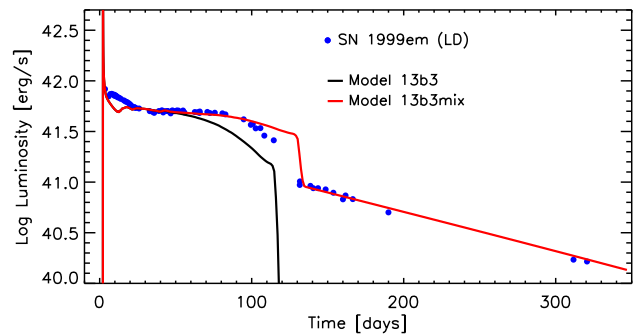}
\caption{Light curve of model 13b3 (black line) and 13b3mix (red line), in which $\rm \sim 0.022~M_\odot$ of $\rm ^{56}Ni$ is deposited and homogeneously mixed up to a mass coordinate marking half of the H-rich envelope. The blue dots refer to the observed bolometric light curve in the LD case (see text).\label{fig:fit1999emLDM}}
\end{figure}

Let us start by analyzing model 13b3. Figure \ref{fig:fit1999emLDM} shows the comparison between the observations (blue dots) and the light curve of the model (black line). While the $L_{30}$ is in good agreement with the observed one, the model does not show a radioactive tail because of the large remnant mass ($M_{\rm rem}=2.22~{\rm M_\odot}$, see Table \ref{tabexpl}) that implies a negligible amount of $\rm ^{56}Ni$ ejected. For this reason we assume that some amount of $\rm ^{56}Ni$ is mixed from the innermost zones outward in mass during the explosion, before the occurrence of the fallback, and simulate such a phenomenon simply by depositing and mixing homogeneously the amount of $\rm ^{56}Ni$ required to fit the radioactive tail. We perform such a $\rm ^{56}Ni$ deposition and homogeneous mixing soon after ($\rm \sim 10~days$) the shock breakout. It is important to note at this point, that the $\rm ^{56}Ni$ synthesized in the innermost zones before the occurrence of the fallback is much higher than $\rm \sim 0.022~M_\odot$ and therefore that it is reasonable to assume that a small fraction of such a $\rm ^{56}Ni$ can be mixed upward in mass before the fallback goes to completion. By the way, let us remind that the outer edge of the zone where the $\rm ^{56}Ni$ is homogeneously mixed corresponds to the mass coordinate marking half of the H-rich envelope
Figure \ref{fig:fit1999emLDM} shows that the light curve of the model (red line) in which $\rm \sim 0.022~M_\odot$ is deposited and homogeneously mixed (13b3mix hereinafter) reproduces fairly well both the $L_{30}$ and the radioactive tail, but it is substantially brighter in the late stages of the plateau phase.  Note also that, in both cases, there is a discrepancy between the observed and the theoretical light curve in the first $\rm \sim 20~days$. More specifically, the luminosity of the theoretical light curve decreases much faster than the observed one. This is a well known problem that has been addressed in a number of quite recent papers \citep{moriya+17,moriya+18,morozova+17,morozova+18,morozova+20,paxton+18}. In all these studies it has been shown that the presence of a dense circumstellar material around the star should produce a better agreement between the theoretical and observed light curve in the first $\sim10-20$ days. Since we do not address this problem in the present work, we will focus only on the light curve at times later than $\rm \sim 20~days$ and leave this subject for a future paper. 

\begin{figure}[ht!]
\epsscale{1.16}
\plotone{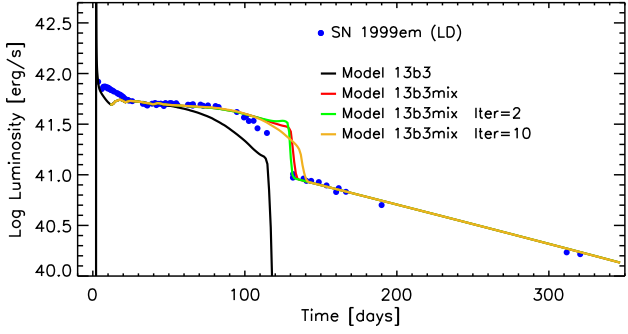}
\caption{Dependence of the light curve behavior on the mixing efficiency: model 13b3 (black line) is the reference model; model 13b3mix (red line) is the same as the reference model but in which $\rm \sim 0.022~M_\odot$ of $\rm ^{56}Ni$ is deposited and homogeneously mixed up to a mass coordinate marking half of the H-rich envelope; models "13b3mix Iter=2" (green line) and "13b3mix Iter=10" (orange line) are the same as 13b3mix but in which the number of iterations of the boxcar is 2 and 10, respectively. The blue dots refer to the observed bolometric light curve reported in the LD case (see text).\label{fig:varitest13b_mixchim}}
\end{figure}
Figure \ref{fig:varitest13b_mixchim} shows that the faster decline of the observed light curve, in the transition phase from the plateau to the radioactive tail, can be better reproduced by assuming a more efficient mixing of the chemical composition. The light curve of the model where we increase the number of iterations in the boxcar parameters (orange line in Figure \ref{fig:varitest13b_mixchim}) is closer to the observations but still rather brighter.  
\begin{figure}[ht!]
\epsscale{1.16}
\plotone{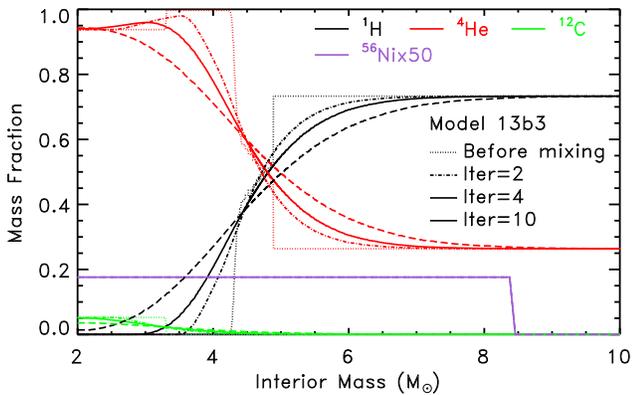}
\caption{Interior profiles of selected isotopes (see legenda) of model 13b3 obtained with different choices of the number of boxcar iterations: no mixing (dotted lines), Iter=2 (dot-dashed lines), Iter=4 (solid lines, the reference value of all the models reported in Table \ref{tabexpl}), Iter=10 (dashed lines). In all the models the $\rm ^{56}Ni$ is homogeneously mixed from the inner edge of the ejecta up to half (in mass) of the H-rich envelope.
\label{fig:mix_chim_iter}}
\end{figure}
Figure \ref{fig:mix_chim_iter} shows the effect of changing the number of box car iterations on the interior composition of model 13b3. It is evident how the transition from the H- to the He-rich zone becomes progressively smoother as the number of boxcar iterations increases. Note that in the model with Iter=10 the hydrogen is mixed down to the base of the ejecta. A more efficient mixing implies also a longer plateau phase and therefore the radioactive tail begins at later times, in this case, compared to the observations.
The opposite effect is obtained by increasing the zone where $\rm ^{56}Ni$ is homogeneously mixed. The larger the zone the earlier the end of the plateau phase and the smoother the transition from the plateau to the radioactive tail (Figure \ref{fig:varitest13b_mixni}). Note that a spread of the $\rm ^{56}Ni$ over a wider zone would determine a slight increase of $L_{30}$.

\begin{figure}[ht!]
\epsscale{1.16}
\plotone{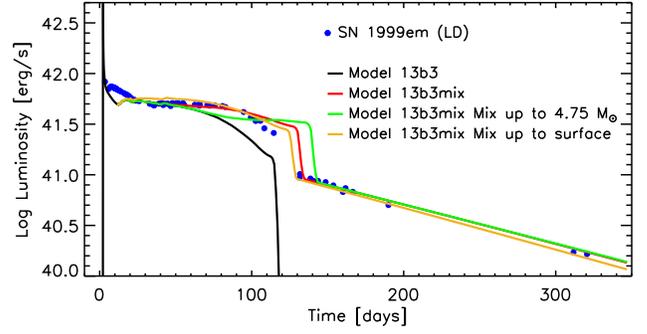}
\caption{Dependence of the light curve behavior on the outer edge of the zone where $\rm ^{56}Ni$ is homogeneously mixed: the black and red lines refer to the models 13b3 and 13b3mix (see text); the green and orange lines refer to the models where $\rm ^{56}Ni$ is homogeneously mixed up to a mass coordinate of $\rm 4.75~M_\odot$ and the surface, respectively. The blue dots refer to the observed bolometric light curve reported in the LD case (see text).\label{fig:varitest13b_mixni}}
\end{figure}

\begin{figure}[ht!]
\epsscale{1.16}
\plotone{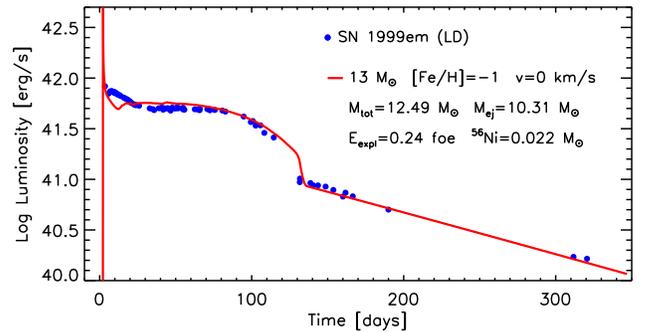}
\caption{Light curve of a non rotating model with initial mass $\rm 13~M_\odot$ and initial metallicity [Fe/H]=-1 (red line): the total mass of this star at the presupernova stage is $\rm M_{tot}=12.49~M_\odot$; after the explosion the ejected mass is $\rm M_{ej}=10.31~M_\odot$ with a total explosion (mainly kinetic) energy of $\rm E_{expl}=0.24~foe$; the $\rm ^{56}Ni$ ejected is $\rm 0.022~M_\odot$. The blue dots refer to the observed bolometric light curve reported in the LD case (see text).\label{fig:fit1999emLDMfinal}}
\end{figure}

By combining a more efficient mixing (Iter=10) with a more extended zone where $\rm ^{56}Ni$ is homogeneously mixed (up to the surface) we obtain a good fit to the observations (model 13b3best, Figure \ref{fig:fit1999emLDMfinal}). Although, in this case the $L_{30}$ increases slightly, this is still compatible with the observed one. A similar, or even better, fit to the observations can be certainly obtained with a different choices of the mixing parameters or by tuning better the explosion energy, but, given all the uncertainties affecting both the observations and the models, we think that the fit  shown in Figure \ref{fig:fit1999emLDMfinal} can be considered satisfactory.

\begin{figure}[ht!]
\epsscale{1.16}
\plotone{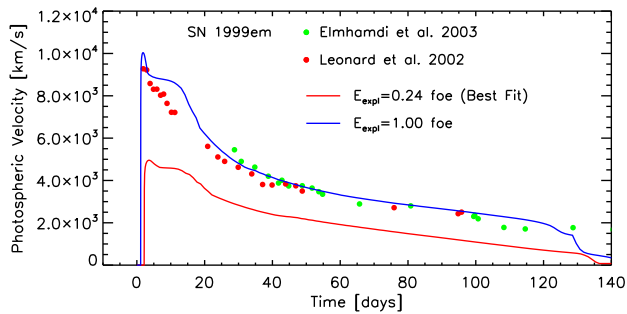}
\caption{Radial velocities at maximum absorption of FeII lines measured by \cite{leonard+02} (red filled dots) and at maximum absorption of ScII lines provided by \cite{elm03} (green filled dots). The red line refers to the model 13b3best while the blue line to the model with a final explosion energy of 1 foe. \label{fig:fit1999emLDMvphotfinal}}
\end{figure}

The good fit to the light curve, however, does not implies a good fit to the observed photospheric velocity. Figure \ref{fig:fit1999emLDMvphotfinal} shows that the photospheric velocity of the model that reproduces the observed light curve of SN 1999em (13b3best) is substantially lower than the observed one, especially at early times, which confirms the difference of the structure of the more external layers between the presupernova model and the real progenitor star. A better agreement is obtained for higher explosion energies. Figure \ref{fig:fit1999emLDMvphotfinal} shows that the model with a final explosion energy of 1 foe reproduces fairly well the observations, however its bolometric luminosity is substantially higher than the observed one. This problem has been already found and discussed by other studies like, e.g., \cite{utrobin17} (see their Figure 6b) and \cite{morozova+20} (see their Figure 3, right panel), and we find similar results. However \cite{paxton+18} have shown that the evaluation of the velocity where the Sobolev optical depth of the Fe {\RomanNumeral{2}} is equal to 1 provides a much better match to the observations than the photospheric velocity. We do not address this problem in the present work but since it is clearly important to find a simultaneous fit to both the light curve and the expansion velocity, we will address this issue in a forthcoming paper.

\begin{figure}[ht!]
\epsscale{1.16}
\plotone{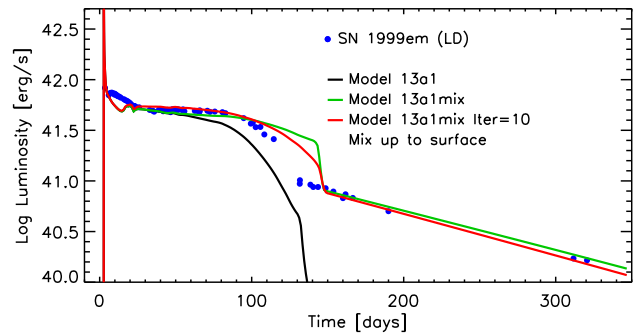}
\caption{Light curves a non rotating model with initial mass $\rm 13~M_\odot$, initial metallicity [Fe/H]=0 and explosion  energy of of 0.2 foe. Black line: reference model (13a1, first line in Table \ref{tabexpl}. Green line: model 13a1mix, i.e.m, same as 13a1 where $\rm 0.022~M_\odot$ of $\rm ^{56}Ni$ are deposited and homogeneously mixed up to half of the H-rich envelope. Red line: model "13a1mix Iter=10", i.e., same as 13a1mix with a more efficient mixing (Iter=10 in the boxcar parameters) and where $\rm ^{56}Ni$ is homogeneously mixed up to the surface. The blue dots refer to the observed bolometric light curve reported in the LD case (see text).\label{fig:fit1999emLDMfinal2}}
\end{figure}

Let us now analyze the comparison between the LD case and the model 13a1. As for the model 13b3, in this case the remnant mass is large enough ($\rm M_{rem}=3.00~M_\odot$) that a negligible amount $\rm ^{56}Ni$ is ejected. Therefore, also in this case, we deposit in the model $\rm 0.022~M_\odot$ of $\rm ^{56}Ni$. The light curve obtained in this case (13a1mix, green line in Figure \ref{fig:fit1999emLDMfinal2}) shows a plateau phase that lasts longer and a luminosity in the transition phase between the plateau and the radioactive tail that is higher than the observed ones. As it has been already been mentioned above, a combination of a more efficient mixing of the composition and a more extended zone where $\rm ^{56}Ni$ is homogeneously mixed produces a shorter plateau and a smoother transition to the radioactive tail and therefore it should produce, in this case, a better agreement with the observations. The red line in Figure \ref{fig:fit1999emLDMfinal2} is obtained assuming the same parameters adopted for the model 13b3best, i.e., a homogeneous mixing of $\rm ^{56}Ni$ up to the surface coupled to a very efficient mixing of the chemical composition (Iter=10). In spite of this more extended and vigorous mixing, Figure \ref{fig:fit1999emLDMfinal2} shows that, in this case (red line), the plateau phase is still longer and brighter in the late stages compared to the observed one. Since this is the maximum efficiency of mixing that we can assume, we must conclude that the model 13a1 cannot reproduce the light curve of SN1999em (LD).

Summarizing the results discussed so far, we conclude that a non rotating star with initial mass $\rm 13~M_\odot$ and metallicity [Fe/H]=-1 is compatible with the progenitor of SN 1999em, when we adopt the lower distance ($\rm 7.83~Mpc$) to the host galaxy NGC 1637. However, a lower progenitor mass and/or a lower explosion energy cannot be excluded.

In the higher distance case ($\rm 11.7~Mpc$, HD), the observed ${\rm Log} L_{30}$ is $\simeq 42.06$. Looking at Table \ref{tabexpl} and Figure \ref{fig:multil30_Z}, for all the progenitor models of the series 'a' and 'b' with initial mass lower than $\rm 25~M_\odot$ there exist explosions providing values of the ${\rm Log} L_{30}$ that bracket the observed value. Therefore, at variance with the LD case, now the mass of the progenitor star spans larger values due to the higher intrinsic luminosity of the supernova. The selected models are the following: 
the 13a2 and 13a3 (with $E_{\rm expl}=2.50\cdot 10^{50}~{\rm erg}$ and $E_{\rm expl}=5.34\cdot 10^{50}~{\rm erg}$, respectively); 
the 15a3 and 15a4 (with $E_{\rm expl}=2.74\cdot 10^{50}~{\rm erg}$ and $E_{\rm expl}=5.88\cdot 10^{50}~{\rm erg}$, respectively); 
the 13b4 (with $E_{\rm expl}=5.29\cdot 10^{50}~{\rm erg}$ ); 
the 15b2 and 15b3 (with $E_{\rm expl}=2.44\cdot 10^{50}~{\rm erg}$ and $E_{\rm expl}=5.91\cdot 10^{50}~{\rm erg}$, respectively); 
and the 20b3 and 20b4 (with $E_{\rm expl}=2.90\cdot 10^{50}~{\rm erg}$ and $E_{\rm expl}=5.93\cdot 10^{50}~{\rm erg}$, respectively). 
Due to the coarse grid in the explosion energies, we computed additional explosions varying the explosion energy in order to obtain a ${\rm Log} L_{30}$ which is closer to the observed one. A substantial fallback occurs in all of the previous models, therefore in all of them $\rm 0.05~M_\odot$ of $\rm ^{56}Ni$ is deposited and homogeneously mixed in order to reproduce the observed radioactive tail. As we have discussed above, once both the ${\rm Log} L_{30}$ and the radioactive tail are reproduced, the shape of the light curve in the transition phase from the plateau to the radioactive tail depends mainly on the efficiency of the mixing of the chemical composition and on the region where $\rm ^{56}Ni$ is homogeneously deposited. In order to be more systematic, once all the other parameters are fixed (mass, metallicity, explosion energy, total amount of $\rm ^{56}Ni$ deposited), we computed various explosions by changing the two parameters that control the efficiency of the chemical mixing and the extension of the zone where $\rm ^{56}Ni$ is homogeneously mixed. In particular we named these explosions with the following rule: xxxZe.ee\_n, where xxx refers to the mass (e.g., 013, 015, etc.); Z is the series in metallicity (e.g., a, b, c, d); e.ee means the explosion energy in foe (e.g. 0.50 means 0.50 foe) while n refers to the various simulations with different choices of the mixing parameters (see Table \ref{tab:tests}), i.e., the number of iterations of the boxcar (Iter) and the outer mass coordinate of the zone where the $\rm ^{56}Ni$ is homogeneously mixed ($\rm M_{\rm out}$). The effect of changing the number of boxcar iterations (Iter=2, 4 and 10) on the chemical composition for all the models mentioned above is similar to the one already shown in Figure \ref{fig:mix_chim_iter}.

\begin{deluxetable}{ccl}
\tablewidth{0pt}
\tablecaption{Calculations with different chioces of the mixing parameters\label{tab:tests}}
\tablehead{
\colhead{Simulation ID} & \colhead{Iter} & \colhead{$\rm M_{\rm out}$} 
}
\startdata
1  &   2   &  H/He interface \\  
2  &   2   &  Half of the H-rich envelope \\  
3  &   2   &  Surface \\  
4  &   4   &  H/He interface \\  
5  &   4   &  Half of the H-rich envelope \\  
6  &   4   &  Surface \\  
7  &  10   &  H/He interface \\  
8  &  10   &  Half of the H-rich envelope \\  
9  &  10   &  Surface \\  
\enddata                                              
 \end{deluxetable}       

\begin{figure*}[ht!]
\epsscale{1.0}
\plotone{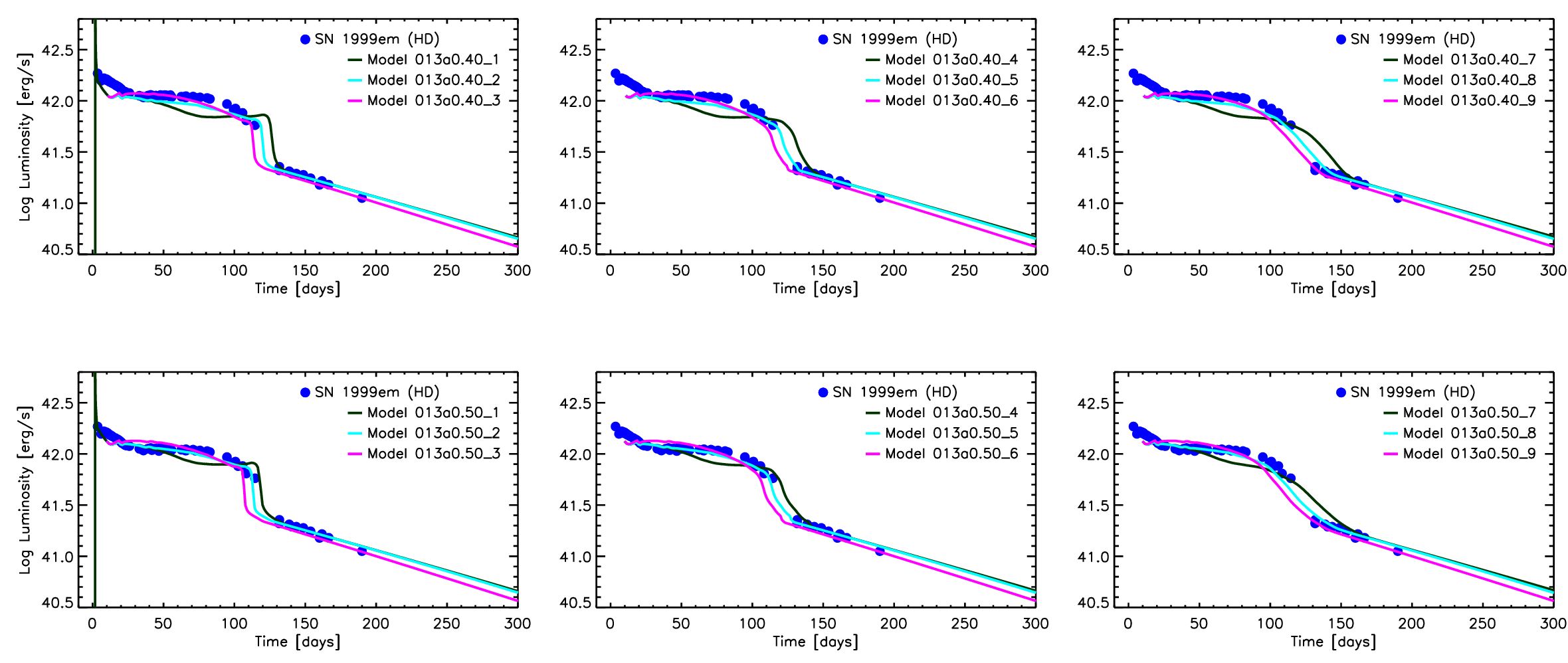}
\caption{Light curves for model 13a obtained for different values of the explosion energies (0.40 and 0.50 foes in the upper and lower rows, respectively) and for different values of the mixing parameters (see Table \ref{tab:tests}). In all the cases $\rm 0.05~M_\odot$ of $\rm ^{56}Ni$ is deposited and homogeneously mixed.\label{fig:013a0.40-0.50_param}}
\end{figure*}

\begin{figure*}[ht!]
\epsscale{1.0}
\plotone{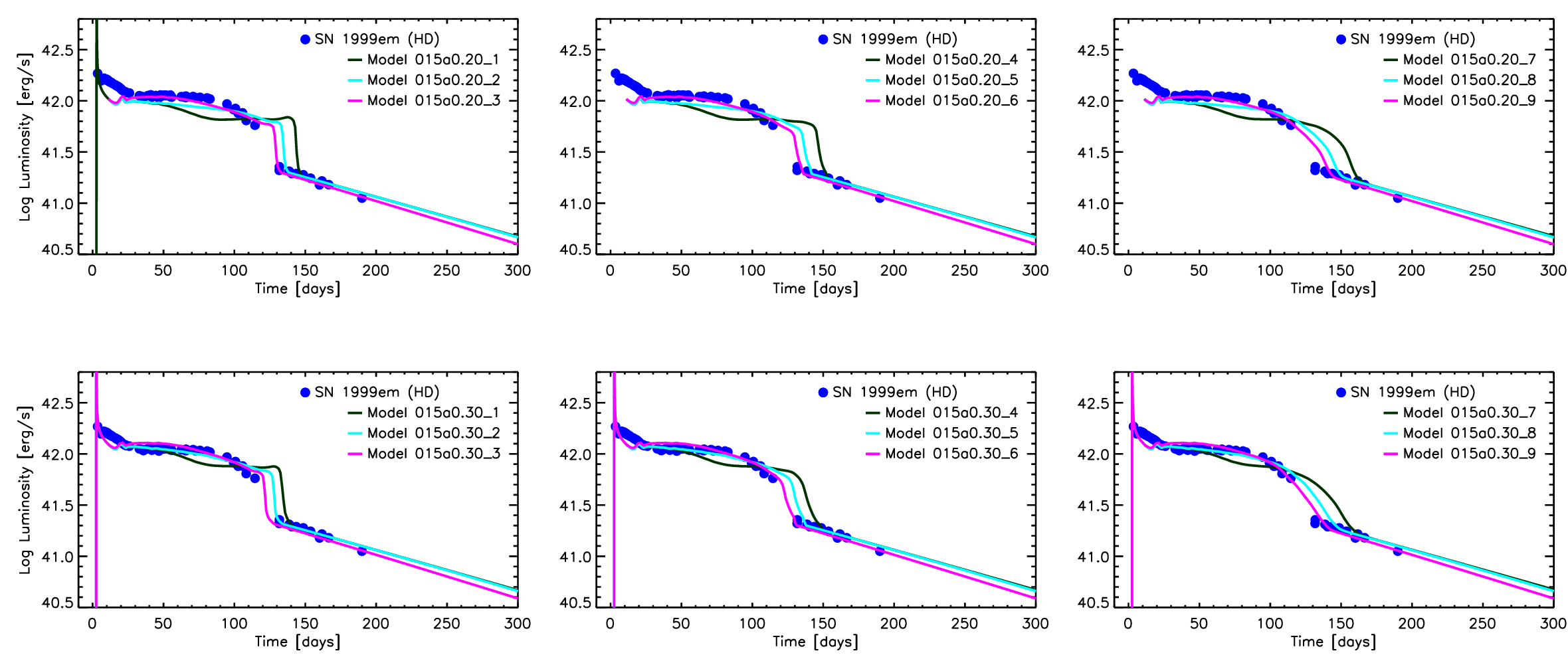}
\caption{Same as Figure \ref{fig:013a0.40-0.50_param}) but for the model 15a.\label{fig:015a0.20-0.30_param}}
\end{figure*}

\begin{figure*}[ht!]
\epsscale{1.0}
\plotone{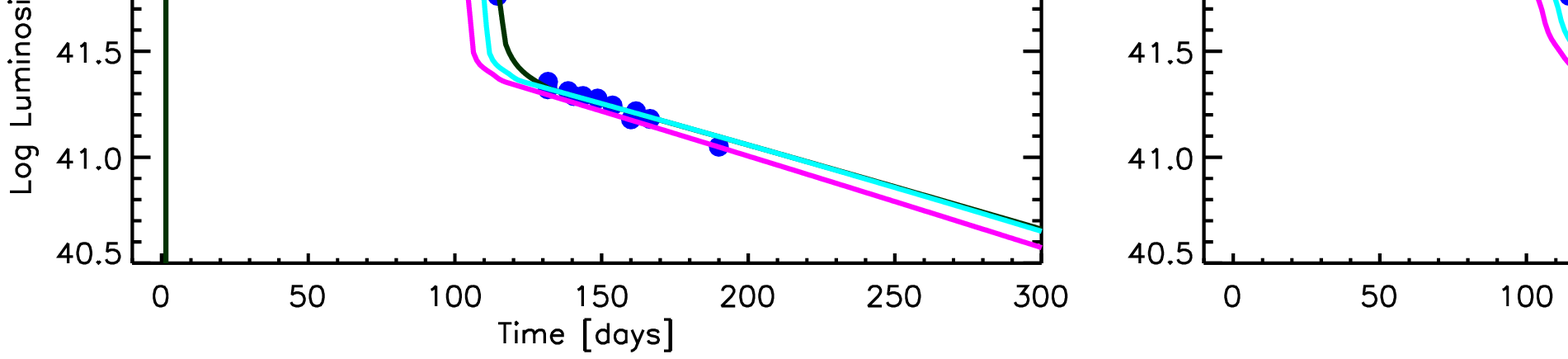}
\caption{Same as Figure \ref{fig:013a0.40-0.50_param}) but for the model 13b.\label{fig:013b0.50-0.50_param}}
\end{figure*}

\begin{figure*}[ht!]
\epsscale{1.0}
\plotone{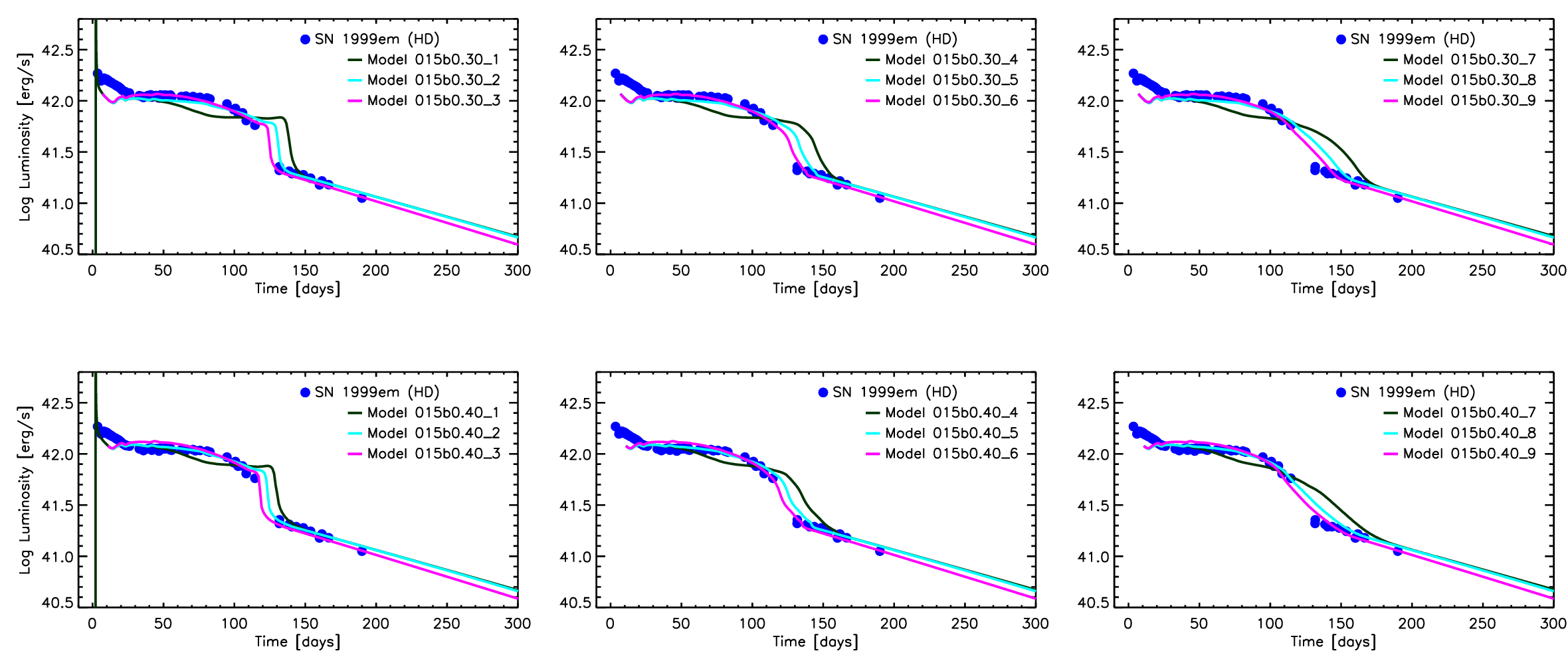}
\caption{Same as Figure \ref{fig:013a0.40-0.50_param}) but for the model 15b.\label{fig:015b0.30-0.40_param}}
\end{figure*}

\begin{figure*}[ht!]
\epsscale{1.0}
\plotone{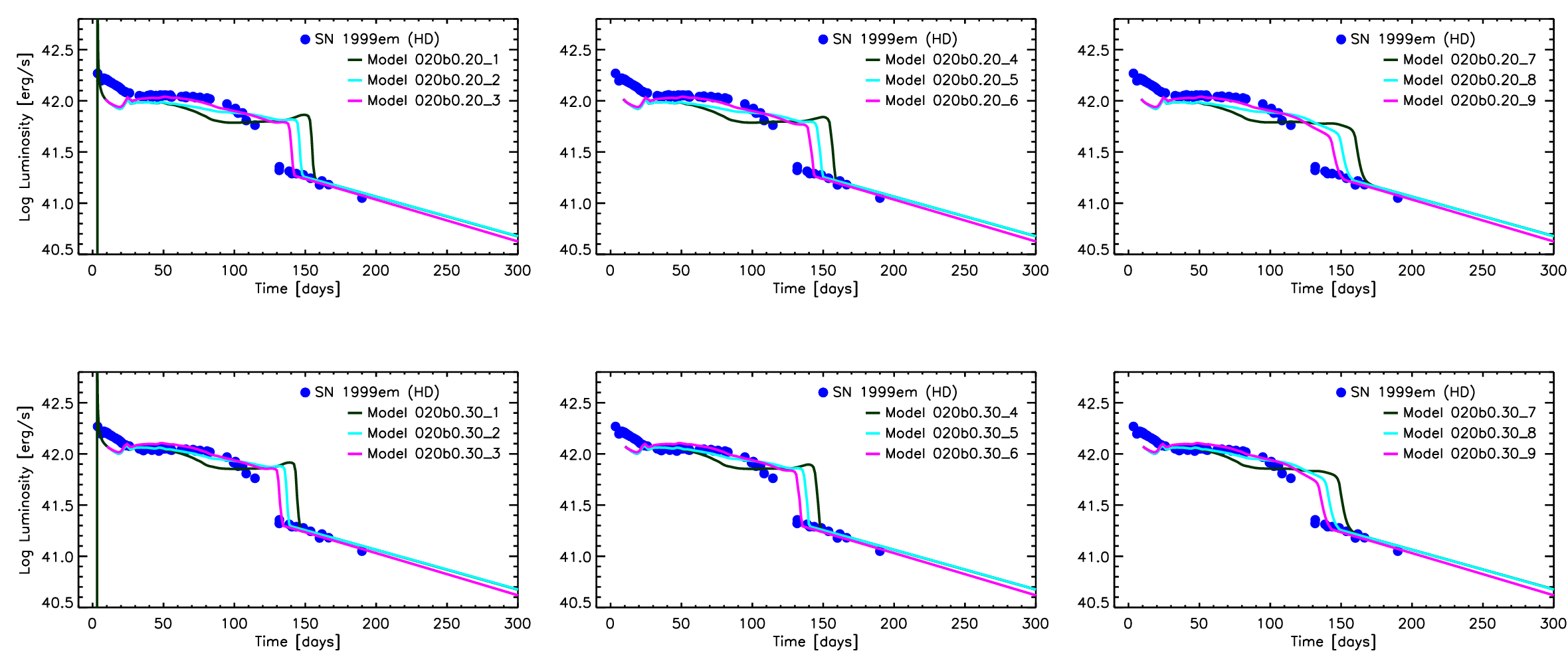}
\caption{Same as Figure \ref{fig:013a0.40-0.50_param}) but for the model 20b.\label{fig:020b0.20-0.30_param}}
\end{figure*}

Figures \ref{fig:013a0.40-0.50_param}-\ref{fig:020b0.20-0.30_param} show the results for all the models that fit simultaneously the observed ${\rm Log} L_{30}$ and the radioactive tail with different choices of the mixing parameters (as reported in Table \ref{tab:tests}). 

Looking at these Figures, we first conclude that all the explosions where the $\rm ^{56}Ni$ is mixed up to the H/He interface (black lines) must be excluded. In these cases, in fact, during the plateau phase, all the light curves show an initial decrease (until day 70-80) followed by a phase where the luminosity is constant or slightly increasing, due to the energy provided by the $\rm ^{56}Ni$, that eventually ends when the radioactive tail sets in. The observed light curve does not show such a behavior but, on the contrary, it is almost flat until day $\sim 80$ and then shows a smooth transition toward the radioactive tail.

Inspection of Figure \ref{fig:020b0.20-0.30_param} reveals that the model 20b must be definitely excluded as a possible progenitor for 1999em. In fact, in all the cases studied, the light curves are brighter in the transition phase and the plateau is longer than the observed ones.

The model 13b must be also excluded since in all the cases the luminosity of the plateau between days 70 and 90 is lower than the observed one.

In all the other cases, i.e., 13a, 15a and 15b there exists at least one case, or even more than one, that is compatible with the observations, within all the theoretical and observation uncertainties. In general, the models that better reproduce the shape of the observed light curve in the transition phase are those with a moderate mixing of the chemical composition (middle panels in Figures \ref{fig:013a0.40-0.50_param}, \ref{fig:015a0.20-0.30_param}, \ref{fig:015b0.30-0.40_param} and with the region where the $\rm ^{56}Ni$ is homogeneously mixed extending up to half of the H-rich envelope. In spite of this general rule there are cases where a less extended mixing of the chemical composition or a more extended zone where $\rm ^{56}Ni$ is mixed cannot be excluded. 

\begin{figure}[ht!]
\epsscale{1.16}
\plotone{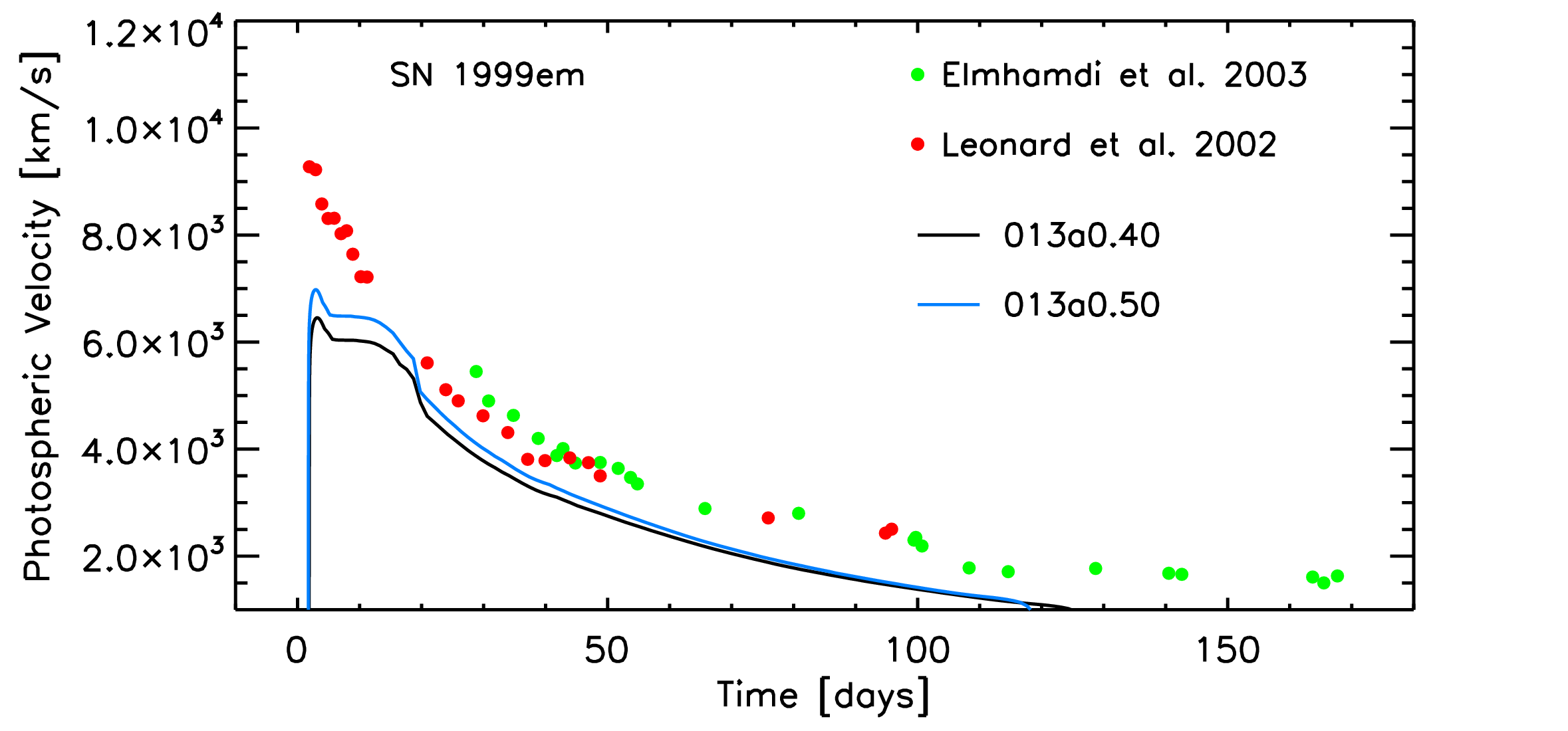}
\caption{Radial velocities at maximum absorption of FeII lines measured by \cite{leonard+02} (red filled dots) and at maximum absorption of ScII lines provided by \cite{elm03} (green filled dots). The black and blue lines refer to the models 0.13a0.40 and 0.13a0.50 (see text), respectively.
\label{fig:013a0.40-0.50_vphot}}
\end{figure}

\begin{figure}[ht!]
\epsscale{1.16}
\plotone{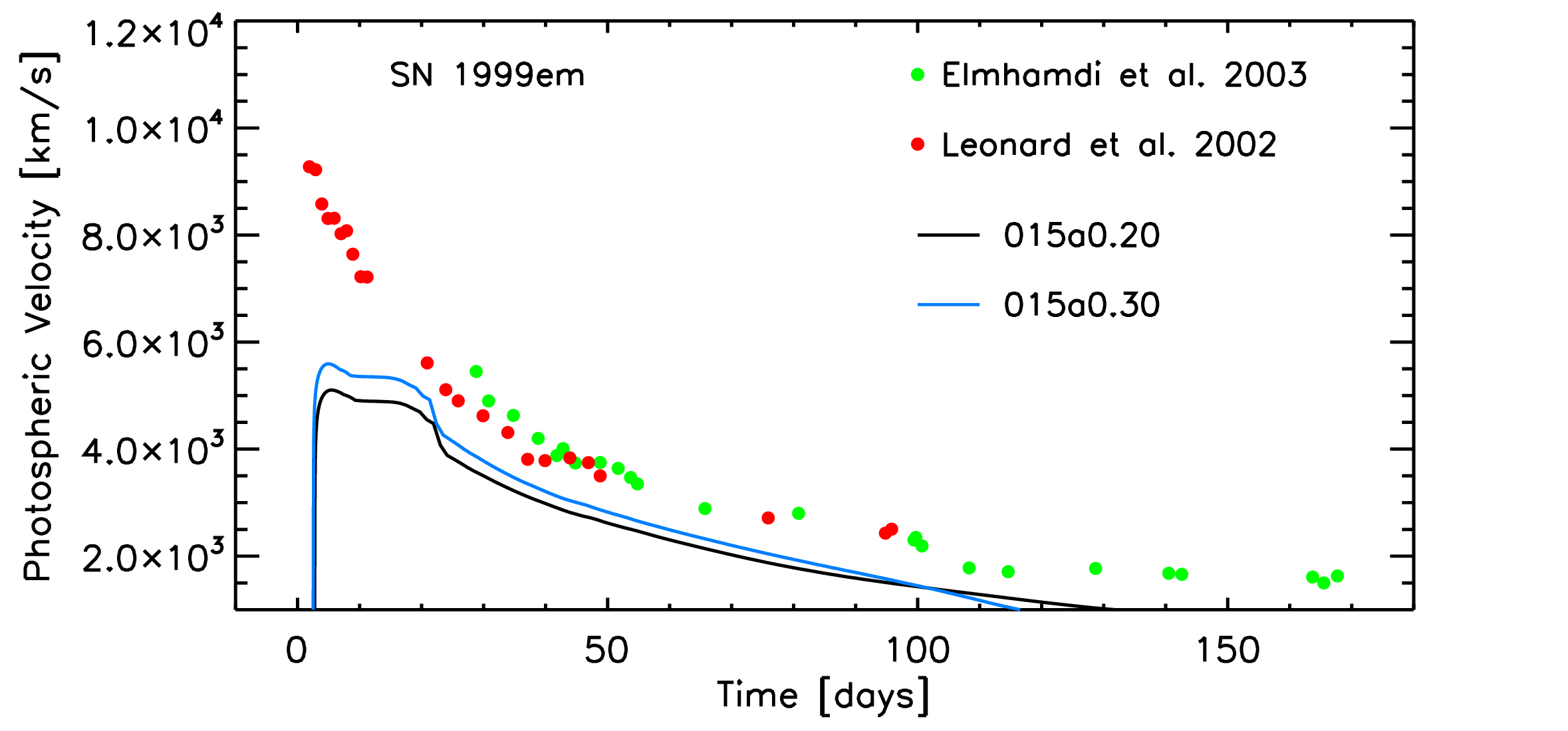}
\caption{Same as Figure \ref{fig:013a0.40-0.50_vphot} but for the models 0.15a0.20 and 015a0.30. respectively.\label{fig:015a0.20-0.30_vphot}}
\end{figure}

\begin{figure}[ht!]
\epsscale{1.16}
\plotone{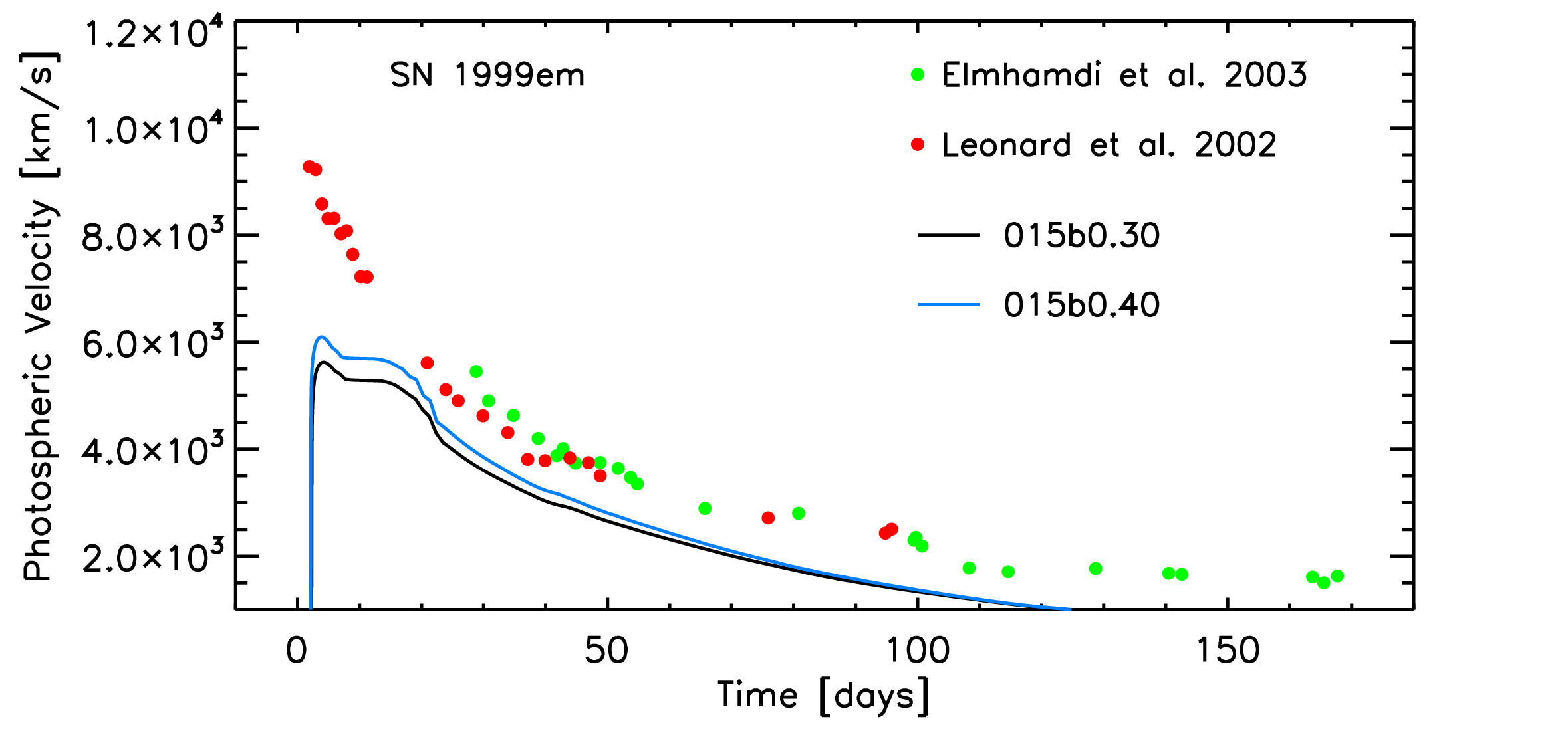}
\caption{Same as Figure \ref{fig:013a0.40-0.50_vphot} but for the models 0.13b0.20 and 013b0.30. respectively.
\label{fig:015b0.30-0.40_vphot}}
\end{figure}

Information on the photospheric velocity could provide an additional constraint on the progenitor star. Therefore, we show in Figures \ref{fig:013a0.40-0.50_vphot}, \ref{fig:015a0.20-0.30_vphot} and \ref{fig:015b0.30-0.40_vphot} the comparison between the predicted and the observed photospheric velocities for all the models 13a, 15a and 15b, corresponding to those reported in Figures \ref{fig:013a0.40-0.50_param}, \ref{fig:015a0.20-0.30_param} and \ref{fig:015b0.30-0.40_param}. 
In the above mentioned figures, each line refers to a model computed with a given explosion energy, regardless of the mixing parameters. The reason is that the photospheric velocity does not depend on the mixing parameters but only on the explosion energy. In this case a discrepancy between observed and predicted photospheric velocity similar to the LD case is found. However, the higher distance to SN1999em implies a higher ${\rm Log} L_{30}$ and therefore a explosion energy for the same progenitor mass. As a consequence, in this case, the discrepancy mentioned above reduces compared to the LD case (Figure \ref{fig:fit1999emLDMvphotfinal}). 

Since there is no strict rule on the basis of which we can definitely say which is the model that best fit the observed light curve we leave this exercise to the reader and draw a more general conclusion. In particular, we conclude that, in the HD case, the following models: $\rm 13~M_\odot$ (with [Fe/H]=0 and $E_{\rm expl}\sim 5\cdot 10^{50}~{\rm erg}$), $\rm 15~M_\odot$ (with [Fe/H]=0 and $E_{\rm expl}\sim 3\cdot 10^{50}~{\rm erg}$) and $15~M_\odot$ (with [Fe/H]=-1 and $E_{\rm expl}\sim 4\cdot 10^{50}~{\rm erg}$) are, in principle, all compatible with the progenitor star of SN1999em. 

\section{Conclusions} \label{sec:conclusions}
In this paper we presented and described in detail the latest version of the {\scshape{Hyperion}} code. {\scshape{Hyperion}} is designed to calculate the explosive nucleosynthesis, the remnant mass and the bolometric light curve associated to the explosion of a massive star. The core of {\scshape{Hyperion}} is based on a previous hydro code, which has been extensively used for explosive nucleosynthesis calculations \citep{lc06, lc12, cl13, cl17, lc18}. It is based on a PPM scheme with a Riemann solver \citep{cw84} coupled to a fully automated nuclear network including 339 nuclear species and more than 3000 nuclear reactions \citep{lc18}. With respect to the previous version, {\scshape{Hyperion}} includes the radiation transport in the flux limited diffusion approximation, and therefore allows the calculation of the bolometric light curve.

By means of this code we computed a set of explosions, and associated explosive nucleosynthesis and bolometric light curves, for a subset of non rotating presupernova models, which retain their H-rich envelope, taken from the database published in \cite{lc18}. All the explosions are induced by depositing instantaneously some amount of thermal energy within the Fe core. The energy deposited is chosen in order to have a given final explosion energy. All the simulations are followed until $\rm 10^7~s$. In this way, the physical and chemical properties of the progenitor star, e.g., the envelope mass, the total radius, the interior profiles of the temperature, density, chemical composition and so on, are not treated as free parameters but, on the contrary, they are the result of the evolution of the star as a function of initial mass and metallicity (in general they depend also on the initial rotation velocity).

As a first check of {\scshape{Hyperion}}, we have deeply analyzed and described in detail the results obtained for a typical case, i.e., a solar metallicity non rotating $\rm 15~M_\odot$ model with final explosion energy of 1 foe. All the phases characterizing the light curve have been discussed in detail and special attention has been devoted to the luminosity "bump" that characterizes the light curve in the transition phase between the plateau and the radioactive tail. Since this feature has never been observed in SN IIP light curves, we studied in detail such a phenomenon and concluded that this characteristic is mainly due to the drop in the opacity within the He core when He recombines. In order to minimize such a sharp variation of the opacity we made additional tests to verify the sensitivity of the "bump" to the mixing of both the density and the chemical composition. In fact, it is quite probable that in more realistic 3D calculations both these quantities could be significantly smoothed with respect to the 1D simulations. The result of these tests was that a proper combination of the smoothing of the density gradient and of the mixing of both the chemical composition and of the $\rm ^{56}Ni$, the "bump" in the light curve disappears. It must be noted, however, that we are dealing with a feature that corresponds to a very small variation of the luminosity (of the order of $\sim 5\%$).

The full set of calculations allowed us to study the main outcomes of the explosions as a function of the progenitor mass, initial metallicity and explosion energy. In particular, we focused on the remnant mass, the ejected amount of $\rm ^{56}Ni$, the luminosity after 30 days and the duration of the plateau of the light curve. In general, as the explosion energy decreases the remnant mass increases and, as a consequence, the amount of $\rm ^{56}Ni$ decreases as well. This is the consequence of the fact that the larger the initial mass the larger the binding energy of the core and also that the $\rm ^{56}Ni$ is produced in the innermost zones of the exploding mantle. For this reason, for each progenitor mass and initial metallicity we found a critical value of the explosion energy below which the light curve does not show the radioactive tail. The other important result was that larger remnant masses are obtained for lower metallicities. The reason being that as the metallicity decreases, the dramatic reduction of the mass loss implies larger CO cores and therefore larger binding energies for the same progenitor mass.

The luminosity of the plateau, evaluated 30 days after the shock breakout, ${\rm Log} L_{30}$, varies between $\sim 41.6$ and $\sim 42.7$ in all the range of parameters. We found that, for any initial metallicity, ${\rm Log} L_{30}$ increases significantly with the explosion energy. On the other hand, for any fixed explosion energy, ${\rm Log} L_{30}$ decreases with decreasing the initial metallicity. Note that, the luminosity evaluated at early times is almost independent on the amount and degree of mixing of $\rm ^{56}Ni$.

The length of the plateau depends on both the explosion energy and the $\rm ^{56}Ni$ ejected, therefore it shows a non monothonic behavior in the whole range of explosion energies, progenitor masses and initial metallicities. In general the length of the plateau decreases with increasing the explosion energy as long as the $\rm ^{56}Ni$ ejected is lower than $\rm \sim 10^{-3}~M_\odot$. For higher values of this last quantity the plateau duration start increasing as the explosion energy progressively increases until a maximum value is reached after which it starts decreasing again.

As a first application of this code we presented a fit to the observed bolometric light curve of SN1999em. We chose SN1999em because it is one of the most widely studied SN IIP and it is often considered as a template for this kind of supernovae. Since the distance to the supernova host galaxy, i.e. NGC 1637, is still under debate, we studied the two extreme cases where the distance is assumed 7.83 Mpc (LD) and 11.8 Mpc (HD), respectively. We presented our fitting strategy that can be summarized through the following steps: (1) we select the progenitors with both the metallicity and the ${\rm Log}L_{30}$ closer to the observed values; iterations on the explosion energy could be necessary to refine the fit to the observed ${\rm Log}L_{30}$; (2) we change the amount of $\rm ^{56}Ni$ ejected in order to fit the radioactive tail; the basic assumption in this step is that in more realistic 3D simulations $\rm ^{56}Ni$ rich bubbles are pushed outward in mass before the occurrence of the fallback; (3) we study the efficiency and extension of the mixing of both the chemical composition and the $\rm ^{56}Ni$ in order to reproduce the transition phase between the plateau and the radioactive tail; also in this case the assumption at the basis of this step is that we expect a substantial degree of mixing during multidimensional simulations. The result of the fitting procedure was that in the LD case we exclude all the progenitors with mass larger than $\rm 13~M_\odot$ and with metallicities  $\rm [Fe/H]\geq 0$ and $\rm [Fe/H]\leq -2$. Note that metallicites $\rm [Fe/H]\leq -2$ are excluded a priori because the metallicity of NGC 1637 has been estimated of the order of $\rm [Fe/H]\sim-0.33$. Therefore we conclude that a non rotating star with mass $\rm M=13~M_\odot$ and with metallicity  $\rm [Fe/H]=-1$ is compatible with the progenitor of SN 1999em (in the LD case). Progenitors with metallicities in the range $\rm 0>[Fe/H]>-1$, or with an initial mass lower than $\rm 13~M_\odot$ cannot be studied because of the coarse grid of presupernova models in both the metallicities and the initial masses. The analysis of the radial velocities shows the existence of a discrepancy between the fit to the light curve and the fit to the photospheric velocity. In particular, the energy required to fit the light curve, and in particular the value of ${\rm Log}L_{30}$, is substantially lower (by about a factor of 2) than the one required to fit the photospheric velocity.
This problem has been already found by other authors, e.g., \cite{utrobin17}, when the adopted presupernova model is the result of the stellar evolution calculations. \cite{paxton+18}, however, have shown that the evaluation of the velocity where the Sobolev optical depth of the Fe {\RomanNumeral{2}} is equal to 1 provides a much better agreement to the observations. In the HD case, the supernova is intrinsically more luminous and therefore the progenitor mass can be as high as $\rm 15~M_\odot$. In particular, we find that models with metallicity [Fe/H]=0, in the mass range $\rm 13-15~M_\odot$, and [Fe/H]=-1, with mass $\rm 15~M_\odot$, are all compatible with the progenitor of SN 1999em. Also in this case we find a discrepancy between the fit to the light curve and the fit to the photospheric velocity. However, due to the higher intrinsic luminosity, such a discrepancy is slightly reduced in this case compared to the LD case.
In both cases (LD and HD) the predicted progenitor mass is compatible with other estimates available in literature based on the preexplosion images of the supernova site \citep{smartt+02} or on an analysis similar to the one described in this paper \citep{utrobin17}.

The results shown in this paper are encouraging and a similar analysis of a more extended set of bolometric light curves will be shortly presented in a companion paper.

\bigskip

We thank Stefano Benetti, Lina Tomasella and Nando Patat for providing the observed bolometric light curve data of SN 1999em and for useful and clarifying discussions. This work has been partially supported by the Italian grants "GRAWITA" (P.I. E. Brocato), “Premiale 2015 MITiC” (P.I. Bianca Garilli) and “Premiale 2015 FIGARO” (P.I. Gianluca Gemme) and by the “ChETEC” COST Action (CA16117), supported by COST (European Cooperation in Science and Technology).

\end{document}